\documentclass[preprint,3p,twocolumn,authoryear]{elsarticle}

\usepackage[position=top]{subfig}

\usepackage{amssymb}
 \usepackage{amsthm}
 \usepackage{amsmath}
 
\usepackage{algorithm}
\usepackage{algpseudocode}

\usepackage{url}

\usepackage{xcolor}

\DeclareMathOperator*{\argmin}{arg\,min}
\DeclareMathOperator{\prox}{prox}

\usepackage{lineno}

\journal{Computerized Medical Imaging and Graphics}

\begin{document}

\begin{frontmatter}



\title{Non-convex primal-dual algorithm for image reconstruction in spectral CT}


\author[1]{Buxin~Chen}

\author[1]{Zheng~Zhang}

\author[1]{Dan~Xia}

\author[1]{Emil~Y.~Sidky}

\author[1,2]{Xiaochuan~Pan\corref{cor1}}
\ead{xpan@uchicago.edu}

\cortext[cor1]{Corresponding author}

\address[1]{Department of Radiology, The University of Chicago, Chicago, IL 60637, USA.}
\address[2]{Department of Radiation and Cellular Oncology, The University of Chicago, Chicago, IL 60637, USA}

%

\begin{abstract}
The work seeks to develop an algorithm for image reconstruction by directly inverting the non-linear data model in spectral CT. Using the non-linear data model, we formulate the image-reconstruction problem as a non-convex optimization program, and develop a non-convex primal-dual (NCPD) algorithm to solve the program. We devise multiple convergence conditions and perform verification studies numerically to demonstrate that the NCPD algorithm can solve the non-convex optimization program and under appropriate data condition, can invert the non-linear data model. Using the NCPD algorithm, we then reconstruct monochromatic images from simulated and real data of numerical and physical phantoms acquired with a standard, full-scan dual-energy configuration. The result of the reconstruction studies shows that the NCPD algorithm can correct accurately for the non-linear beam-hardening effect. Furthermore, we apply the NCPD algorithm to simulated and real data of the numerical and physical phantoms collected with non-standard, short-scan dual-energy configurations, and obtain monochromatic images comparable to those of the standard, full-scan study, thus revealing the potential of the NCPD algorithm for enabling non-standard scanning configurations in spectral CT, where the existing indirect methods are limited.
\end{abstract}




\end{frontmatter}


\section{Introduction}
%
%
%
%
%
%

In CT imaging with polychromatic X-rays, the beam-hardening effect necessarily leads to a non-linear data model~\citep{alvarez_energy-selective_1976}.
Works exist on investigating indirect methods or direct algorithms for image reconstruction in dual-energy and spectral CT based on the non-linear data model.
An indirect reconstruction method first converts non-linear data into linear data from which an image is reconstructed subsequently with an algorithm, such as the filtered-back projection (FBP) algorithm, designed for inverting a linear data model~\citep{alvarez_energy-selective_1976, roessl_k-edge_2007, zou_analysis_2008, zhang_model-based_2014, sawatzky_proximal_2014, rigie_joint_2015}. It generally requires multiple measurements made for each of the X-ray paths with a full set of multiple spectra for performing data conversion.
The measurement condition required by the indirect methods may not be satisfied by non-standard scanning configurations of interest, thus limiting their application to configurations with overlapping rays. Conversely, direct algorithms reconstruct an image directly by the inversion of the non-linear data model in a discrete form\footnote{We simply use terms ``data model'' and ``data model in a discrete form'' interchangeably hereinafter, without causing confusion.}~\citep{cai_full-spectral_2013, long_multi-material_2014, nakada_joint_2015, zhao_extended_2015, barber_algorithm_2016, chen_image_2017, mory2018comparison}, and they may reconstruct images from data measured only with a subset of multiple spectra for some X-ray paths, thus allowing for enabling non-standard configurations~\citep{chen_image_2017, chen2018algorithm}.

The objective of the work is to develop a direct algorithm for image reconstruction by  
formulating it as a solution to a non-convex optimization program involving the non-linear data model in spectral CT. While some of the previous works~\citep{cai_full-spectral_2013, long_multi-material_2014, nakada_joint_2015, zhao_extended_2015} 
have reported the development of direct algorithms for image reconstruction in spectral CT, it is unclear whether the direct algorithms can (mathematically or numerically) accurately solve their non-convex optimization programs and invert the non-linear data model. Meanwhile, there has been significant advancement in solving convex optimization programs, which may be exploited for developing algorithms to solve non-convex optimization programs.

There has also been an effort in adapting existing algorithms to non-convex optimization, such as the inertial proximal algorithm for non-convex optimization~\citep{ochs2014ipiano} and the primal-dual (PD) hybrid gradient method for nonlinear operators~\citep{valkonen2014primal}. It remains to be investigated whether the algorithms can be applied to inverting the non-linear data model in spectral CT. 

We have recently developed iterative algorithms by adapting a first-order primal-dual (PD) algorithm~\citep{chambolle_first-order_2010, sidky_convex_2012} to solve convex optimization programs of different forms relevant to image reconstruction in tomographic imaging systems\citep{zhang2016artifact, xia2016optimization, zhang2016investigation}. We have also been investigating to adapt instances of the PD algorithm to non-convex optimization programs for the application to image reconstruction in spectral CT, including our previous work~\citep{barber_algorithm_2016, barber2016mocca} that involved forming a convex quadratic local bounding function to the non-convex data discrepancy term and adapting instances of the PD algorithm to provide a descent step for that quadratic approximation. 

We continue to seek alternate adaptations of the instances of the PD algorithm to non-convex optimization programs that are simpler in an algorithmic sense~\citep{chen_non-convex_2018}. This work first presents a derivation of a new instance of the PD algorithm, referred to as the convex PD (CPD) algorithm and then an intuitive adaptation of the CPD algorithm, with minimally modified update steps, to a non-convex PD (NCPD) algorithm for a non-convex optimization program. A significant difference
between the NCPD algorithm and our previous work \citep{barber_algorithm_2016, barber2016mocca} is that the step-size parameters are fixed and computed ahead of time in the former, while they have to be calculated at every iteration for the latter.  

The issue of non-convexity in the work is dealt with by ``convexifying" the non-convex optimization program, using a global linear expansion of the non-linear data model about a set of stationary basis images selected. 
This strategy has been applied to adapting the ASD-POCS algorithm~\citep{sidky_image_2008} to a non-convex optimization in spectral CT~\citep{chen_image_2017, chen2018algorithm}. We investigate here the strategy for adapting the new CPD instance of the general PD algorithm to the NCPD algorithm for image reconstruction in spectral CT. Specifically, the NCPD algorithm is based upon the CPD algorithm, thus eliminating issues associated with the ASD-POCS algorithm: (1) it is only a heuristic algorithm for convex optimization programs and (2) it can be applied only to convex optimizations of a specific form with data-$\ell_2$-constrained TV-minimization.

Following the derivation and demonstration of the NCPD algorithm in Sections~\ref{Sec:methods} and~\ref{Sec:methods-1} for solving the non-convex program and under appropriate data condition, for inverting the non-linear data model in spectral CT,  we then perform studies on reconstruction of the monochromatic images by using the NCPD algorithm from full-scan simulated- and real-data in spectral CT in Sections~\ref{sec:simulated-data-recon} and~\ref{sec:real-data-recon} to show accurate correction of the non-linear beam-hardening effect, along with a study in Section~\ref{sec:real-short-data-recon} that demonstrates the potential of the NCPD algorithm for enabling non-standard scanning configurations in spectral CT.

\section{Methods: derivation of the NCPD algorithm}\label{Sec:methods}
In spectral CT, a monochromatic image is of interest  that represents the spatial distribution of X-ray linear-attenuation coefficient at a particular energy within the subject scanned.
Let vector $\mathbf{f}_m$ of size $I$ denote the monochromatic image with element $f_{m i}$ within voxel $i$ at discrete energy bin $m$, where $m\in{1,\cdots,M}$, $i\in{1, \cdots,I}$, and $M$ and $I$ are total numbers of energy bins and voxels, respectively. We use vector $\mathbf{b}_k$ of size $I$ to denote basis image $k$ with element $b_{ki}$ within voxel $i$, $k\in{1,\cdots,K}$, and $K$ is the total number of basis images considered. An aggregated basis image vector $\mathbf{b}$ of size $K_c=K I$ can be formed by concatenating $\mathbf{b}_k$.

We assume that monochromatic image $\mathbf{f}_m$ can be expressed as
\begin{equation}\label{eq:basis-decomposition}
\mathbf{f}_{m}(\mathbf{b})=\sum_k {\mu}_{m k} \, \mathbf{b}_{k},
\end{equation}
where decomposition coefficients ${\mu}_{m k}$ are assumed to be known.
Therefore, the reconstruction task of $\mathbf{f}_{m}$ is tantamount to that of basis image $\mathbf{b}$.

Let $g_{s j}^{[\mathcal{M}]}$ denote measured data, after air-scan normalization and logarithmic transform, for ray $j$ with spectrum $s$, where $j\in{1,\cdots,J_{s}}$, $J_{s}$ is the total number of rays illuminated with spectrum $s$, $s\in{1,\cdots,S}$, and $S$ is the total number of different spectra involved.  Using vector $\mathbf{g}_{s}^{[\mathcal{M}]}$ of size $J_{s}$ to denote measured data for spectrum $s$, we can form aggregated vector $\mathbf{g}^{[\mathcal{M}]}$ of size $J=\sum_s J_{s}$, composed by concatenating $\mathbf{g}_{s}^{[\mathcal{M}]}$ of size $J_{s}$, and simply refer to $\mathbf{g}^{[\mathcal{M}]}$ as measured data. The task of image reconstruction considered in the work is to determine basis image $\mathbf{b}$ from knowledge of  measured data $\mathbf{g}^{[\mathcal{M}]}$.

\subsection{Non-linear data model}

Using Eq.~\eqref{eq:basis-decomposition}, we can write the non-linear data model in a discrete form  in spectral CT as \citep{chen_image_2017}
\begin{equation} \label{eq:g-disc}
    g_{s j}^{[NL]}(\mathbf{b})
    =
    \! - \! \ln \!
    \sum_m \, q_{s j m}
    \exp \!\left(\!
        \!-\sum_k {\mu}_{m k} \, \sum_i a_{s ji} b_{ki}
    \!\right)\!,
\end{equation}
where $g_{s j}^{[NL]}(\mathbf{b})$ denotes the non-linear-model data for ray $j$ with spectrum $s$;  $q_{s jm}$ normalized X-ray spectrum\footnote{For simplicity, we refer to $q_{s jm}$ as X-ray spectrum, even though it should be understood conceptually as the combination of X-ray-source spectrum and detector-spectrum response.} $s$ at energy bin $m$ for ray $j$, i.e., $\sum_m \, q_{s j m}=1$; 
$a_{s j i}$ the discrete X-ray transform specified by ray $j$ and voxel $i$ with spectrum $s$.

As in \citet{zhao_extended_2015, chen_image_2017}, we perform a Taylor-series expansion of Eq. \eqref{eq:g-disc} at $\bar{\mathbf{b}}=\mathbf{0}$: 
\begin{equation} \label{eq:g-disc-Tayor}
    g_{s j}^{[NL]}(\mathbf{b}) = 
    \sum_k \mu_{s j k} \sum_i a_{s ji} b_{ki} + \Delta g_{s j}^{[NL]}(\mathbf{b}),
\end{equation}
where the zeroth-order term is zero at $\bar{\mathbf{b}}=0$, and $\mu_{sjk}$, the spectrum-weighted average attenuation coefficient independent of energy, and $\Delta g_{s j}^{[NL]}(\mathbf{b})$, the high-order, non-linear term in the expansion, are given by
\noindent 
\begin{equation} \label{eq:Q-Tayor-0}
\mu_{s j k}=\sum_m \, q_{s j m} \,\mu_{m k},
\end{equation}
\begin{equation} \label{eq:delta-g-Tayor}
\begin{aligned}
    \Delta g_{s j}^{[NL]}(\mathbf{b}) 
    =  g_{s j}^{[NL]}(\mathbf{b}) -
    \sum_k \mu_{s j k} \sum_i a_{s ji} b_{ki}.
\end{aligned}
\end{equation}

For spectrum $s$, we form vectors $\mathbf{g}_{s}^{[NL]}(\mathbf{b})$ and $\Delta\mathbf{g}_{s}^{[NL]}(\mathbf{b})$ of size $J_s$, respectively, with elements $g_{s j}^{[NL]}(\mathbf{b})$ and $\Delta g_{s j}^{[NL]}(\mathbf{b})$. Subsequently, we can also form aggregated vectors  $\mathbf{g}^{[NL]}(\mathbf{b})$ and $\Delta \mathbf{g}^{[NL]}(\mathbf{b})$ of size $J$, respectively, by concatenating $\mathbf{g}_{s}^{[NL]}(\mathbf{b})$ and $\Delta\mathbf{g}_{s}^{[NL]}(\mathbf{b})$.
With the aggregated vectors defined, we rewrite Eq. \eqref{eq:g-disc} in a matrix-vector form as
\begin{equation} \label{eq:g-NL-vec}
     \mathbf{g}^{[NL]}(\mathbf{b})
    =\mathcal{H}\, \mathbf{b}  + \Delta  \mathbf{g}^{[NL]}(\mathbf{b}),
\end{equation}
where matrix $\mathcal{H}$ of size $(J\times K_c)$ is given by
\begin{equation} \label{eq:H}
\mathcal{H} =
    \begin{pmatrix}
		\mathcal{R}_{11 } \, \mathcal{A}_1 &
        \mathcal{R}_{12}\, \mathcal{A}_1 &
        \cdots & \mathcal{R}_{1 K} \, \mathcal{A}_1  \\
        \mathcal{R}_{2 1}\, \mathcal{A}_2  &
        \mathcal{R}_{2 2}\, \mathcal{A}_2  &
        \cdots & \mathcal{R}_{2 K} \, \mathcal{A}_2  \\
        \vdots & \vdots & \ddots \\
        \mathcal{R}_{S 1}\, \mathcal{A}_{S} &
        \mathcal{R}_{S 2}\, \mathcal{A}_{S} &
        \cdots & \mathcal{R}_{S K}\, \mathcal{A}_{S}
        \end{pmatrix},
\end{equation}
$\mathcal{R}_{s k}$ is a diagonal matrix of size $(J_{s}\times J_{s})$ with diagonal element $\mu_{s j k}$,
and $\mathcal{A}_{s}$ denotes a matrix of size $(J_{s}\times I)$ with element $a_{s j i}$.

The objective of the work is to develop the NCPD algorithm for reconstructing basis image $\mathbf{b}$, and subsequently monochromatic image $\mathbf{f}_{m}(\mathbf{b})$, from measured data $\mathbf{g}^{[\mathcal{M}]}$ by inversion of the non-linear data model in Eq. \eqref{eq:g-NL-vec}.

\subsection{Non-convex optimization program}
Using the non-linear data model in Eq.~\eqref{eq:g-NL-vec}, we formulate the image reconstruction problem in spectral CT as an optimization program as 
 \vspace{-0.0cm} 
\begin{equation} \label{eq:opt-constr-NL}
\begin{aligned}
 	{\mathbf{b}}^\star &= \argmin_\mathbf{b}
    D_{\mathbf{g}^{[NL]}}(\mathbf{b}) \,\, \\
    &\mathrm{s.\,t.} \,\,
   || ( |\nabla \mathbf{f}_{m'}(\mathbf{b}) | )||_1 \leq \!\gamma \,\, {\&}
   \,\,  \mathbf{f}_{m'}(\mathbf{b}) \succeq 0,
\end{aligned}
\end{equation}
where data term $D_{\mathbf{g}^{[NL]}}(\mathbf{b})$ is given by 
\begin{equation}\label{eq:data-div-NL}
\begin{aligned}
D_{\mathbf{g}^{[NL]}}(\mathbf{b}) & \!=\!  || \mathbf{g}^{[\mathcal{M}]}  - \mathbf{g}^{[NL]}(\mathbf{b})  ||_2^2/2 \\
&\! =\!  ||  (\mathbf{g}^{[\mathcal{M}]} - \Delta  \mathbf{g}^{[NL]}(\mathbf{b}) )- \mathcal{H}\, \mathbf{b}  ||_2^2/2,
\end{aligned}
\end{equation}
$|| ( |\nabla \mathbf{f}_{m'}(\mathbf{b}) | )||_1$ 
the total variation (TV) of monochromatic image $\mathbf{f}_{m'}$, $\nabla$ a matrix of size $3I\times I$ for spatial finite difference, $\succeq \mathbf{0}$  a vector-element-wise non-negativity constraint of monochromatic image, and $\gamma$ a non-negative parameter.

The optimization program in Eq.~\eqref{eq:opt-constr-NL} is similar in form to the ones used in our previous work~\citep{zhang2016artifact, xia2016optimization, zhang2016investigation}, however, the TV and non-negativity constraints are applied to the monochromatic image, which is a linear combination of the basis images, instead of the basis images themselves, at a given energy. Such constraints are used because the monochromatic image directly represents physical quantity, i.e., the X-ray linear attenuation, of application interest. In the numerical studies of the work, both the TV and non-negativity constraints are enforced on a monochromatic image at an energy of 100 keV.

Because  the Hessian of data term $D_{\mathbf{g}^{[NL]}}(\mathbf{b})$ can be shown to have negative eigenvalues, $D_{\mathbf{g}^{[NL]}}(\mathbf{b})$ and consequently the optimization program in  Eq. \eqref{eq:opt-constr-NL} are non-convex. Basing on the CPD algorithm developed in~\ref{sec:app-a} and~\ref{sec:app-b}, we derive below the NCPD algorithm for solving the non-convex optimization program in  Eq. \eqref{eq:opt-constr-NL}.

\subsection{The NCPD algorithm}\label{sec:NCPD-algo}
The NCPD algorithm is developed heuristically below because no algorithms exist, to the best of our knowledge, that solve mathematically exactly the non-convex program in Eq. \eqref{eq:opt-constr-NL}. In particular, the NCPD algorithm is adapted from the CPD algorithm presented in~\ref{sec:app-a} and~\ref{sec:app-b} for solving a convex program in Eq. \eqref{eq:opt-constr-ln} involving the linear data model in Eq. \eqref{eq:linear-data-model}. It can be observed that Eqs. \eqref{eq:opt-constr-NL} and \eqref{eq:opt-constr-ln}  are identical in form except for the difference between terms  $\Delta  \mathbf{g}^{[NL]}(\mathbf{b})$ in Eq. \eqref{eq:g-NL-vec} and $\Delta \mathbf{g}_c$ in Eq. \eqref{eq:linear-data-model}. It is the similarity in form of the two programs that motivates the adaptation of the NCPD algorithm from the mathematically exact and numerically accurate CPD algorithm derived.

It can be observed that if $\Delta  \mathbf{g}^{[NL]}(\mathbf{b})$ is replaced with constant term $\Delta\mathbf{g}_c$, Eq. \eqref{eq:opt-constr-NL} becomes convex and can thus be solved by use of the CPD algorithm in~\ref{sec:app-a} and~\ref{sec:app-b}.  Motivated by this observation, we devise below the heuristic NCPD algorithm for numerically solving  Eq. \eqref{eq:opt-constr-NL}
by using $\Delta  \mathbf{g}^{[NL]}(\mathbf{b}^{(n)})$, computed with $\mathbf{b}^{(n)}$ at current iteration in Eq. \eqref{eq:delta-g-Tayor}, as an estimate of $\Delta\mathbf{g}_c$ in line 6 of Algorithm 1 in~\ref{sec:app-a}, i.e.,
\begin{equation}\label{eq:CPD-line-6-change}
6: \mathbf{p}^{(n+1)} = 
	\frac{\mathbf{p}^{(n)} \!-\!
\sigma_{m'}[ \mathbf{g}^{[\mathcal{M}]}
\!-\! \Delta\mathbf{g}^{[NL]}(\mathbf{b}^{(n)}) \!-\! \mathcal{H} \bar{\mathbf{b}}^{(n)}]} {1+\sigma_{m'}}.
\end{equation}

We thus obtain {\bf the NCPD algorithm with pseudo codes identical to those listed in Algorithm 1 in~\ref{sec:app-a} except that line 6 is now given by Eq. \eqref{eq:CPD-line-6-change} above.}
The heuristic component in the NCPD algorithm is thus  limited only to Eq. \eqref{eq:CPD-line-6-change} because the CPD algorithm is based upon a mathematically rigorous footing.

As a part of the NCPD algorithm development, we devise below convergence conditions, which are used for numerically demonstrating that the NCPD algorithm can solve the non-convex optimization in Eq. \eqref{eq:opt-constr-NL} and that under appropriate data condition, can invert the non-linear data model in Eq. \eqref{eq:g-NL-vec}.

\subsection{Convergence conditions of the NCPD algorithm} 
It should be pointed out that different convergence conditions, as discussed below, can be designed, depending on whether measured data are consistent (i.e., $\mathbf{g}^{[\mathcal{M}]}=\mathbf{g}^{[NL]}(\mathbf{b}^{[\rm truth]})$)
 or inconsistent (i.e., $\mathbf{g}^{[\mathcal{M}]}\neq \mathbf{g}^{[NL]}(\mathbf{b}^{[\rm truth]})$) with the data model. 

\subsubsection{General convergence conditions}
For measured data that are either consistent or inconsistent with the data model, we first devise general convergence conditions of the NCPD algorithm as
\begin{equation}\label{eq:conv-conds-nlinear-2}
	d\widetilde{D}_{\mathbf{g}}^{(n)} \rightarrow 0
   	\quad
	\widetilde{D}_\text{TV}^{(n)}\rightarrow 0
	\quad
	d\widetilde{D}_\mathbf{b}^{(n)} \rightarrow 0,
\end{equation}
as $n\rightarrow \infty$, where the dimensionless metrics are given by 
\begin{equation}\label{eq:Dg-Dtv-Df-metrics-NL}
\begin{aligned}
  	d\widetilde{D}_{\mathbf{g}}^{(n)}
    		& ={\vert D_{\mathbf{g}^{[NL]}}(\mathbf{b}^{(n)})-D_{\mathbf{g}^{[NL]}}(\mathbf{b}^{(n-1)})\vert}/{||\mathbf{g}^{[\mathcal{M}]}||_2}\\
    \widetilde{D}_\text{TV}^{(n)}
		&= {|\,|| ( |\nabla \mathbf{f}_{m'}(\mathbf{b}^{(n)}) | )||_1-\gamma|}/{\gamma}\\
	d\widetilde{D}_\mathbf{b}^{(n)}
		&= {||\mathbf{b}^{(n)}-\mathbf{b}^{(n-1)}||_2}/{||\mathbf{b}^{(n-1)}||_2}.
\end{aligned}
\end{equation}
It can be observed that the convergence conditions in Eq. \eqref{eq:conv-conds-nlinear-2} are similar to those in Eq. \eqref{eq:conv-conds-linear-2} of the CPD algorithm in~\ref{sec:app-a}.  

As shown in Eq. \eqref{eq:conv-conds-linear-1} in~\ref{sec:app-a}, the CPD algorithm also has mathematical convergence conditions as 
\begin{equation}\label{eq:conv-conds-nlinear-1}
	\widetilde{\text{cPD}}^{(n)} \rightarrow 0
	\quad  
	\widetilde{\text{T}}^{(n)}\rightarrow 0
	\quad
	\widetilde{\text{S}}^{(n)}\rightarrow 0,
\end{equation}
where
dimensionless metrics
are given by
\begin{equation}\label{eq:normal-PD-T-S-metrics}
\begin{aligned}
 	\hskip -.3cm
 	\widetilde{\text{cPD}}^{(n)}=\frac{{\text{cPD}}^{(n)}}{{\text{cPD}}^{(1)}},
\,\,\,
	\widetilde{\text{T}}^{(n)}=\frac{{\text{T}}^{(n)}}{{\text{T}}^{(1)}},
	\,\,\,
	{\rm and}
\,\,\,
	\widetilde{\text{S}}^{(n)}=\frac{{\text{S}}^{(n)}}{{\text{S}}^{(1)}},
\end{aligned}
\end{equation}
with conditional primal-dual (cPD) gap ${\text{cPD}}^{(n)}$~\citep{xia2016optimization},  transversality ${\text{T}}^{(n)}$~\citep{hiriart2013convex}, and dual residual (or splitting gap)  ${\text{S}}^{(n)}$~\citep{goldstein2013adaptive} defined in Eq.~\eqref{eq:PD-T-S-metrics} in~\ref{sec:app-a}. 

While the metrics can still be computed at each iteration of the NCPD algorithm, it is unclear whether, due to the heuristic step in Eq. \eqref{eq:CPD-line-6-change}, the conditions in Eq. \eqref{eq:conv-conds-nlinear-1} remain mathematical convergence conditions of the NCPD algorithm. Noticing that the NCPD algorithm becomes the CPD algorithm as $\Delta\mathbf{g}^{[NL]}(\mathbf{b}^{(n)})$ stabilizes to becoming constant,  we use them instead as {\it empirical convergence conditions} for the NCPD algorithm. 

\subsubsection{Data-convergence condition}
For measured data consistent with the non-linear data model, i.e., $\mathbf{g}^{[\mathcal{M}]}=\mathbf{g}^{[NL]}(\mathbf{b}^{[\rm truth]})$, we can devise an additional convergence condition as 
\begin{equation}\label{eq:conv-conds-nlinear-3}
\begin{aligned}
	\widetilde{D}_{\mathbf{g}}^{(n)}
		& ={D_{\mathbf{g}^{[NL]}}(\mathbf{b}^{(n)})}/{||\mathbf{g}^{[\mathcal{M}]}||_2}
		\rightarrow 0,
\end{aligned}
\end{equation}
as $n\rightarrow \infty$.
Clearly, for a case in which measured data are inconsistent with the data model, i.e., $\mathbf{g}^{[\mathcal{M}]}\neq\mathbf{g}^{[NL]}(\mathbf{b}^{[\rm truth]})$, this condition cannot be used because data term $D_{\mathbf{g}^{[NL]}}(\mathbf{b}^{(n)})$ is generally non-zero.

\subsubsection{Image-convergence condition}\label{sec:image-conv-condition}
The convergence conditions discussed above are used for demonstrating that the NCPD algorithm can numerically accurately solve the non-convex optimization program in  Eq. \eqref{eq:opt-constr-NL}. However, it remains to be demonstrated that the NCPD algorithm can numerically accurately invert the non-linear data model in Eq. \eqref{eq:g-NL-vec}. The demonstration is performed under the conditions that $\mathbf{g}^{[\mathcal{M}]}= \mathbf{g}^{[NL]}(\mathbf{b})$ and that $\mathbf{b}^\mathrm{[truth]}$ is known, as the NCPD algorithm is developed based upon the non-linear data model. Therefore, we devise a unique image-convergence condition directly to check on the accuracy of recovering $\mathbf{b}^\mathrm{[truth]}$ as 
\begin{equation}\label{eq:conv-conds-nlinear-image}
\begin{aligned}
   	\widetilde{D}_\mathbf{b}^{(n)}
		&= {||\mathbf{b}^{(n)}-\mathbf{b}^{\text{[truth]}}||_2}/{||\mathbf{b}^{[\text{truth}]}||_2}
		\rightarrow 0,
\end{aligned}
\end{equation}
as $n\rightarrow \infty$.

In all of the numerical studies in Sections~\ref{Sec:methods-1}-\ref{sec:real-short-data-recon} below, we obtain the results when the respective convergence conditions are achieved up to four-byte floating-point precision and simply state that the convergence conditions are achieved.

\section{Methods: numerical demonstration of the NCPD algorithm's Convergence}\label{Sec:methods-1}
We conduct quantitative studies below numerically to demonstrate that the NCPD algorithm can solve the non-convex optimization program in Eq. \eqref{eq:opt-constr-NL} and under appropriate data condition, can invert Eq. \eqref{eq:g-NL-vec}, in terms of achieving the respective convergence conditions.

A circular fan-beam scanning configuration is considered that has source-to-center-of-detector and source-to-center-of-rotation distances of 1500 mm and 1000 mm, along with a 400-mm linear detector consisting of 256 1.56-mm detector bins. We obtain $\mu_{m k}$ from the NIST database~\citep{hubbell_tables_2004} for water and bone basis materials, and create two X-ray spectra of 80 kVp and 140 kVp by using the TASMICS worksheet~\citep{hernandez_tungsten_2014}.   

\begin{figure}[htb]
 	\centering
	\subfloat[Water truth]{
 		\centering
 		\includegraphics[width=.12\textwidth]{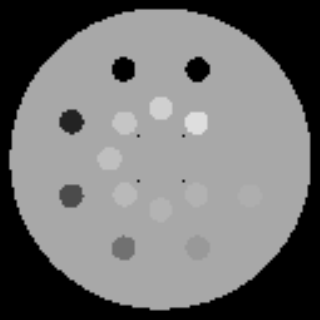}
 		\label{fig:veri-b0}
		}~\hspace{-10pt}~
	\subfloat[Water recon.]{
 		\centering
		\includegraphics[width=.12\textwidth]{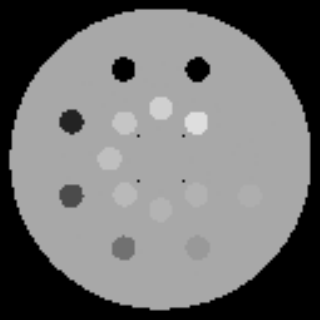}
 		\label{fig:veri-b0}
		}~\hspace{-10pt}~
	\subfloat[Bone truth]{
		\centering
		\includegraphics[width=.12\textwidth]{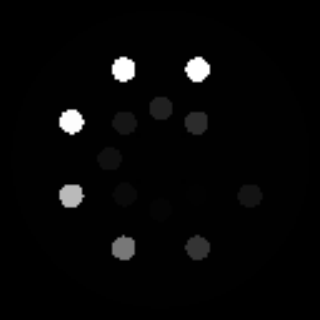}
 		\label{fig:veri-b1}
		}~\hspace{-10pt}~
	\subfloat[Bone recon.]{
		\centering
 		\includegraphics[width=.12\textwidth]{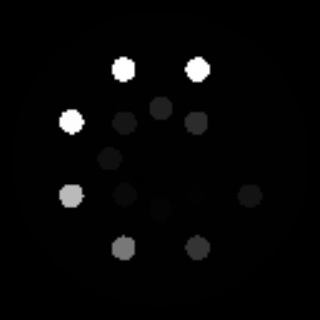}
 		\label{fig:veri-b1}
		}
 	\caption{(a) Truth water $\&$ (c) truth bone basis images, and (b) water $\&$ (d) bone basis images reconstructed by use of the NCPD algorithm, of the disk phantom. Display window: [0.0, 1.5].}
 	\label{fig:veri-truth-image}
 \end{figure}

In the studies, data are generated from a numerical disk phantom with water and bone basis images (i.e., the truth basis images) of size 128$\times$128, shown in columns (a) and (c) in Fig. \ref{fig:veri-truth-image}. Because truth basis image $\mathbf{b}^{[\rm truth]}$ is known, truth-monochromatic image $\mathbf{f}_{m'}^{[\rm truth]}$, and its TV (i.e., truth-TV) value, can thus be calculated from the truth basis image.

\subsection{Numerical demonstration of the NCPD algorithm to solve the non-convex optimization program}\label{sec:NCPD-solving}

We numerically demonstrate below that the NCPD algorithm can solve the non-convex optimization program in Eq. \eqref{eq:opt-constr-NL} for data that are either consistent or inconsistent with the non-linear data model in Eq. \eqref{eq:g-NL-vec}.  

\subsubsection{Numerical demonstration with consistent data}\label{sec:convergence-verification-study}

Using the fan-beam scanning configuration and numerical disk phantom described above in the non-linear data model in Eq. \eqref{eq:g-NL-vec}, we generate measured data  $\mathbf{g}^{[\mathcal{M}]}$ at 160 views uniformly distributed over 2$\pi$, and refer to the data as the full-scan consistent data, i.e., $\mathbf{g}^{[\mathcal{M}]}= \mathbf{g}^{[NL]}(\mathbf{b})$.

\begin{figure}[htb]
 	\centering
    \subfloat{
 		\centering
 		\includegraphics[width=.15\textwidth, trim={15 15 12 12}, clip]{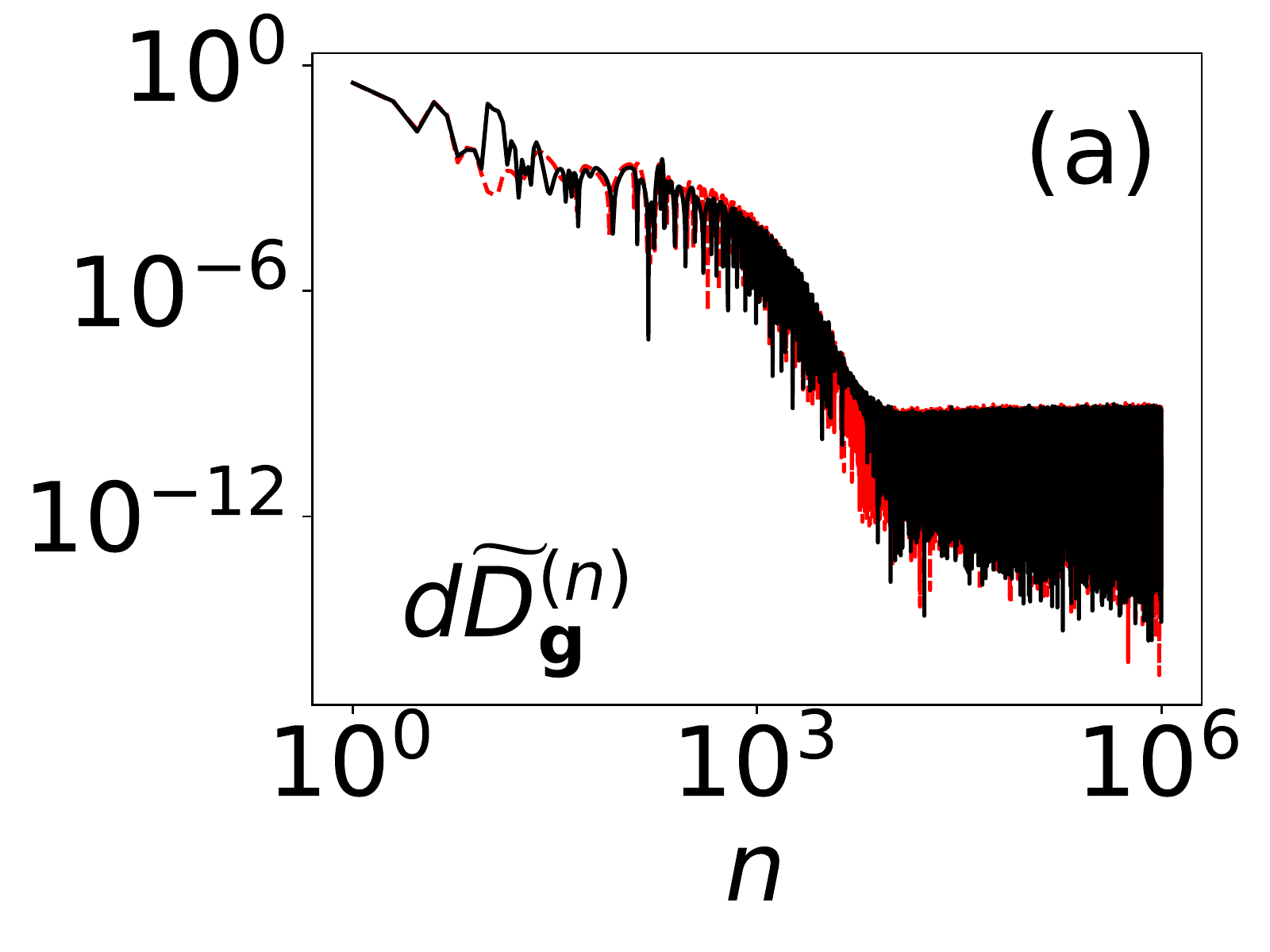}
 		\label{fig:veri-NCPD-data-der}
		}%
	\subfloat{
		\centering
 		\includegraphics[width=.15\textwidth, trim={15 15 12 12}, clip]{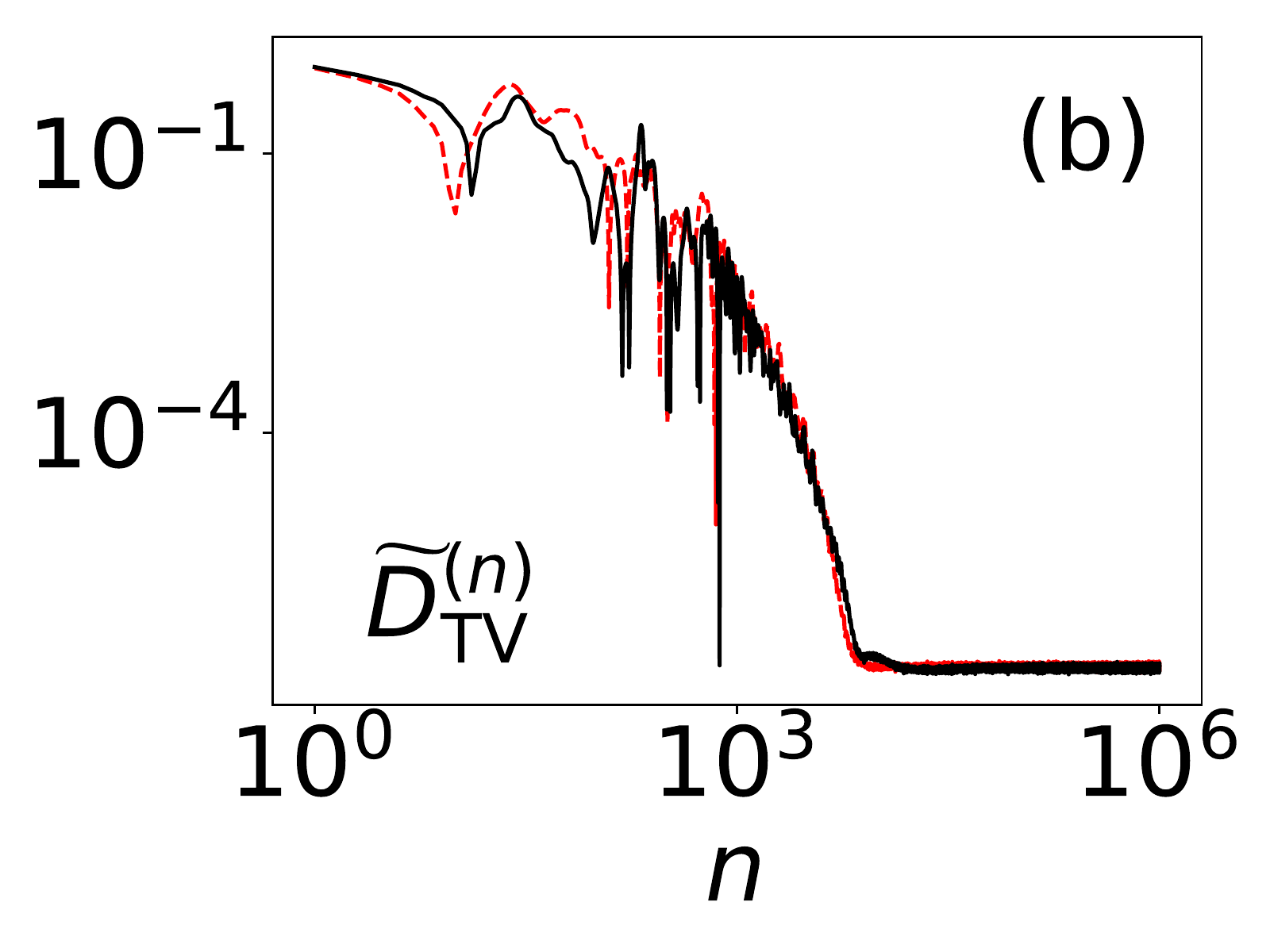}
 		\label{fig:veri-NCPD-tv}
		}
	\subfloat{
 		\centering
 		\includegraphics[width=.15\textwidth, trim={15 15 12 0}, clip]{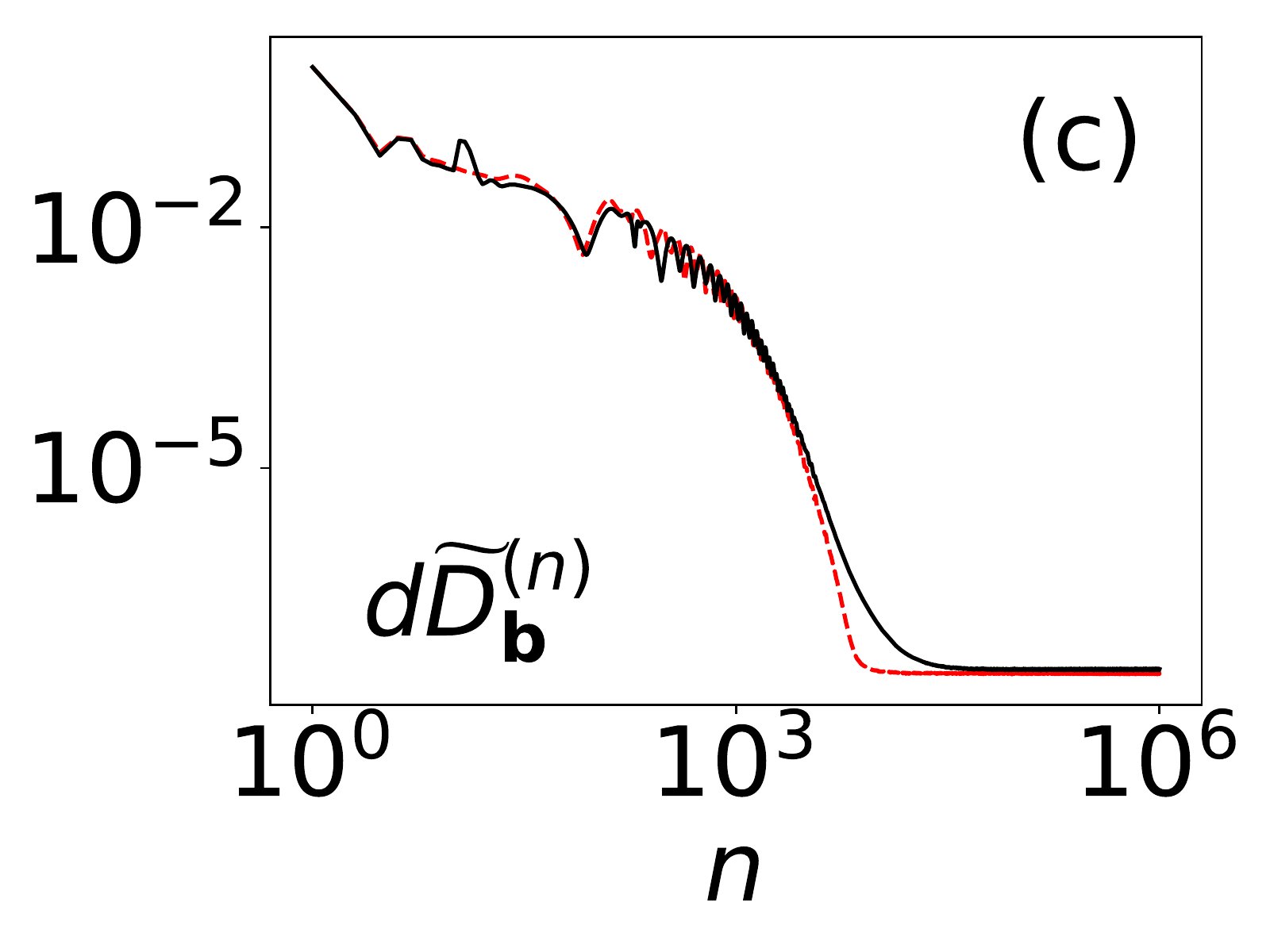}
 		\label{fig:veri-NCPD-img-der}
		}
	\\\vspace{-5pt}
	\subfloat{
 		\centering
 		\includegraphics[width=.15\textwidth, trim={15 15 12 12}, clip]{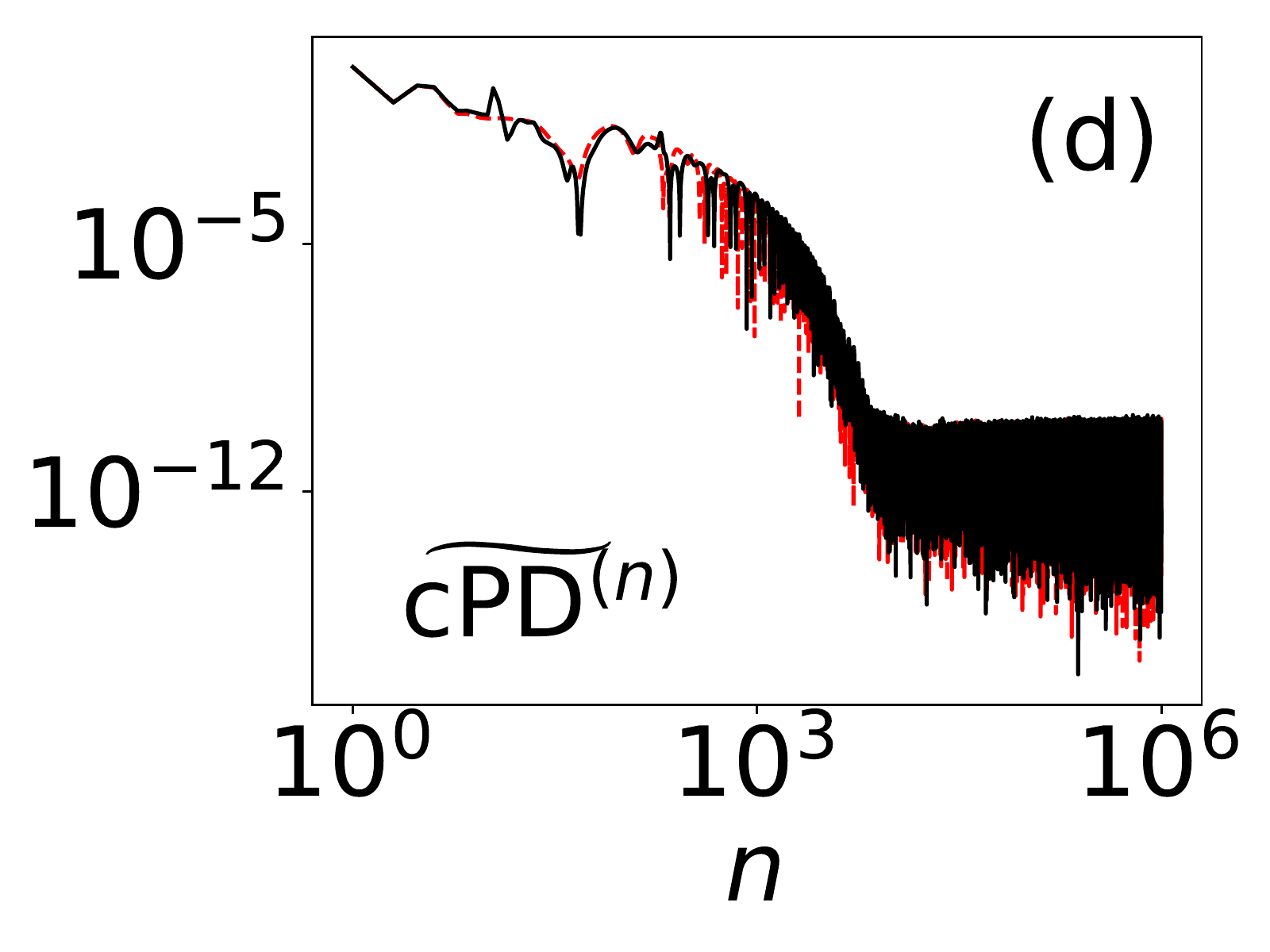}
 		\label{fig:veri-NCPD-cpd}
		}
	\subfloat{
 		\centering
 		\includegraphics[width=.15\textwidth, trim={15 15 12 12}, clip]{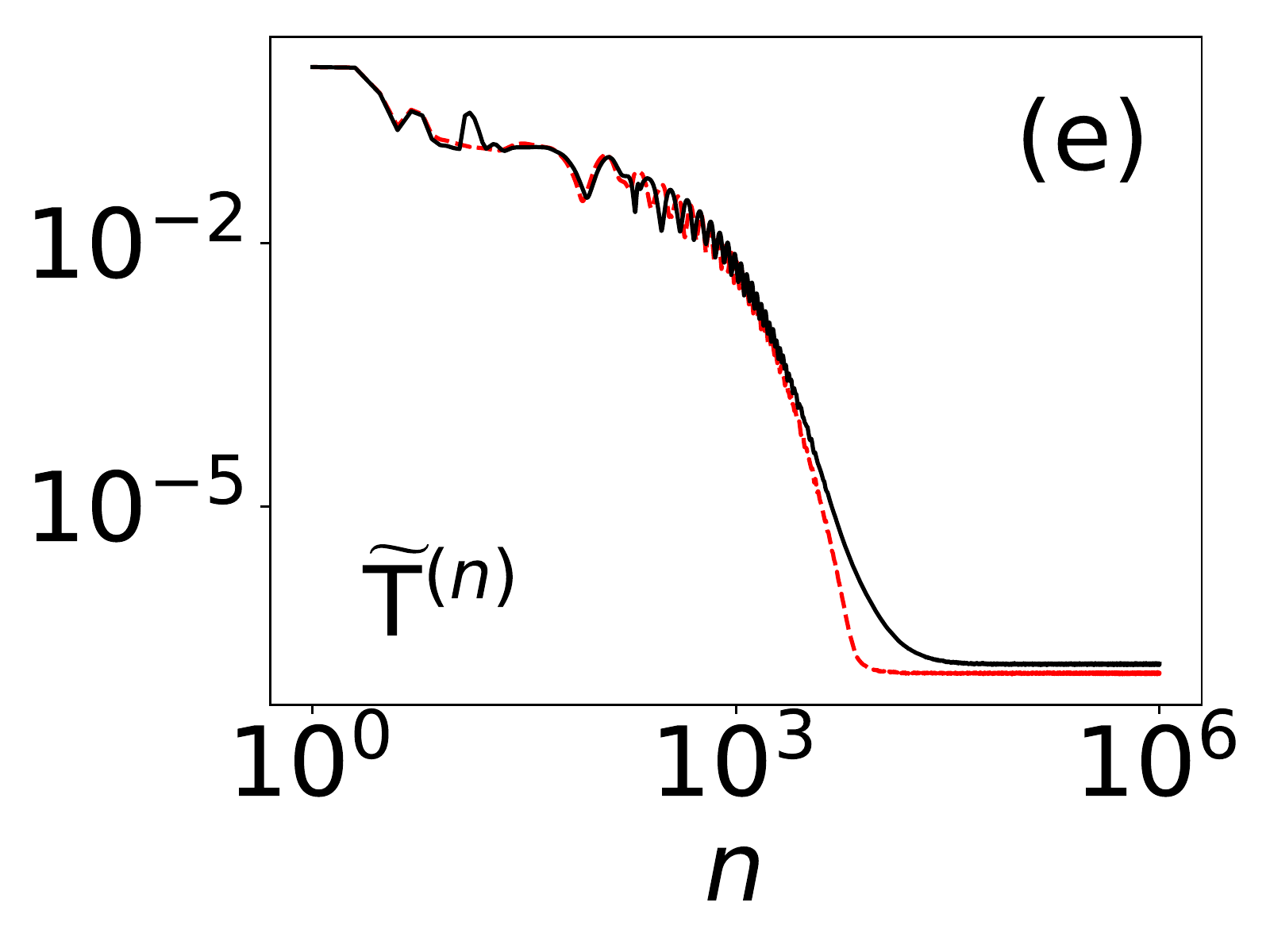}
 		\label{fig:veri-NCPD-transversality}
		}%
    \subfloat{
		\centering
 		\includegraphics[width=.15\textwidth, trim={15 15 12 12}, clip]{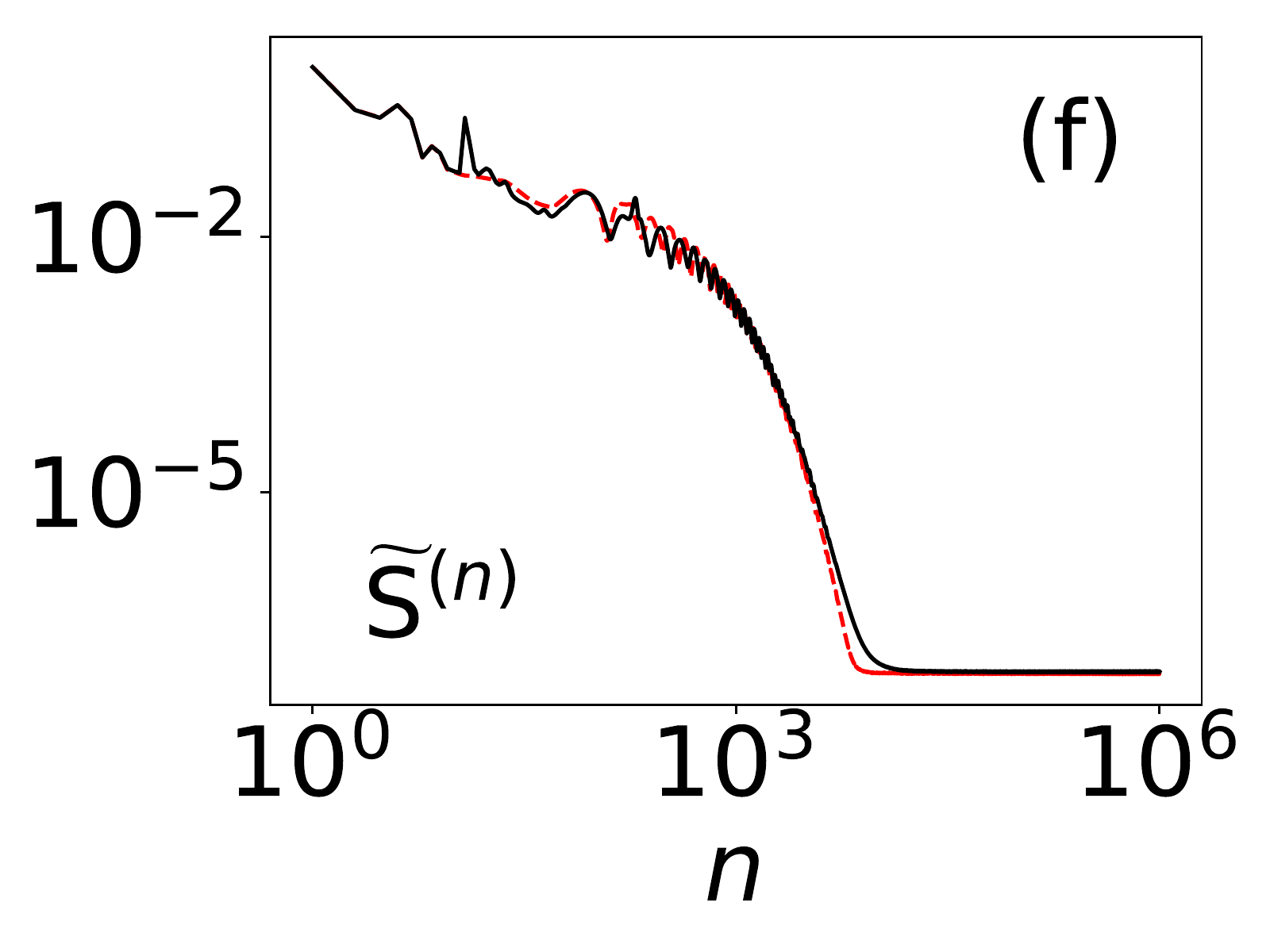}
 		\label{fig:veri-NCPD-splitting}
		}
		\\\vspace{-5pt}
		\subfloat{
 		\centering
 		\includegraphics[width=.15\textwidth, trim={15 15 12 12}, clip]{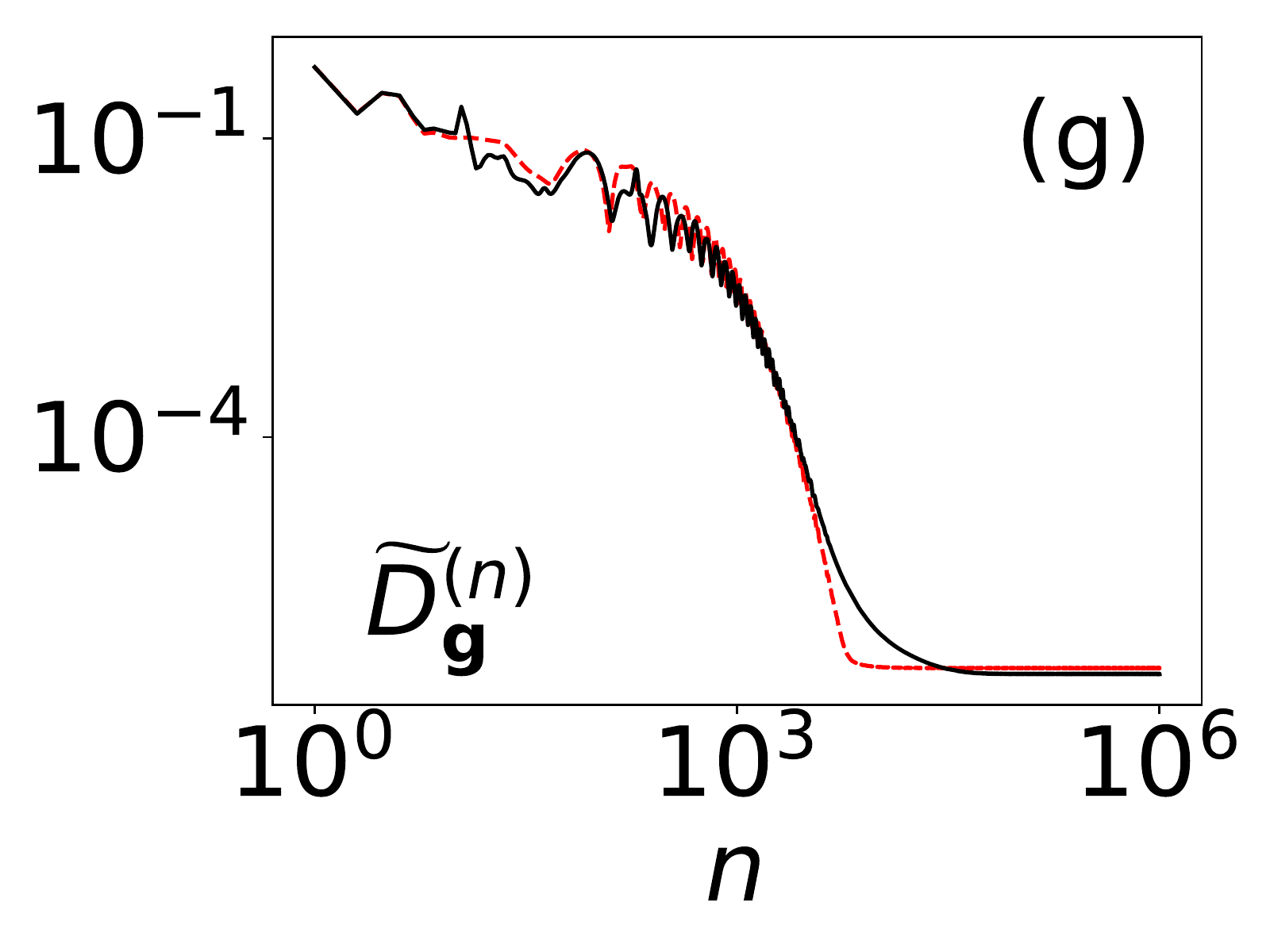}
 		\label{fig:veri-NCPD-data}
		}%
	\subfloat{
 		\centering
 		\includegraphics[width=.15\textwidth, trim={15 15 12 12}, clip]{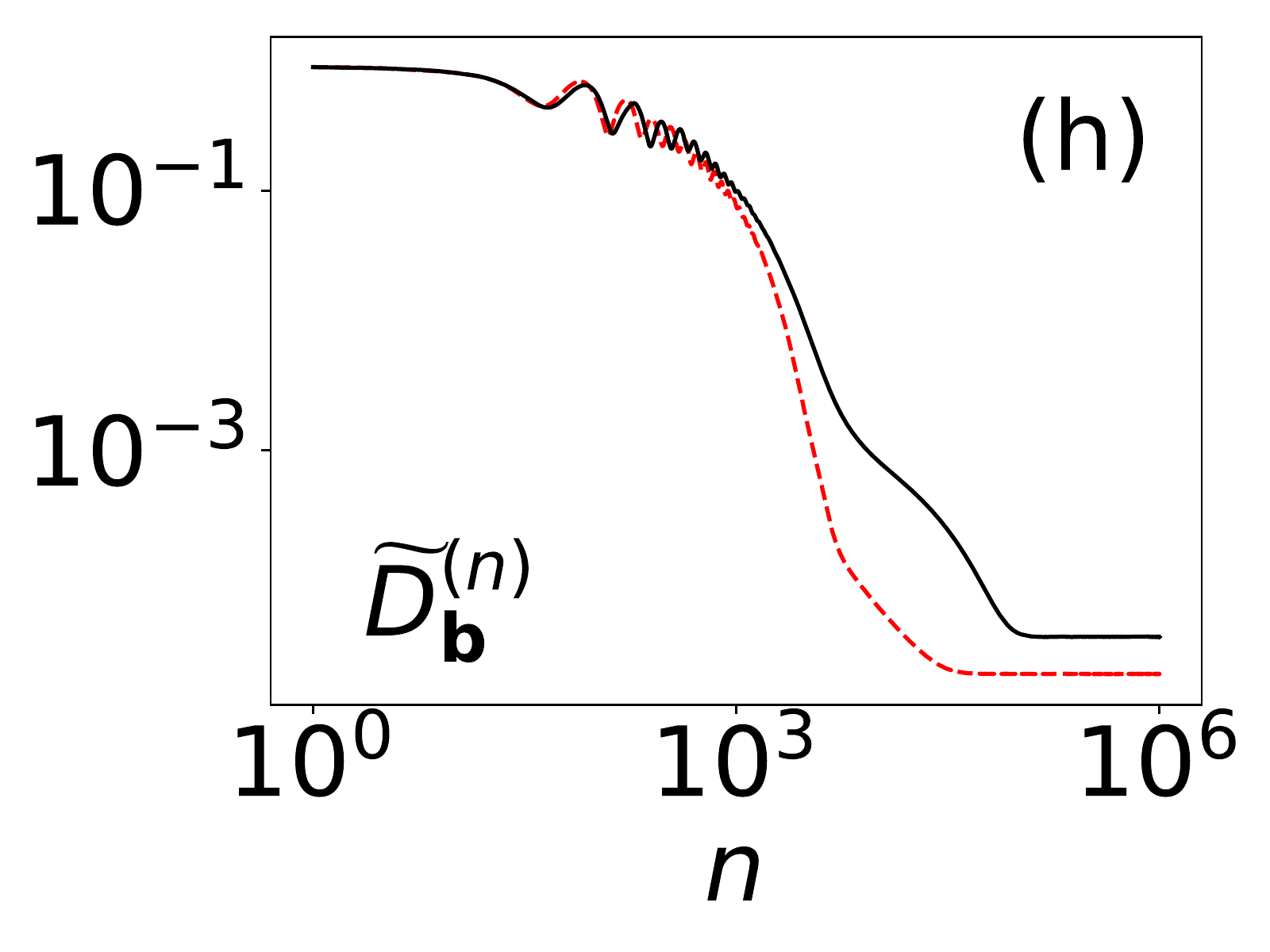}
 		\label{fig:veri-NCPD-img}
		}
 	\caption{
	Convergence metrics
	(a) $d\widetilde{D}_{\mathbf{g}}^{(n)}$,
	(b) $\widetilde{D}_\mathrm{TV}^{(n)}$,
	(c) $d\widetilde{D}_\mathbf{b}^{(n)}$,
	(g) $\widetilde{D}_{\mathbf{g}}^{(n)}$,
	and (h) $\widetilde{D}_\mathbf{b}^{(n)}$,
       and empirical convergence metrics (d) $\widetilde{\text{cPD}}^{(n)}$,
	(e) $\widetilde{\text{T}}^{(n)}$,
	and (f) $\widetilde{\text{S}}^{(n)}$,
	of the NCPD algorithm (solid, dark), as functions of iteration $n$, from full-scan consistent data.
	For comparison, their counterparts for the CPD algorithm (dashed, red) from Fig.~\ref{fig:veri-PD} are also over-plotted.}
 	\label{fig:veri-NCPD}
 \end{figure}

Applying the NCPD algorithm, along with $\gamma$ that is the truth-TV value, to the full-scan consistent data, we compute the convergence conditions in Eqs. \eqref{eq:conv-conds-nlinear-2} and \eqref{eq:conv-conds-nlinear-3} from images reconstructed and plot them as functions of iteration $n$ in panels (a)-(c) and (g) of Fig.~\ref{fig:veri-NCPD} (solid, dark). It can be observed that the convergence metrics consistently decrease, revealing that the NCPD algorithm can solve the non-convex optimization program in Eq. \eqref{eq:opt-constr-NL} by numerically achieving the convergence conditions in Eqs. \eqref{eq:conv-conds-nlinear-2} and \eqref{eq:conv-conds-nlinear-3}. 

We also compute and show in panels (d)-(f) of Fig.~\ref{fig:veri-NCPD} (solid, dark) empirical convergence conditions 
in Eq. \eqref{eq:conv-conds-nlinear-1}.
Considering the NCPD algorithm is adapted from the CPD algorithm, we are interested in comparing their convergence properties. As such, we also inlay in Fig.~\ref{fig:veri-NCPD} the corresponding convergence results (dashed, red) of the CPD algorithm from Fig.~\ref{fig:veri-PD} in~\ref{sec:app-b}. It can be observed that the convergence property of the NCPD algorithm is analogous to that of the CPD algorithm. Therefore, the convergence property of the CPD algorithm provides a nice benchmark for that of the NCPD algorithm.

\subsubsection{Numerical demonstration with inconsistent data}\label{sec:convergence-characterization-study}
We now show that the NCPD algorithm can also solve the non-convex optimization program in Eq. \eqref{eq:opt-constr-NL} for inconsistent data. 
We generate full-scan inconsistent data $\mathbf{g}^{[\mathcal{M}]}$ by adding Poisson noise, corresponding to $10^7$ photons per detector bin in an air scan, to full-scan consistent data $\mathbf{g}^{[NL]}(\mathbf{b}^{[\rm truth]})$ above. Therefore, $\mathbf{g}^{[\mathcal{M}]}\neq \mathbf{g}^{[NL]}(\mathbf{b})$, and Eq. \eqref{eq:conv-conds-nlinear-3} cannot be used as a convergence condition as $\widetilde{D}_{\mathbf{g}}^{(n)}$ does not approach to zero as $n\rightarrow \infty$.

\begin{figure}[htb]
 	\centering
    \subfloat{
 		\centering
 		\includegraphics[width=.15\textwidth, trim={15 15 12 2}, clip]{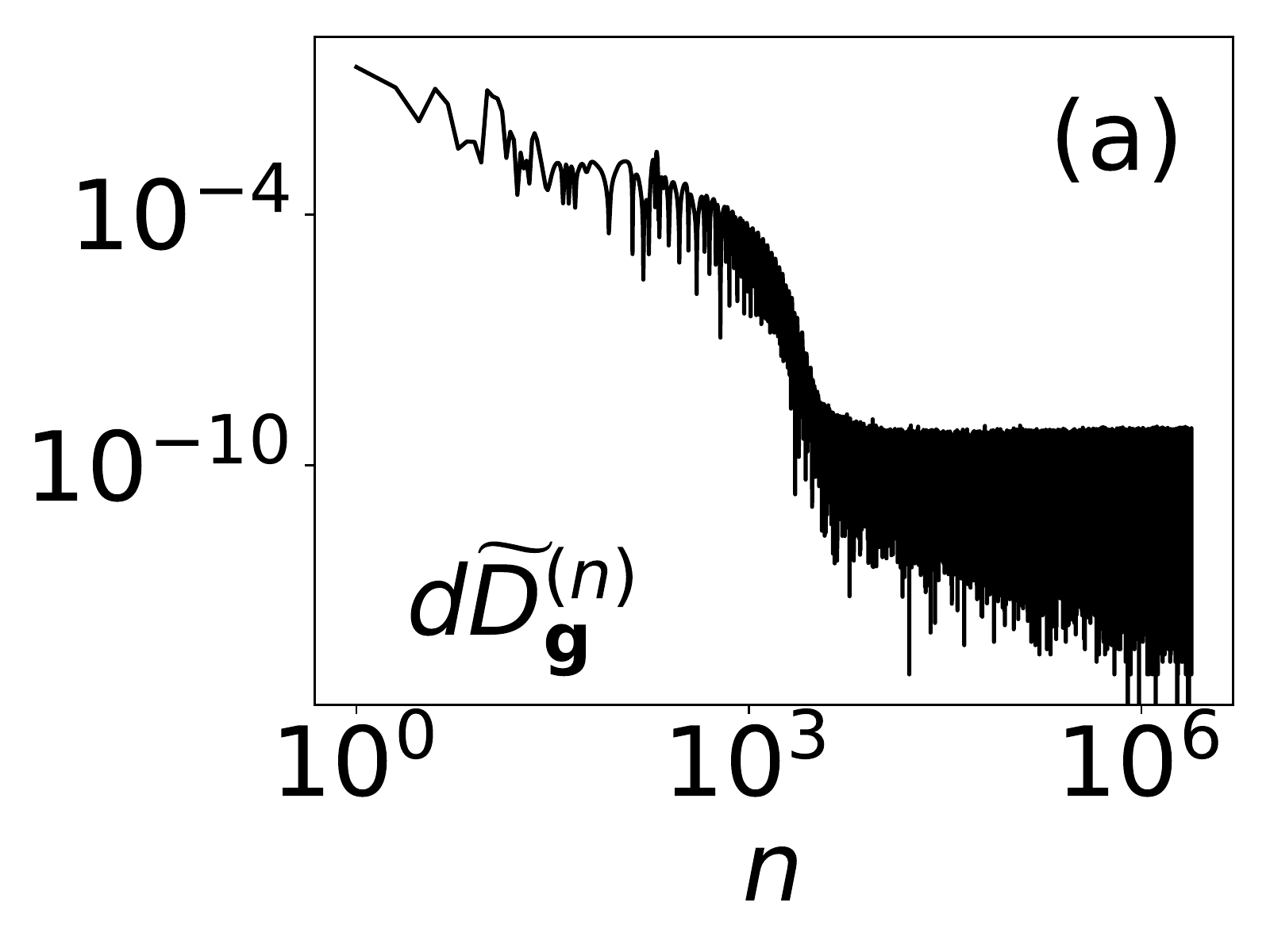}
 		\label{fig:char-noisy-data}
		}%
	\subfloat{
		\centering
 		\includegraphics[width=.15\textwidth, trim={15 15 12 0}, clip]{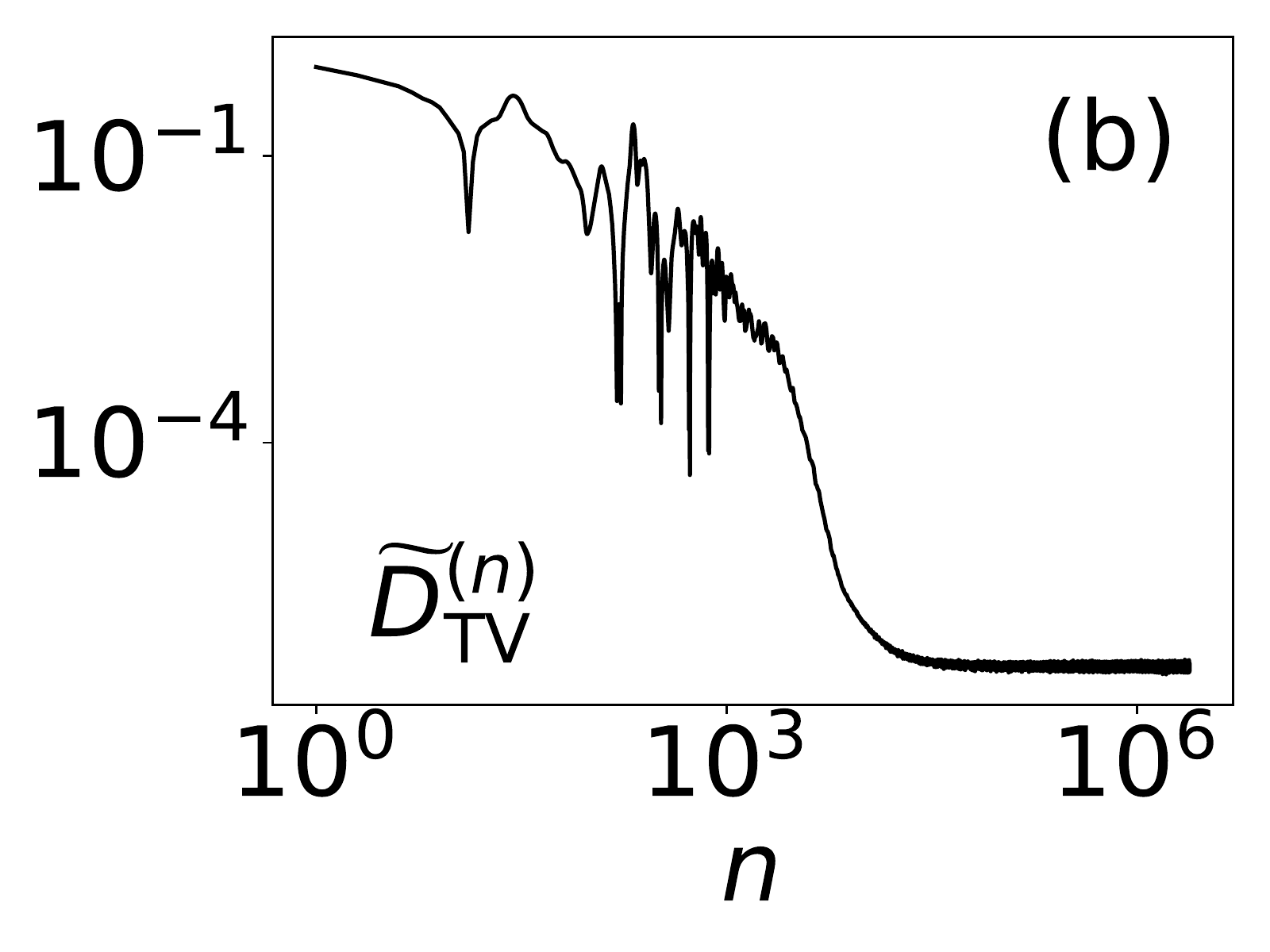}
 		\label{fig:char-noisy-tv}
		}
	\subfloat{
 		\centering
 		\includegraphics[width=.15\textwidth, trim={15 15 12 2}, clip]{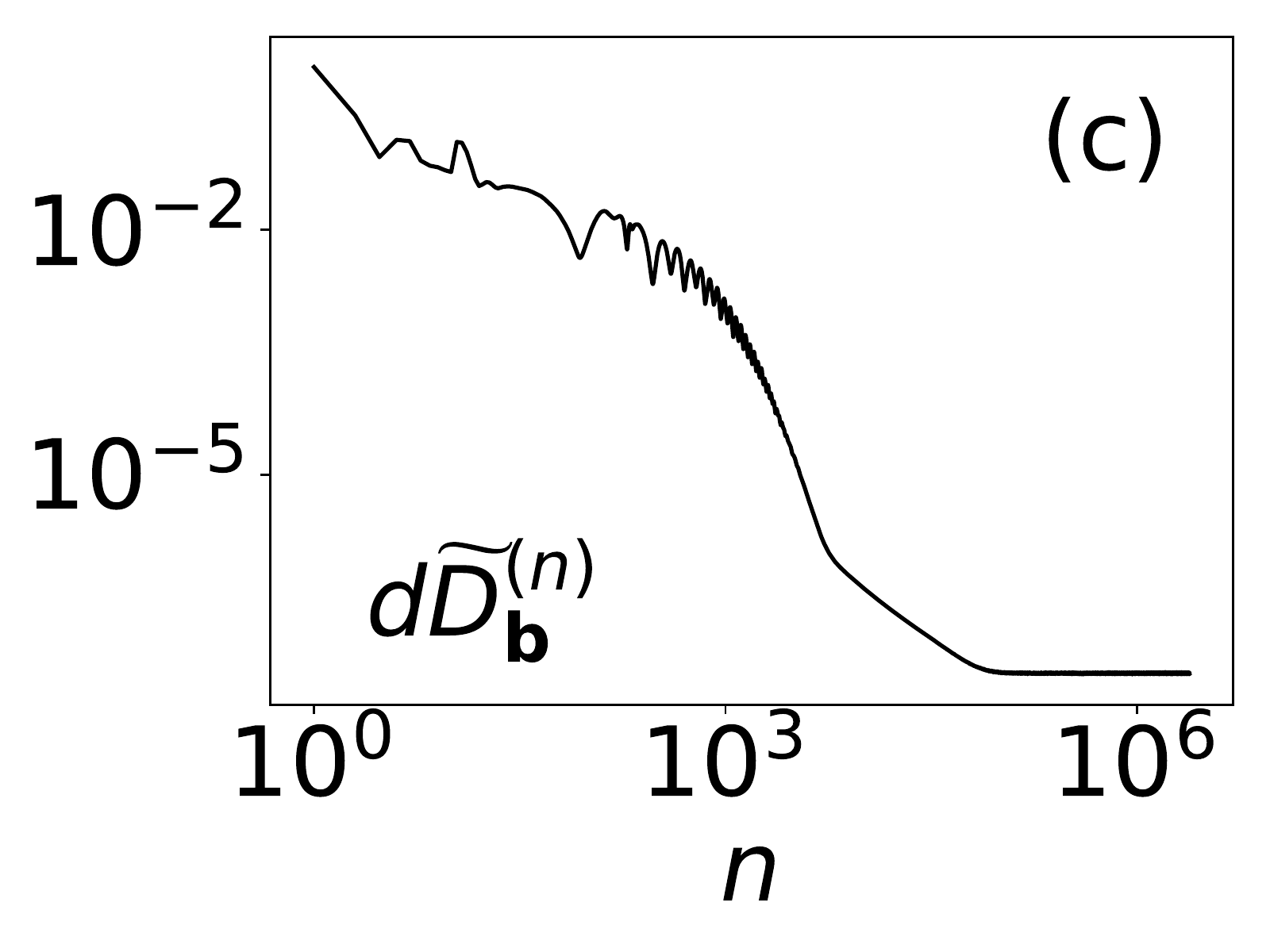}
 		\label{fig:char-noisy-img}
		}
	\\\vspace{-5pt}
	\subfloat{
 		\centering
 		\includegraphics[width=.15\textwidth, trim={15 15 12 12}, clip]{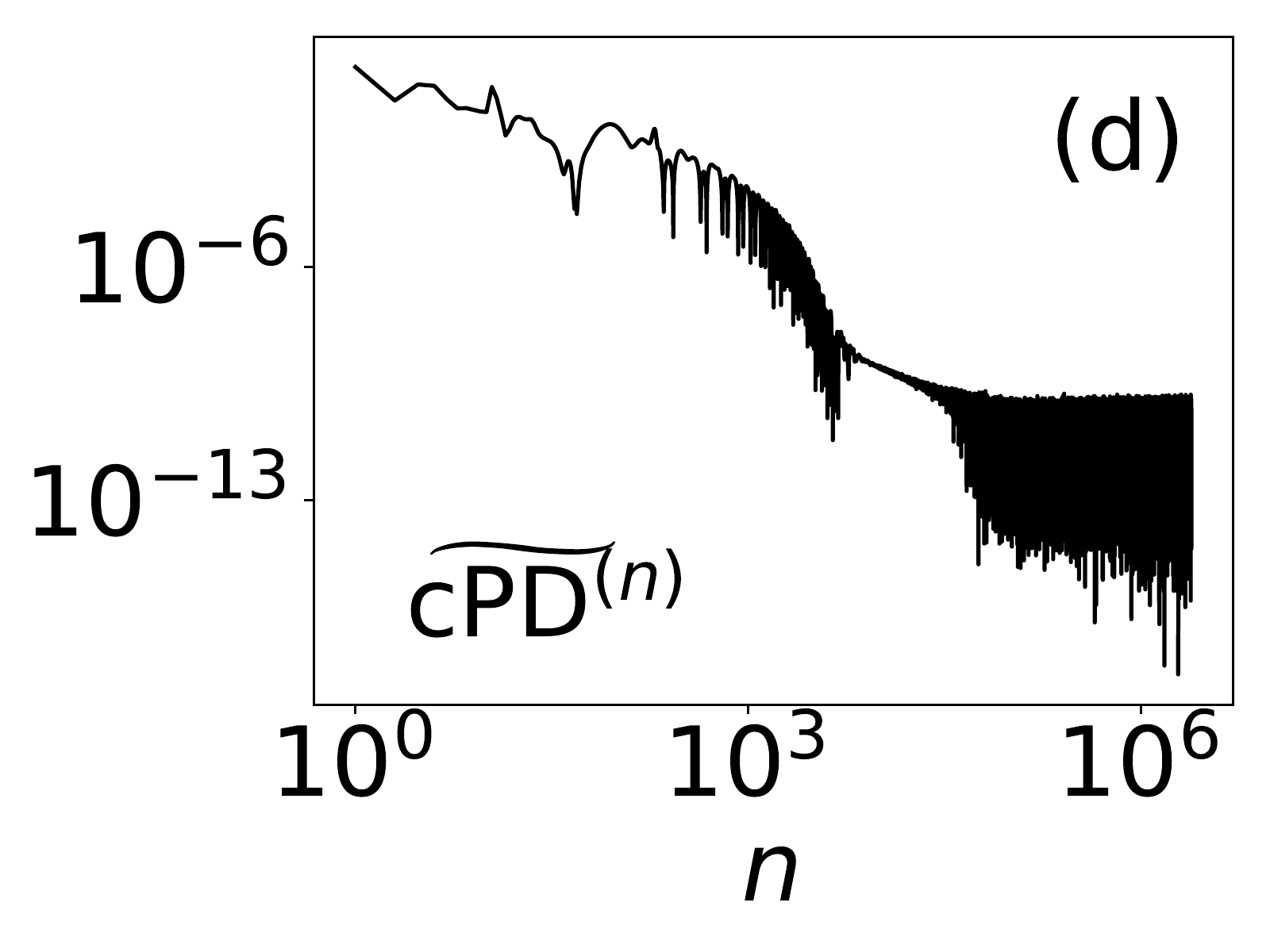}
 		\label{fig:char-noisy-splitting}
		}
	\subfloat{
 		\centering
 		\includegraphics[width=.15\textwidth, trim={15 15 12 0}, clip]{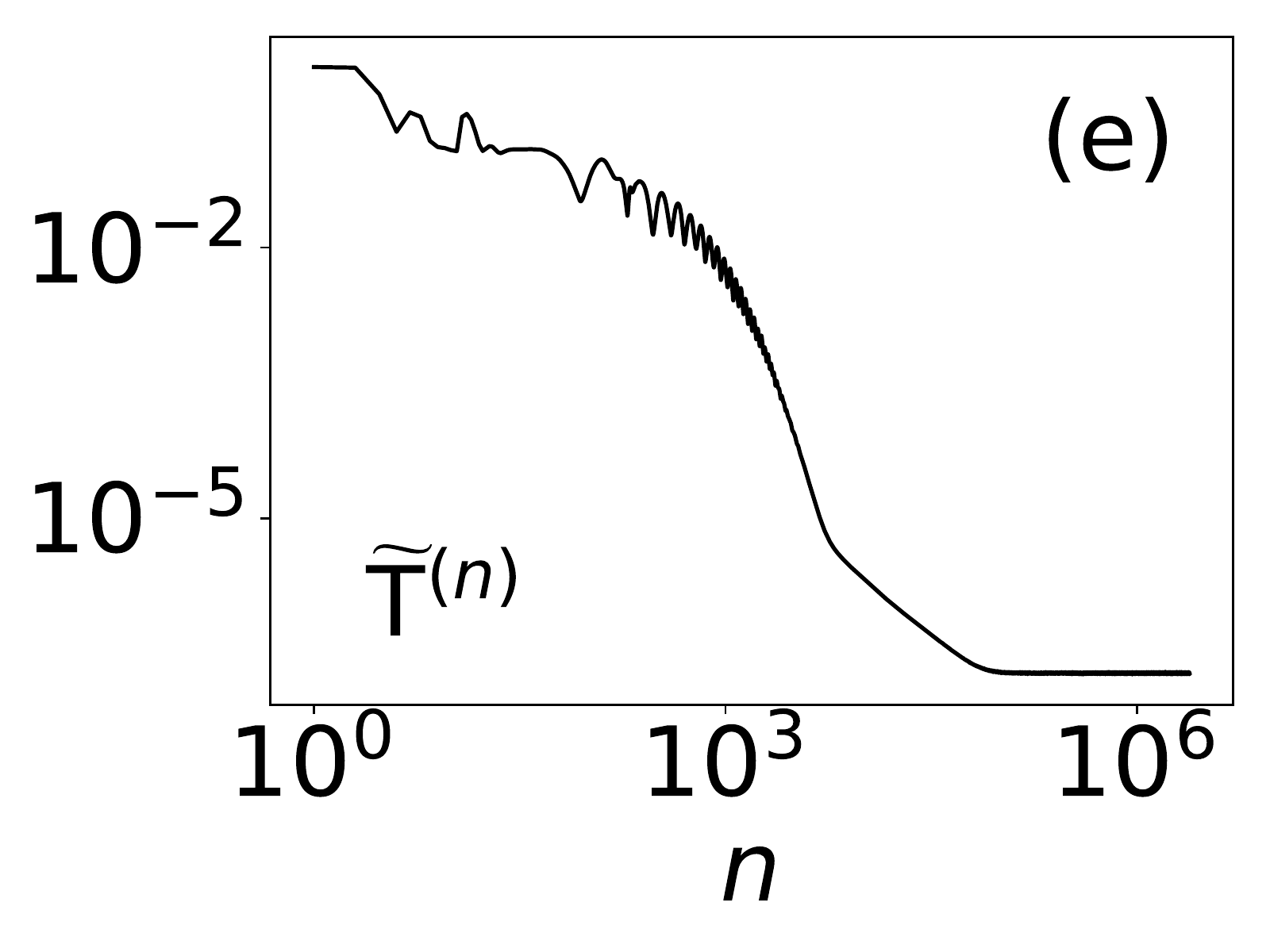}
 		\label{fig:char-noisy-transversality}
		}%
    \subfloat{
		\centering
 		\includegraphics[width=.15\textwidth, trim={15 15 12 12}, clip]{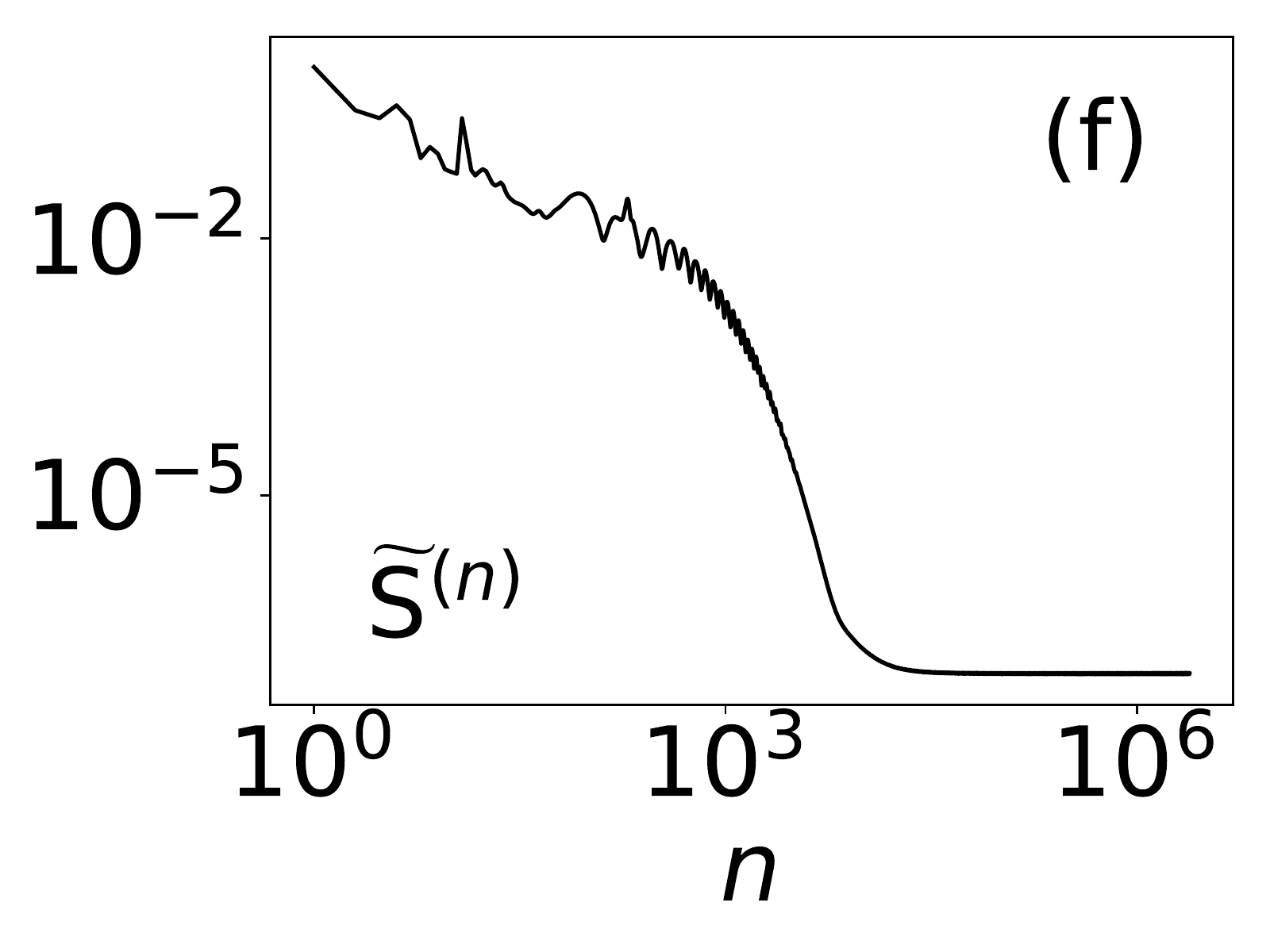}
 		\label{fig:char-noisy-splitting}
		}
 	\caption{Convegence metrics
	(a) $d\widetilde{D}_{\mathbf{g}}^{(n)}$,
	(b) $\widetilde{D}_\text{TV}^{(n)}$, and
	(c) $d\widetilde{D}_\mathbf{b}^{(n)}$, and empirical convergence
	metrics (d) $\widetilde{\text{cPD}}^{(n)}$,
	(e) $\widetilde{\text{T}}^{(n)} $, and 
	(f) $\widetilde{\text{S}}^{(n)}$, of the NCPD algorithm, as functions of $n$, from full-scan inconsistent data of the disk phantom.}
 	\label{fig:characterization-curves-NCPD}
 \end{figure}

Applying the NCPD algorithm, along with $\gamma$ that is $98\%$ of the truth-TV value, 
to the full-scan inconsistent data, we compute the convergence conditions in Eq. \eqref{eq:conv-conds-nlinear-2} and plot them as functions of iteration $n$ in panels (a)-(c) of Fig. \ref{fig:characterization-curves-NCPD}. We also compute and show in panels (d)-(f) of Fig.~\ref{fig:characterization-curves-NCPD} empirical convergence conditions in Eq. \eqref{eq:conv-conds-nlinear-1}.
 
\subsection{Numerical demonstration of the NCPD algorithm to invert the non-linear data model}\label{sec:NCPD-inversion}

Even if the NCPD algorithm is demonstrated to solve the non-convex optimization program in Eq. \eqref{eq:opt-constr-NL} above, evidence remains to be established as to whether the NCPD algorithm can invert the non-linear data model in Eq. \eqref{eq:g-NL-vec}. As discussed in Section~\ref{sec:image-conv-condition} above, we seek the evidence only for consistent data, i.e., $\mathbf{g}^{[\mathcal{M}]}= \mathbf{g}^{[NL]}(\mathbf{b})$ with $\mathbf{b}^\mathrm{[truth]}$ known, because the NCPD algorithm is developed based upon the non-linear data model. 

Applying the NCPD algorithm to the full-scan simulated, consistent data described in Section~\ref{sec:convergence-verification-study} above, we reconstruct basis images $\mathbf{b}^{(n)}$, calculate metric $\widetilde{D}_\mathbf{b}^{(n)}$ in Eq. \eqref{eq:conv-conds-nlinear-image}, and plot it (solid, dark) in panel (h) of Fig. \ref{fig:veri-NCPD}. It can be observed that metric $\widetilde{D}_\mathbf{b}^{(n)}$ consistently decays, revealing that under the full-scan data condition, the NCPD algorithm can invert the non-linear data model in Eq. \eqref{eq:g-NL-vec}. We show in columns (b) and (d) of Fig.~\ref{fig:veri-truth-image} reconstructed water and bone basis images of the numerical disk phantom, which are numerically identical to their respective truth counterparts in columns (a) and (c) of Fig.~\ref{fig:veri-truth-image}. As the verification study shows in Section~\ref{sec:real-short-data-recon}, the NCPD algorithm can also, under short-scan data conditions, invert the non-linear data model in Eq. \eqref{eq:g-NL-vec}.

The verification study not only yields evidence that the NCPD algorithm can invert the non-linear data model in Eq. \eqref{eq:g-NL-vec} but also identifies data conditions under which the evidence can be established. Moreover, the result of a verification study likely provides a ``best case scenario'' in terms of reconstruction accuracy by use of the NCPD algorithm.

\section{Results: monochromatic image reconstruction from simulated data}\label{sec:simulated-data-recon}

We now perform reconstruction studies on monochromatic images from simulated data by using the NCPD and CPD algorithms and demonstrate that the beam-hardening artifacts in the CPD image can be corrected effectively for in the NCPD image. For a benchmark purpose, we also reconstruct monochromatic images from the same data by using an indirect method~\citep{zou_analysis_2008}, in which the FBP algorithm, along with a Hanning kernel and a cut-off frequency of 0.3 cm$^{-1}$, is used for reconstruction of the basis images from the estimated basis projections.

\subsection{Scanning configurations in simulation studies}\label{sec:simulation-configuration} 
In the simulation studies below, we use a circular fan-beam scanning configuration identical to that described in paragraph 2 of Section~\ref{Sec:methods-1}, except that the linear detector now consists of 1024 0.39-mm detector bins. 

We create a numerical disk phantom to mimic the standard DE-472 physical phantom ~\citep{jrtassociates_dual_2015} used in dual-energy CT imaging. The disk phantom contains water and bone (i.e., $K=2$) basis images of $512\times 512$ square pixels of size 0.49 mm. When each of the two spectra is used to generate data from the phantom over a full rotation, we obtain a set of dual-energy data, which are referred to as full-scan data in the work.

\subsection{Monochromatic image reconstruction from simulated-noiseless data}\label{sec:sim-noiseless-data}

Using Eq. \eqref{eq:g-NL-vec} with the configuration and spectra parameters described, we generate from the numerical disk phantom full-scan simulated-noiseless data $\mathbf{g}^{[\mathcal{M}]}$  at 640 views uniformly distributed over 2$\pi$. In the study, because $\mathbf{g}^{[\mathcal{M}]}=\mathbf{g}^{[NL]}(\mathbf{b}^\mathrm{[truth]})$ and because truth image $\mathbf{b}^\mathrm{[truth]}$ is known, convergence conditions in  Eqs. \eqref{eq:conv-conds-nlinear-2}, \eqref{eq:conv-conds-nlinear-3}, and \eqref{eq:conv-conds-nlinear-image} can be calculated and are thus used. In this simulation study, truth-monochromatic image $\mathbf{f}_{m'}^{[\rm truth]}$ is known because truth image $\mathbf{b}^{[\rm truth]}$ is known, and so is its  TV (i.e., truth-TV) value that is used as the value of parameter $\gamma$.

\begin{figure}[htb]
 	\centering 
		\subfloat[NCPD]{
 		\centering
 		\includegraphics[width=.15\textwidth]{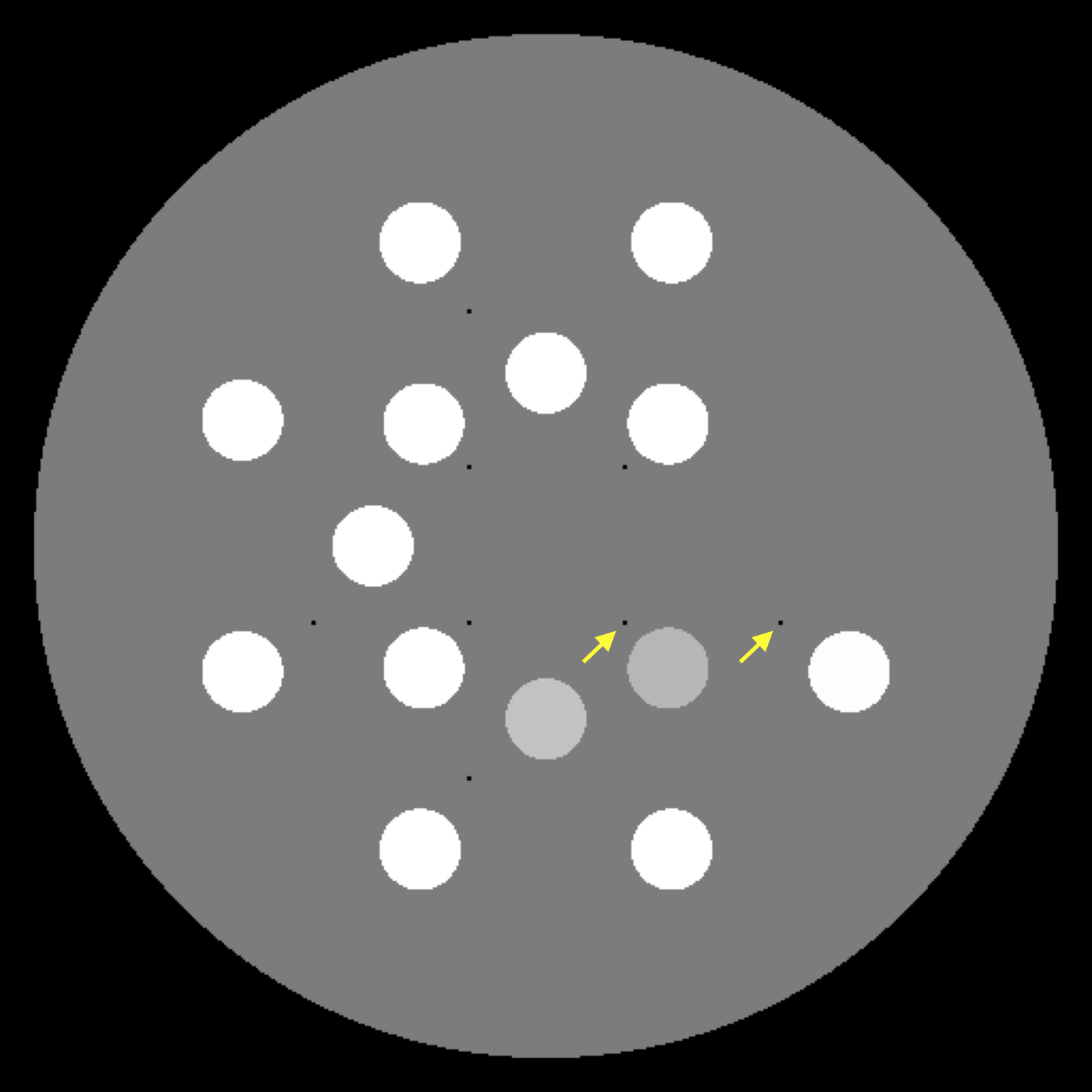}
 		\label{fig:ncpd}
		}
        \subfloat[CPD]{
 		\centering
 		\includegraphics[width=.15\textwidth]{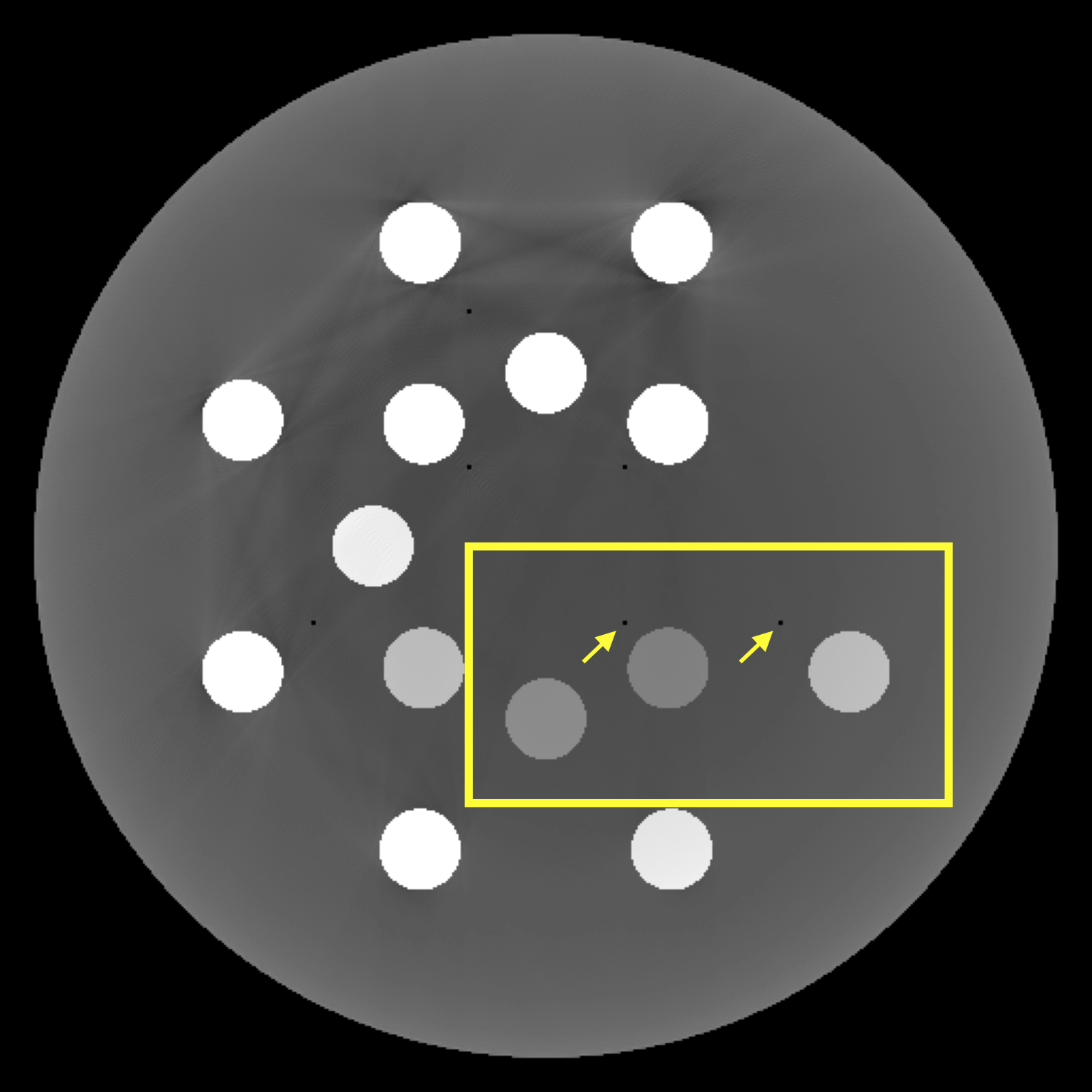}
 		\label{fig:ln}
		}
		\subfloat[Indirect]{
 		\centering
 		\includegraphics[width=.15\textwidth]{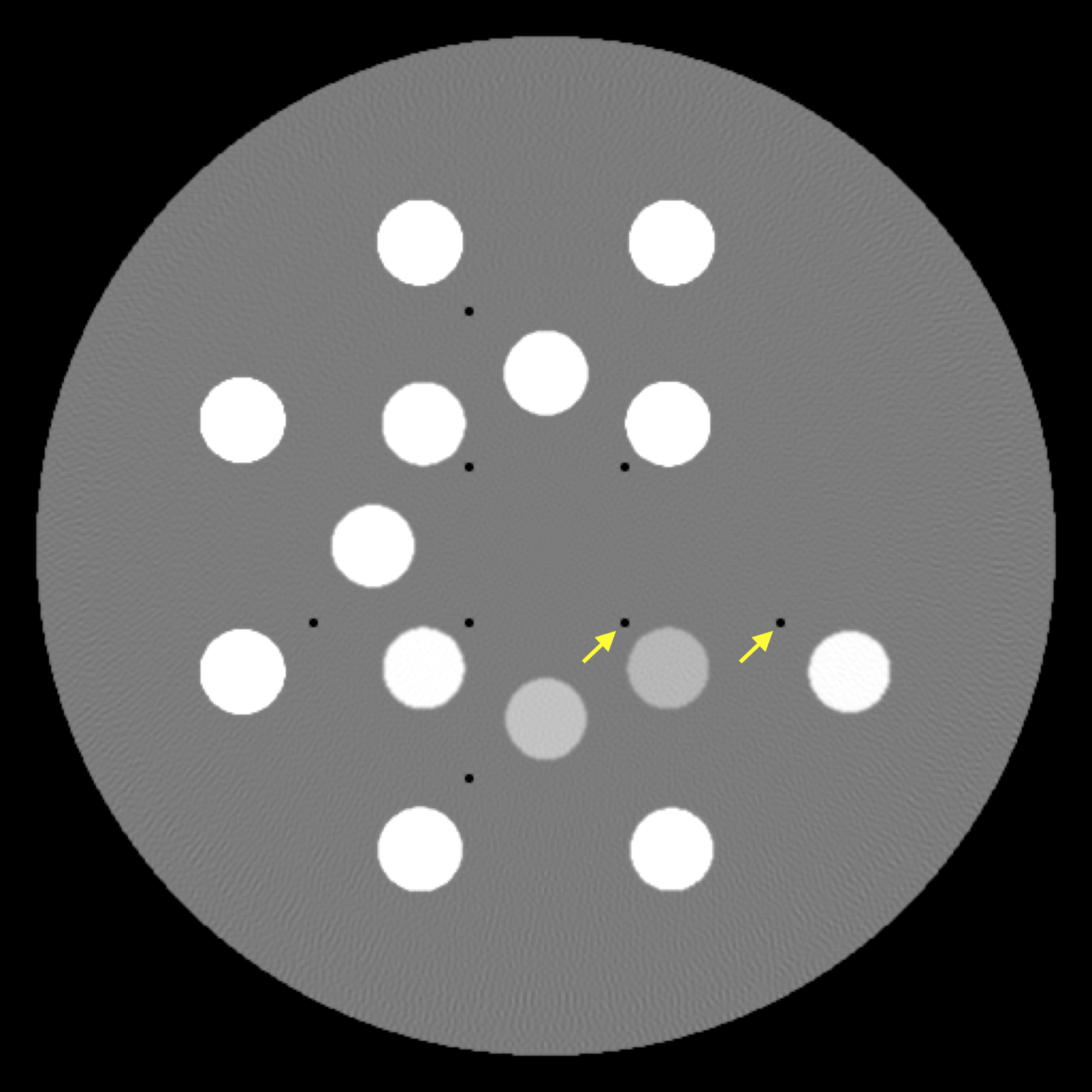}
 		\label{fig:ref}
		}
\vspace{-5pt}
		\subfloat{
 		\centering
 		\includegraphics[width=.15\textwidth, trim={160 100 50 190}, clip]{figures/rslt_for_manuNCCP/verification_full_p6/nonlinear_single_mono100keV.png}
 		\label{fig:ncpd-roi}
		}
        \subfloat{
 		\centering
 		\includegraphics[width=.15\textwidth, trim={160 100 50 190}, clip]{figures/rslt_for_manuNCCP/verification_full_p6/nonlinear_LN_mono100keV.png}
 		\label{fig:ln-roi}
		}
		\subfloat{
 		\centering
 		\includegraphics[width=.15\textwidth, trim={160 100 50 190}, clip]{figures/rslt_for_manuNCCP/verification_full_p6/fbp-hanning_mono100keV.png}
 		\label{fig:ref-roi}
		}
 	\caption{Monochromatic (top row) and their ROI (bottom row) images at energy 100 keV reconstructed by use of (a) the NCPD and (b) the CPD algorithms, and (c) the indirect method, from simulated-noiseless data. The rectangle ROI is shown in column (b). The arrows highlight the pin-hole structures of negative contrast. Display window: [-150, 150] HU.}
 	\label{fig:sim-noiseless}
 \end{figure}

Applying the NCPD algorithm to the full-scan simulated-noiseless data, we reconstruct the basis images 
and subsequently monochromatic image $\mathbf{f}_{m'}$ at energy 100 keV with Eq. \eqref{eq:basis-decomposition}, which are displayed in column (a) of Fig. \ref{fig:sim-noiseless}. We also show in  columns (b) $\&$ (c) of Fig. \ref{fig:sim-noiseless} images reconstructed from the same data by use of the CPD algorithm and the indirect method. It can be observed that the beam-hardening artifacts, including dark bands connecting the inserts and the cupping effect in the background material, in the CPD image can be corrected effectively for by use of the NCPD algorithm. The indirect method appears also to correct effectively for the beam-hardening artifacts. However, it yields an image of spatial resolution lower than does the NCPD algorithm. This can be observed from the edges of the inserts and enlarged pin-hole structures of negative contrast, as highlighted by the arrows in Fig. \ref{fig:sim-noiseless}.

\subsection{Monochromatic image reconstruction from simulated-noisy data}\label{sec:sim-noisy-data}
We create full-scan simulated-noisy data by adding Poisson noise to the full-scan simulated-noiseless data in Section~\ref{sec:sim-noiseless-data}. The added noise level corresponds to $10^7$ photons per detector bin in an air scan at each detector bin prior to the logarithmic step. 
Because measured data $\mathbf{g}^{[\mathcal{M}]}$ now contain noise that is inconsistent with the non-linear data model in Eq. \eqref{eq:g-NL-vec}, i.e.,  $\mathbf{g}^{[\mathcal{M}]}\neq \mathbf{g}^{[NL]}(\mathbf{b}^\mathrm{[truth]})$,  an image is reconstructed when the convergence conditions only in Eq. \eqref{eq:conv-conds-nlinear-2} are achieved.
 
In the studies with noisy data here and real data below, the truth-TV value may be neither appropriate nor generally available, and we need to determine the value of parameter $\gamma$ for a reconstruction.  For example, the value of $\gamma$ could be determined that empirically yields a highest quality/utility metric designed within a $\gamma$-value range of interest. Please see additional discussion of image reconstructions with different $\gamma$ values in Section~\ref{sec:recons-gammas} below. 

\begin{figure}[!t]
 	\centering 
		\subfloat[NCPD]{
 		\centering
 		\includegraphics[width=.15\textwidth]{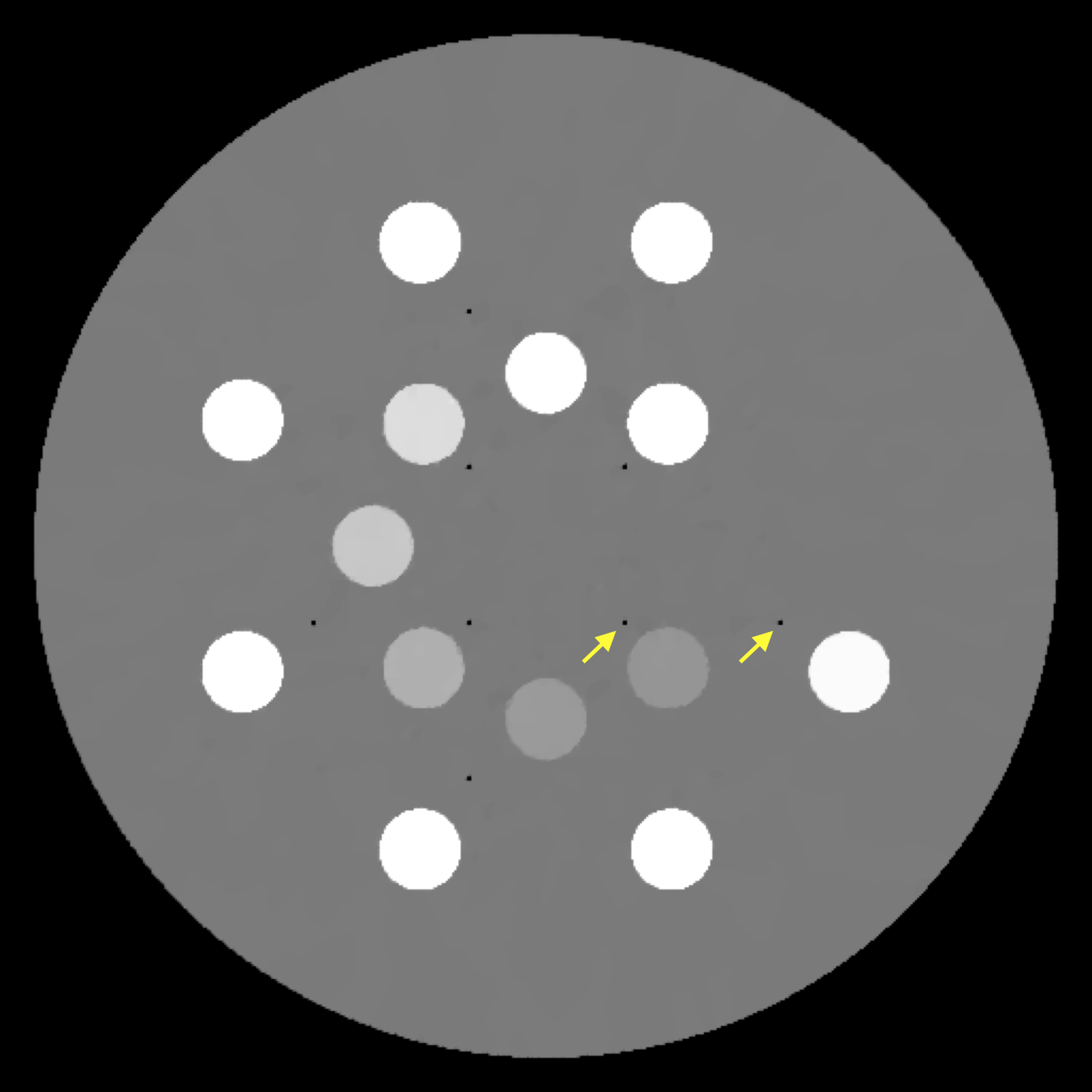}
 		\label{fig:ncpd}
		}
        \subfloat[CPD]{
 		\centering
 		\includegraphics[width=.15\textwidth]{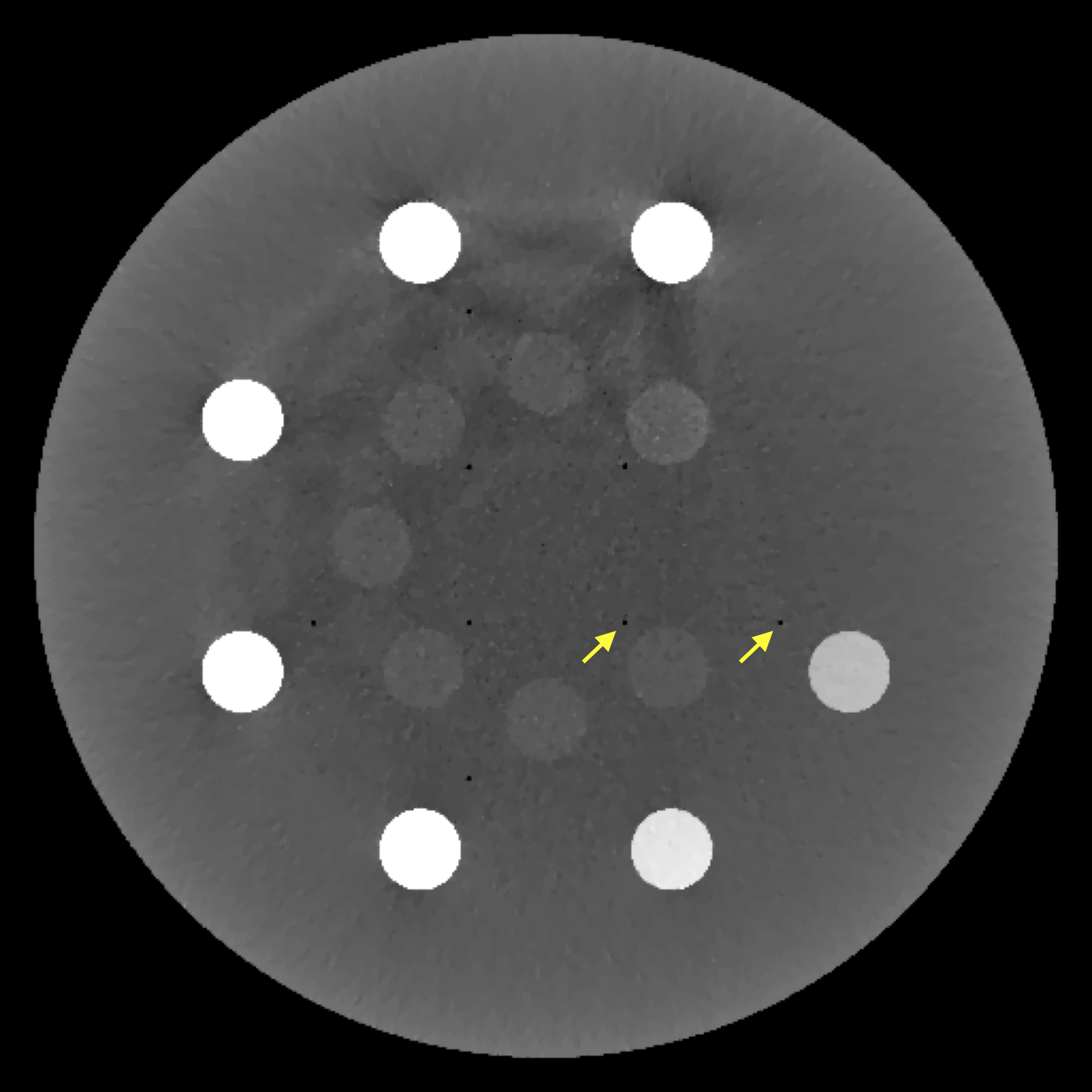}
 		\label{fig:ln}
		}
		\subfloat[Indirect]{
 		\centering
 		\includegraphics[width=.15\textwidth]{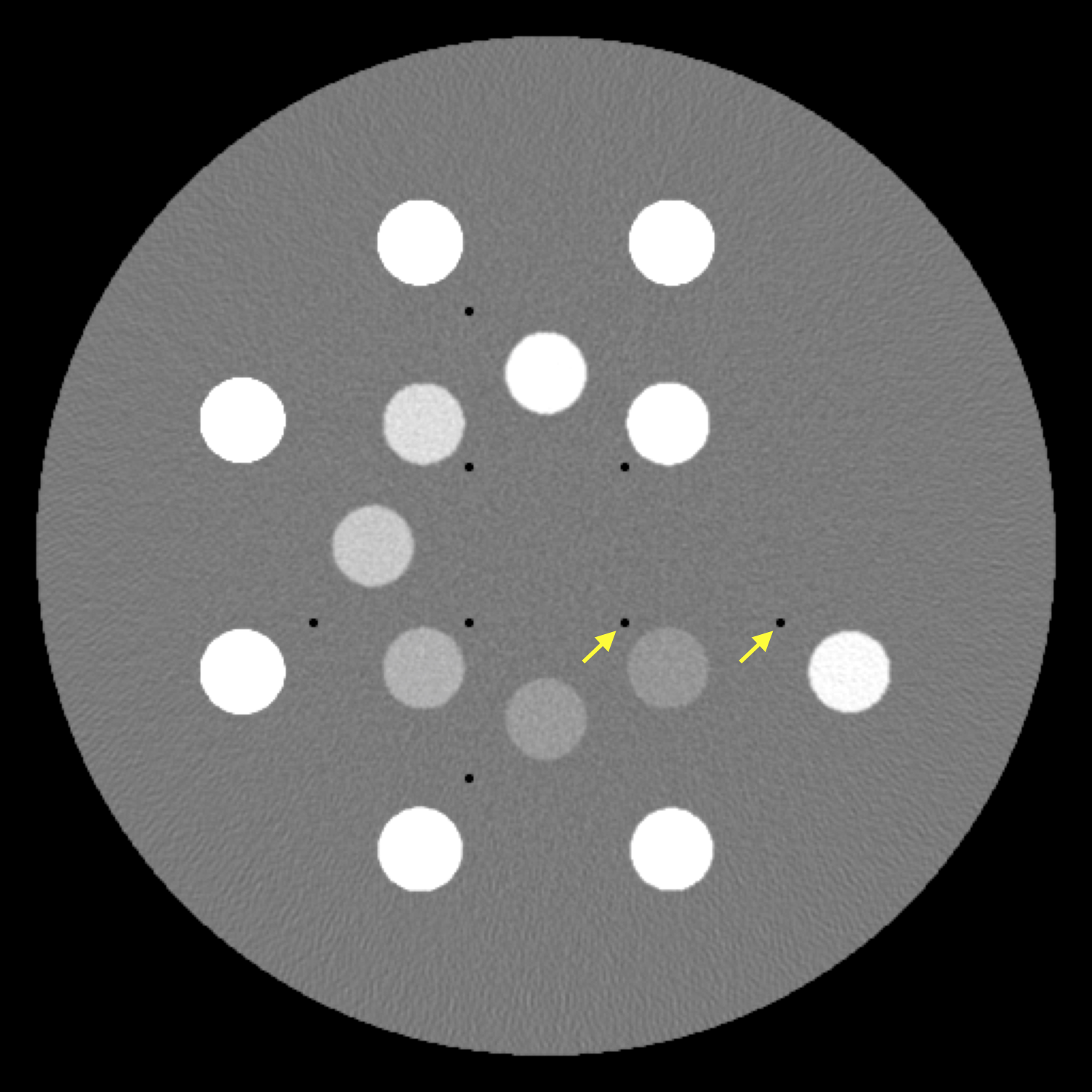}
 		\label{fig:ref}
		}
\vspace{-5pt}
		\subfloat{
 		\centering
 		\includegraphics[width=.15\textwidth, trim={160 100 50 190}, clip]{figures/rslt_for_manuNCCP/characterization/pois1e7_p6/ncpd_mono100keV.png}
 		\label{fig:ncpd-roi}
		}
        \subfloat{
 		\centering
 		\includegraphics[width=.15\textwidth, trim={160 100 50 190}, clip]{figures/rslt_for_manuNCCP/characterization/pois1e7_p6/cpd_mono100keV.png}
 		\label{fig:ln-roi}
		}
		\subfloat{
 		\centering
 		\includegraphics[width=.15\textwidth, trim={160 100 50 190}, clip]{figures/rslt_for_manuNCCP/characterization/pois1e7_p6/fbp-corr-hanning_mono100keV.png}
 		\label{fig:ref-roi}
		}
 	\caption{Monochromatic (top row) and their ROI (bottom row) images at energy 100 keV reconstructed by use of (a) the NCPD and (b) the CPD algorithms, and (c) the indirect method, from full-scan simulated-noisy data. The rectangle ROI is depicted in Fig. \ref{fig:sim-noiseless}. The arrows highlight the pin-hole structures of negative contrast. Display window: [-150, 150] HU.}
 	\label{fig:sim-noisy}
 \end{figure}

In the study, we use the NCPD algorithm to reconstruct, from the full-scan simulated-noisy data, a number of basis images with different values of parameter $\gamma$, and then monochromatic images $\mathbf{f}_{m'}$ at energy 100 keV. Through visual inspection of the monochromatic images, we select subjectively $\gamma=423.7$ that appears to yield a monochromatic image with an optimal combination of spatial resolution and low noise level. 
In column (a) of Fig. \ref{fig:sim-noisy}, we display monochromatic images obtained with $\gamma=423.7$. The images reconstructed by use of the CPD algorithm and indirect method are shown also in columns (b) $\&$ (c)  of Fig. \ref{fig:sim-noisy}.  Again, the beam-hardening artifacts in the CPD image can be corrected effectively for by use of the NCPD algorithm and the indirect method. However, the latter yields images with a noise level comparable to, but spatial resolution lower, than that of the former, as highlighted by the arrows in Fig. \ref{fig:sim-noisy}.

\section{Results: monochromatic image reconstruction from real data}\label{sec:real-data-recon}

In the study with real data below, we collect data from the GAMMEX DE-472 CT phantom by use of a clinical CT scanner~\citep{chen2018algorithm}, which has a curved-detector array of 896 1-mm bins forming a fan angle of 49.2$^\circ$ and a magnification factor of 1.7. The GAMMEX DE-472 CT phantom consists of calcium and iodine inserts~\citep{jrtassociates_dual_2015} and is used widely in dual-energy CT studies.
For each spectrum of 80 kVp and 135 kVp, data are collected at 1200 views uniformly distributed over 2$\pi$. Again, we refer to the dual-energy data collected as {full-scan real data.}
The images are reconstructed on an array of $I=512\times 512$ square pixels of size 0.68 mm. 

\subsection{Monochromatic image reconstruction from full-scan real data}\label{sec:real-full-data-recon}

\noindent \subsubsection{Monochromatic image reconstruction} In the study with real data, the truth-TV value is unknown. Therefore, as discussed above, we use the NCPD algorithm to reconstruct, from the full-scan real data, a number of basis images with different values of parameter $\gamma$, and then monochromatic images $\mathbf{f}_{m'}$ at energy 100 keV. Through visual inspection of the monochromatic images, we select subjectively $\gamma=436.6$ that appears to yield a monochromatic image with an optimal combination of spatial resolution and low noise level. 

\begin{figure}[htb]
 	\centering 
		\subfloat[NCPD]{
 		\centering
 		\includegraphics[width=.15\textwidth]{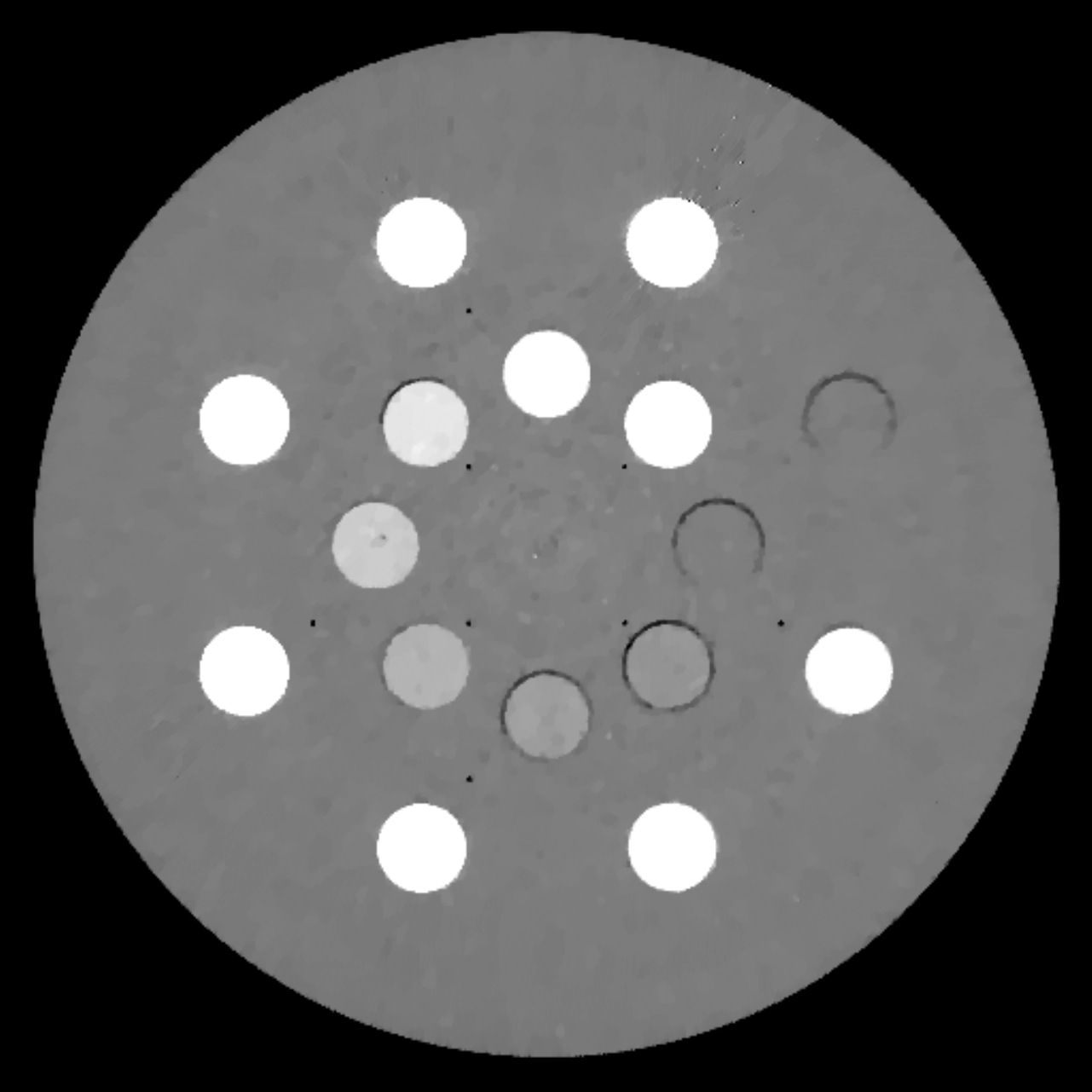}
 		\label{fig:ncpd}
		}
        \subfloat[CPD]{
 		\centering
 		\includegraphics[width=.15\textwidth]{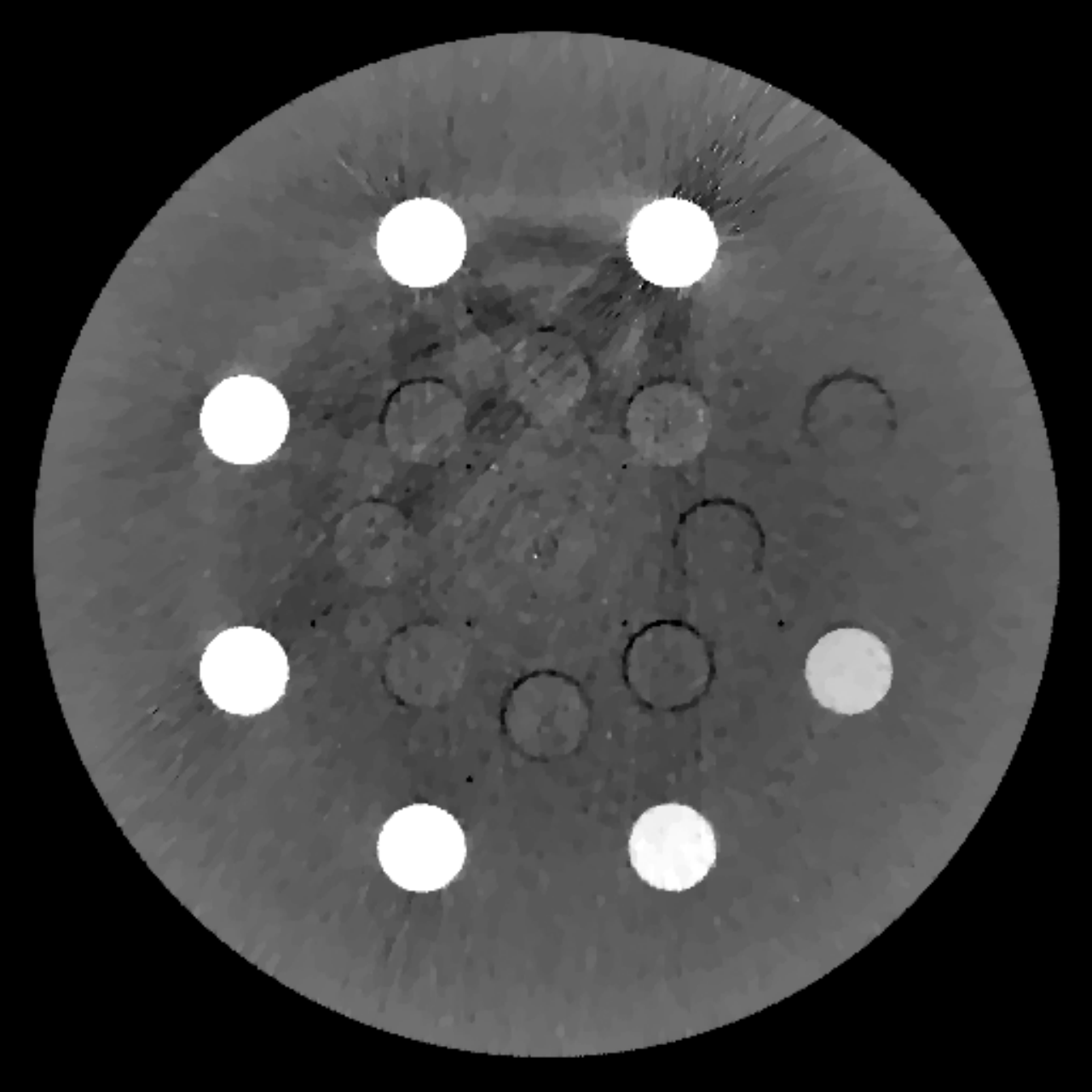}
 		\label{fig:ln}
		}
		\subfloat[Indirect]{
 		\centering
 		\includegraphics[width=.15\textwidth]{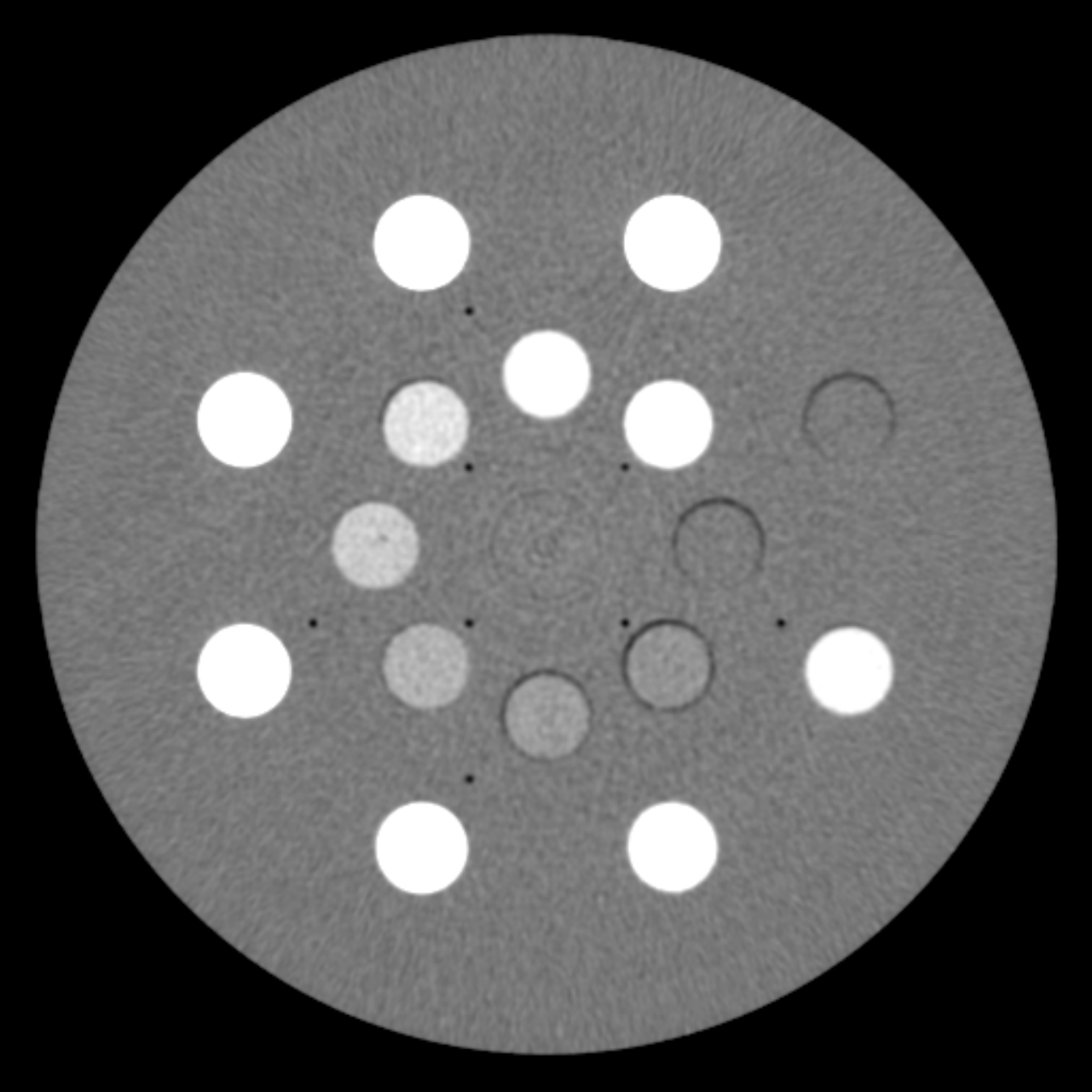}
 		\label{fig:ref}
		}
\vspace{-5pt}
		\subfloat{
 		\centering
 		\includegraphics[width=.15\textwidth, trim={160 100 50 190}, clip]{figures/rslt_for_manuNCCP/characterization/realdata_r/gamma45_mono100keV.pdf}
 		\label{fig:ncpd-roi}
		}
        \subfloat{
 		\centering
 		\includegraphics[width=.15\textwidth, trim={160 100 50 190}, clip]{figures/rslt_for_manuNCCP/short/LN_mono100keV.pdf}
 		\label{fig:ln-roi}
		}
		\subfloat{
 		\centering
 		\includegraphics[width=.15\textwidth, trim={160 100 50 190}, clip]{figures/rslt_for_manuNCCP/characterization/realdata_r/ref_mono100keV.pdf}
 		\label{fig:ref-roi}
		}
 	\caption{Monochromatic (top row) and their ROI (bottom row) images at energy 100 keV reconstructed by use of (a) the NCPD and (b) the CPD algorithms, and (c) the indirect method, from full-scan real data. The rectangle ROI is depicted in Fig. \ref{fig:sim-noiseless}.  Display window: [-150, 150] HU.}
 	\label{fig:real-disk-recons}
 \end{figure}

From the basis image reconstructed with $\gamma=436.6$, we subsequently obtain monochromatic image $\mathbf{f}_{m'}$ at energy 100 keV, which are displayed in  column (a) of Fig. \ref{fig:real-disk-recons}. For visual comparison, we also show the images obtained with the CPD algorithm and the indirect method in columns (b) $\&$ (c) of Fig. \ref{fig:real-disk-recons}, respectively. It can be observed that the beam-hardening artifacts in the CPD image can be corrected effectively for by use of the NCPD algorithm and the indirect method. However, the indirect method yields an image that appears noisier and less sharp than that of the NCPD algorithm.

\noindent \subsubsection{Monochromatic images with different $\gamma$ values}\label{sec:recons-gammas} 
\begin{figure}[htb]
 	\centering
        \subfloat[$\gamma=378.4$]{
 		\centering
 		\includegraphics[width=.15\textwidth]{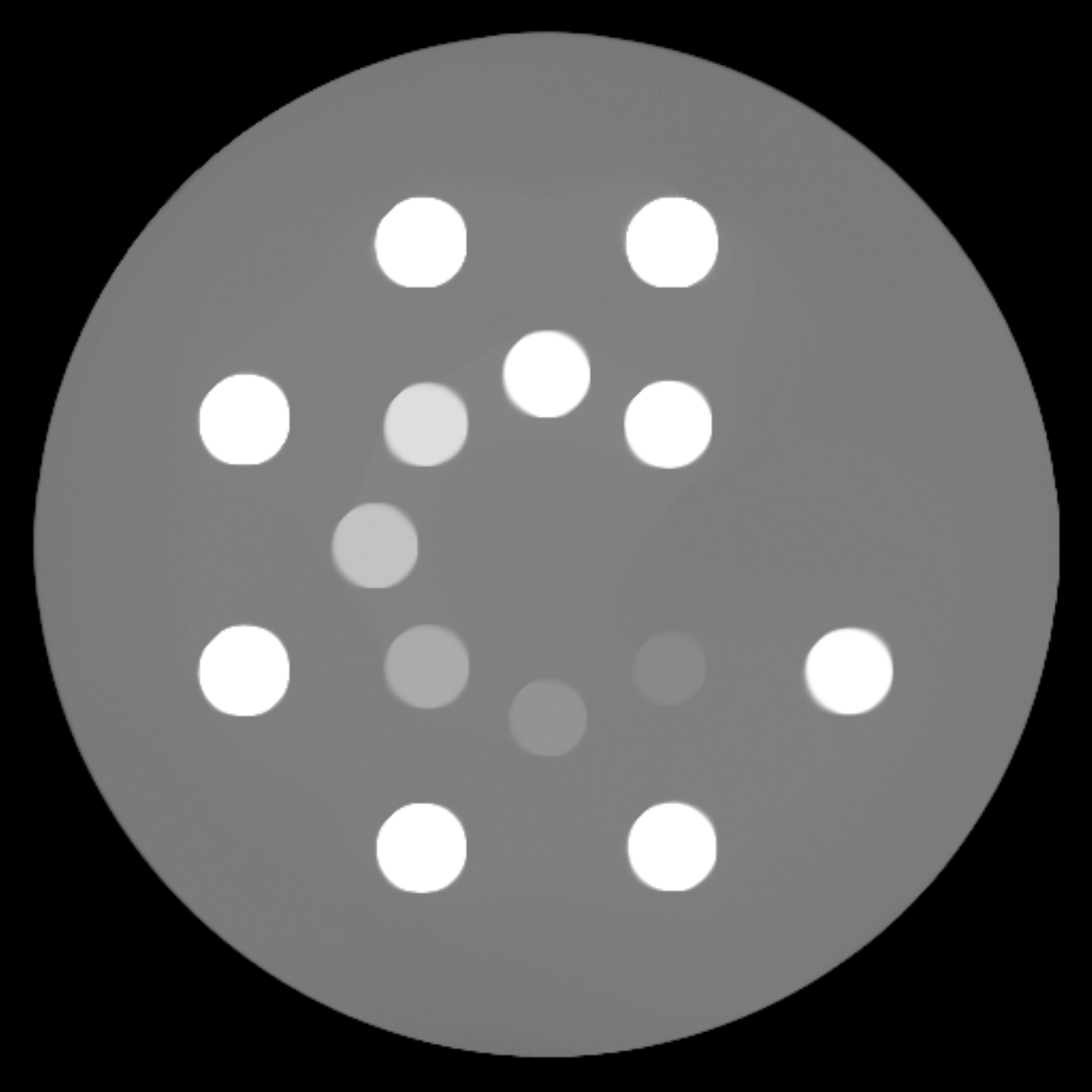}
 		\label{fig:char-realdata-gamma1}
		}
		\subfloat[$\gamma=436.6$]{
 		\centering
 		\includegraphics[width=.15\textwidth]{figures/rslt_for_manuNCCP/characterization/realdata_r/gamma45_mono100keV.pdf}
 		\label{fig:char-realdata-gamma2}
		}
		\subfloat[$\gamma=1455.2$]{
 		\centering
 		\includegraphics[width=.15\textwidth]{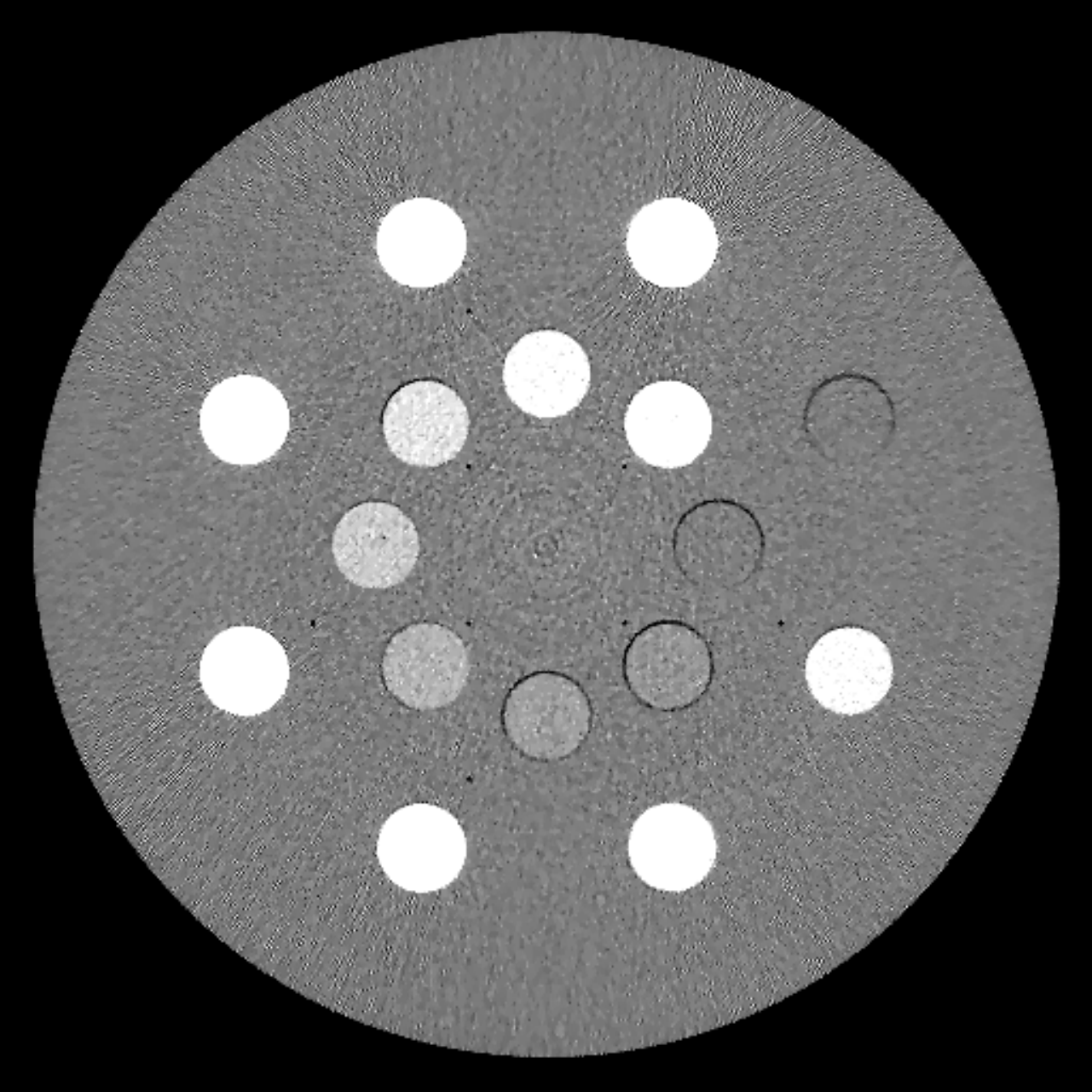}
 		\label{fig:char-realdata-gamma3}
		}
\vspace{-5pt}
        \subfloat{
 		\centering
 		\includegraphics[width=.15\textwidth, trim={160 100 50 190}, clip]{figures/rslt_for_manuNCCP/characterization/realdata_r/gamma39_mono100keV.pdf}
 		\label{fig:char-realdata-gamma1}
		}
		\subfloat{
 		\centering
 		\includegraphics[width=.15\textwidth, trim={160 100 50 190}, clip]{figures/rslt_for_manuNCCP/characterization/realdata_r/gamma45_mono100keV.pdf}
 		\label{fig:char-realdata-gamma2}
		}
		\subfloat{
 		\centering
 		\includegraphics[width=.15\textwidth, trim={160 100 50 190}, clip]{figures/rslt_for_manuNCCP/characterization/realdata_r/gamma150_mono100keV.pdf}
 		\label{fig:char-realdata-gamma3}
		}
 	\caption{Monochromatic (top row) and their ROI (bottom row) images at energy 100 keV obtained with (a) $\gamma=$378.4,  (b) 436.6,
 	and (c) 1455.2, from full-scan real data by use of the NCPD algorithm. The rectangle ROI is depicted in Fig. \ref{fig:sim-noiseless}.  
 	Display window: [-150, 150] HU.}
 	\label{fig:real-mono-gammas}
 \end{figure}

As $\gamma$ plays a role in specifying the solution set of the non-convex program in Eq. \eqref{eq:opt-constr-NL}, we carry out a series of reconstructions by using the NCPD algorithm from the same real data with different values of $\gamma$, and display in Fig. \ref{fig:real-mono-gammas} monochromatic images $\mathbf{f}_{m'}$ reconstructed at energy 100 keV (top row) from three selected $\gamma$ values. 
It can be observed that the images obtained with different values of $\gamma$ are visually different, as expected.  In an application, as a specified quality/utility metric is computed  based generally upon the image reconstructed, different values of $\gamma$ would result in different reconstructed images and thus different values of the metric. This observation also suggests that the parameter could be selected when yielding an empirically high value of a quality/utility metric of interest
\citep{zhang2016artifact, xia2016optimization, zhang2016investigation}.

\noindent \subsubsection{Monochromatic images at different iterations} 
\begin{figure}[htb]
 	\centering
		\subfloat[iter. 30]{
 		\centering
 		\includegraphics[width=.15\textwidth]{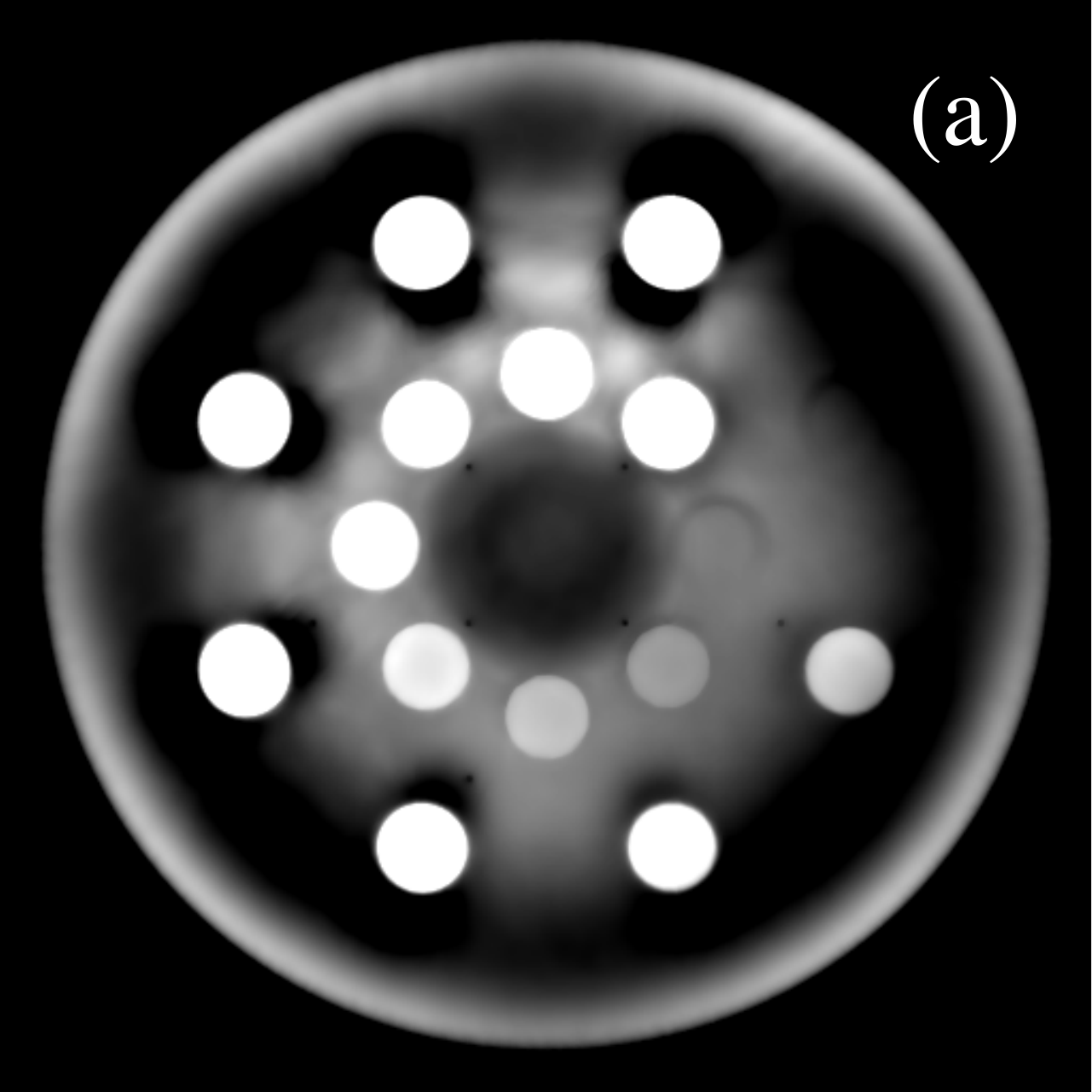}
 		\label{fig:char-realdata-gamma2-iter30}
		}
		\subfloat[iter. 100]{
 		\centering
 		\includegraphics[width=.15\textwidth]{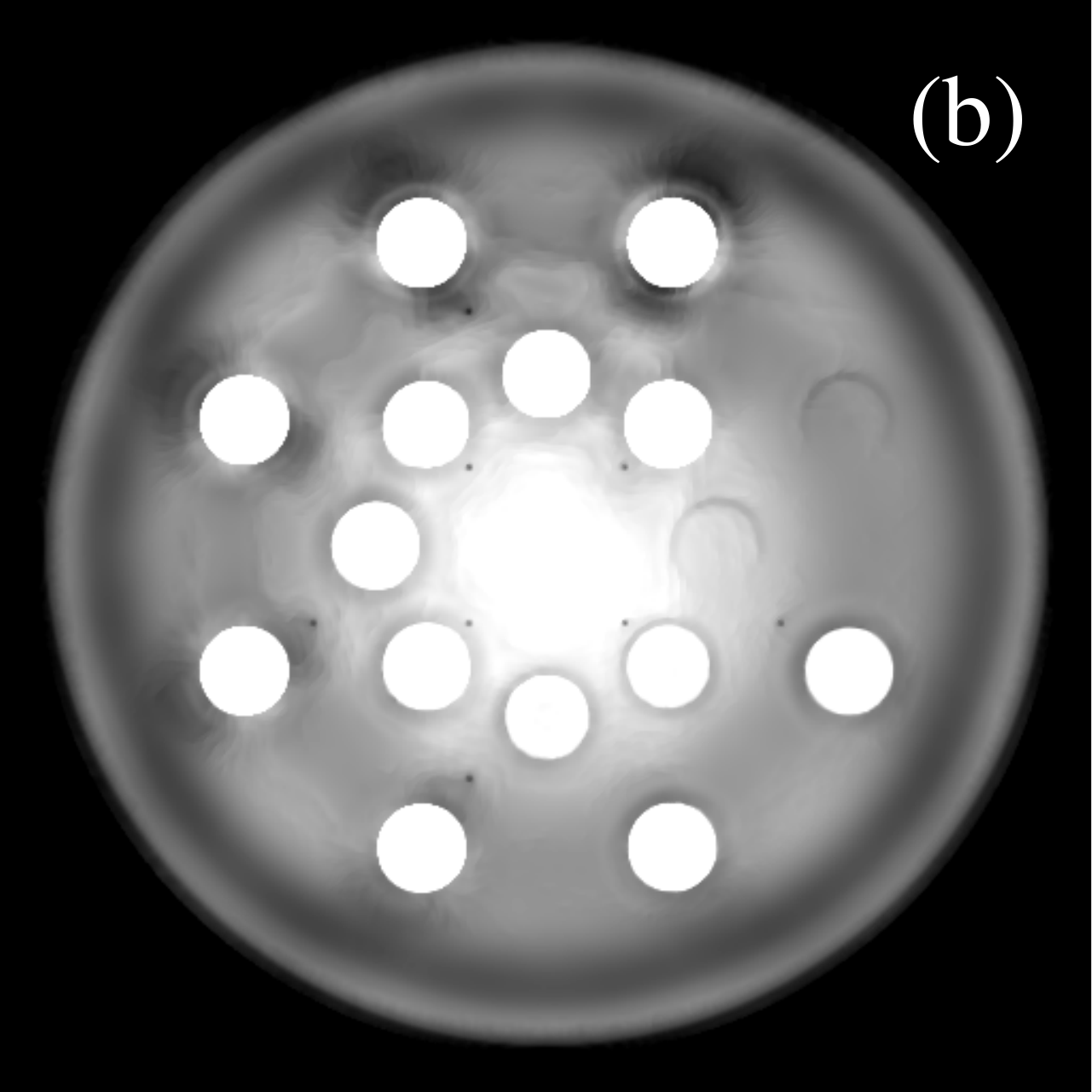}
 		\label{fig:char-realdata-gamma2-iter100}
		}
		\subfloat[iter. 500]{
 		\centering
 		\includegraphics[width=.15\textwidth]{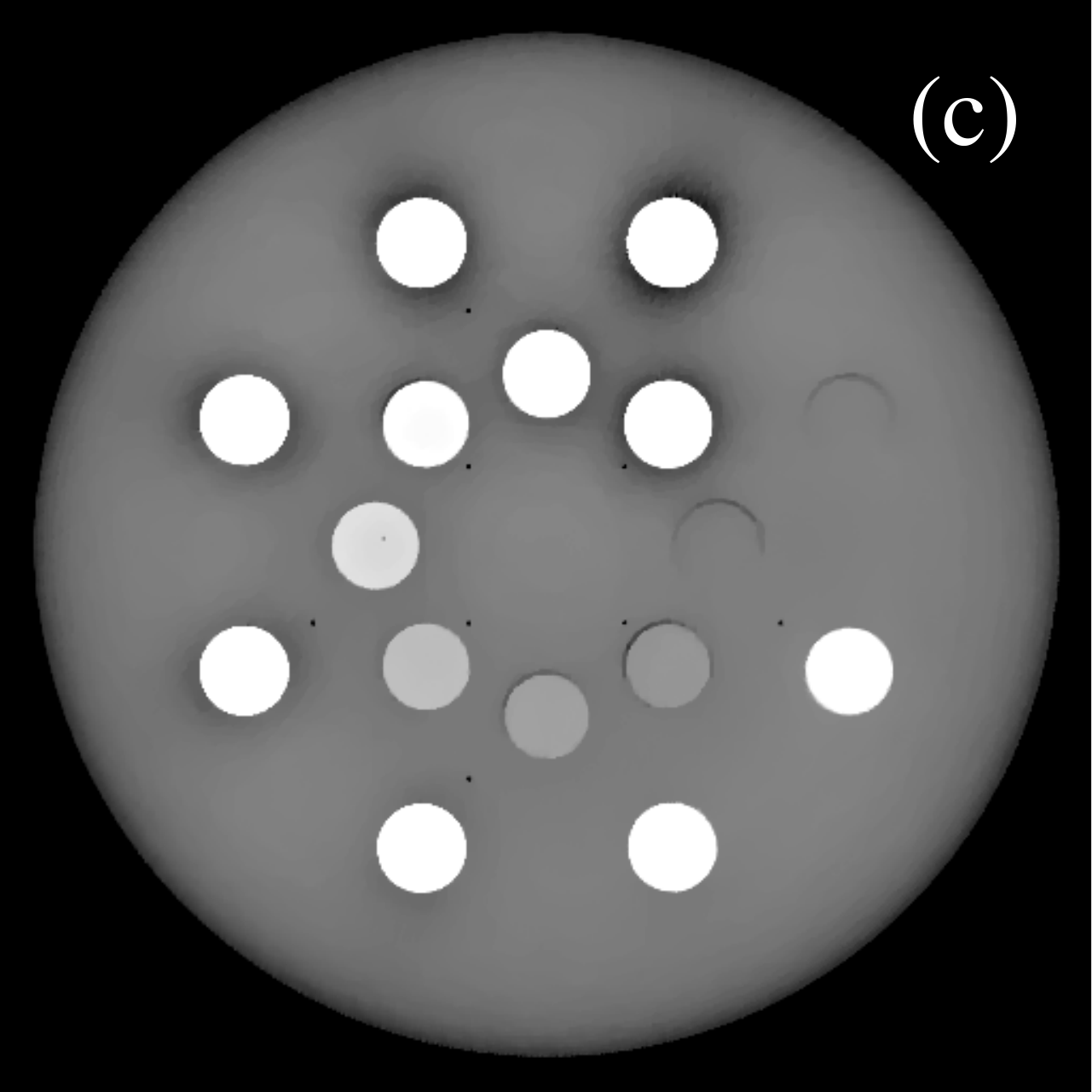}
 		\label{fig:char-realdata-gamma2-iter500}
		}
\vspace{-5pt}
		\subfloat[iter. 1000]{
 		\centering
 		\includegraphics[width=.15\textwidth]{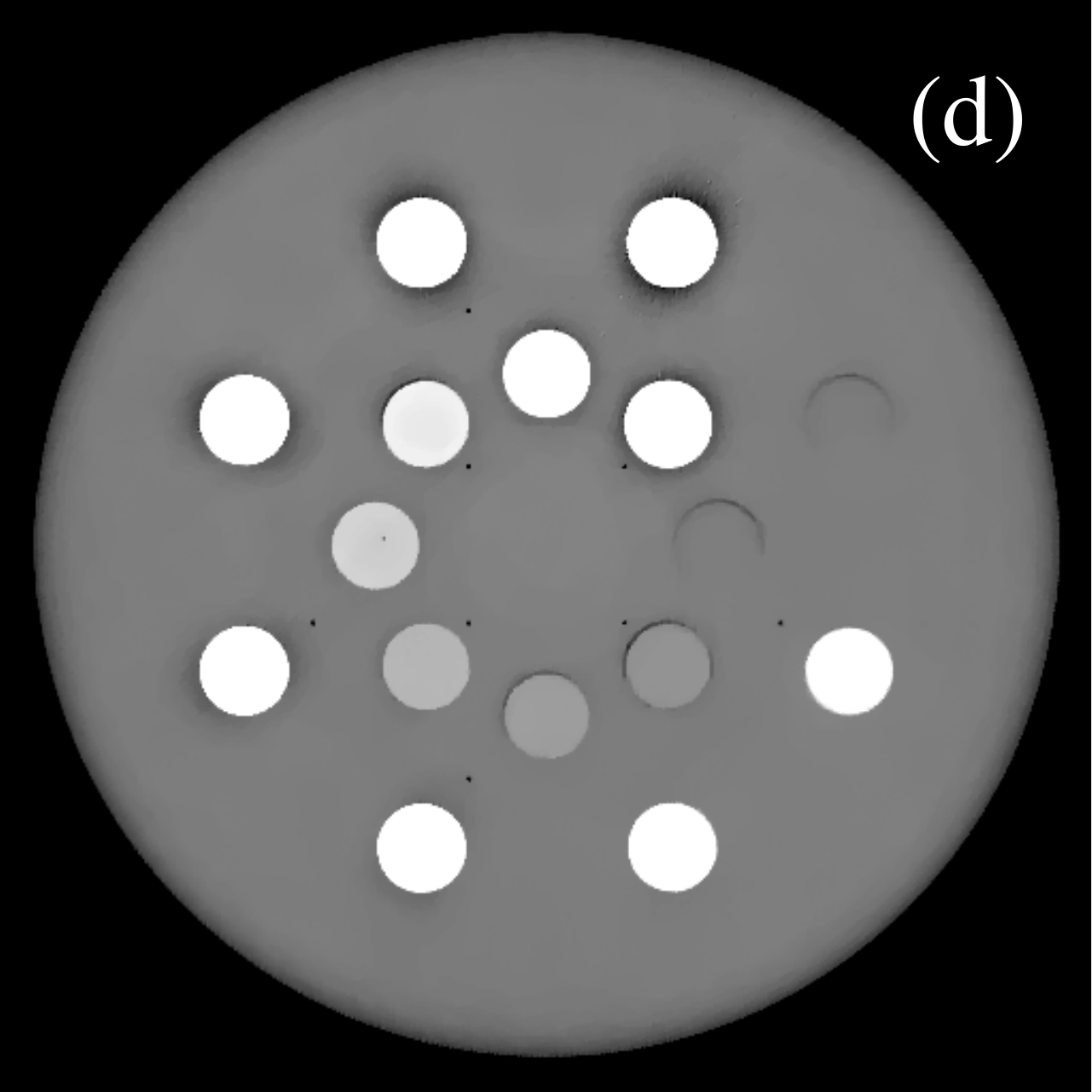}
 		\label{fig:char-realdata-gamma2-iter1000}
		}
		\subfloat[iter. 2000]{
 		\centering
 		\includegraphics[width=.15\textwidth]{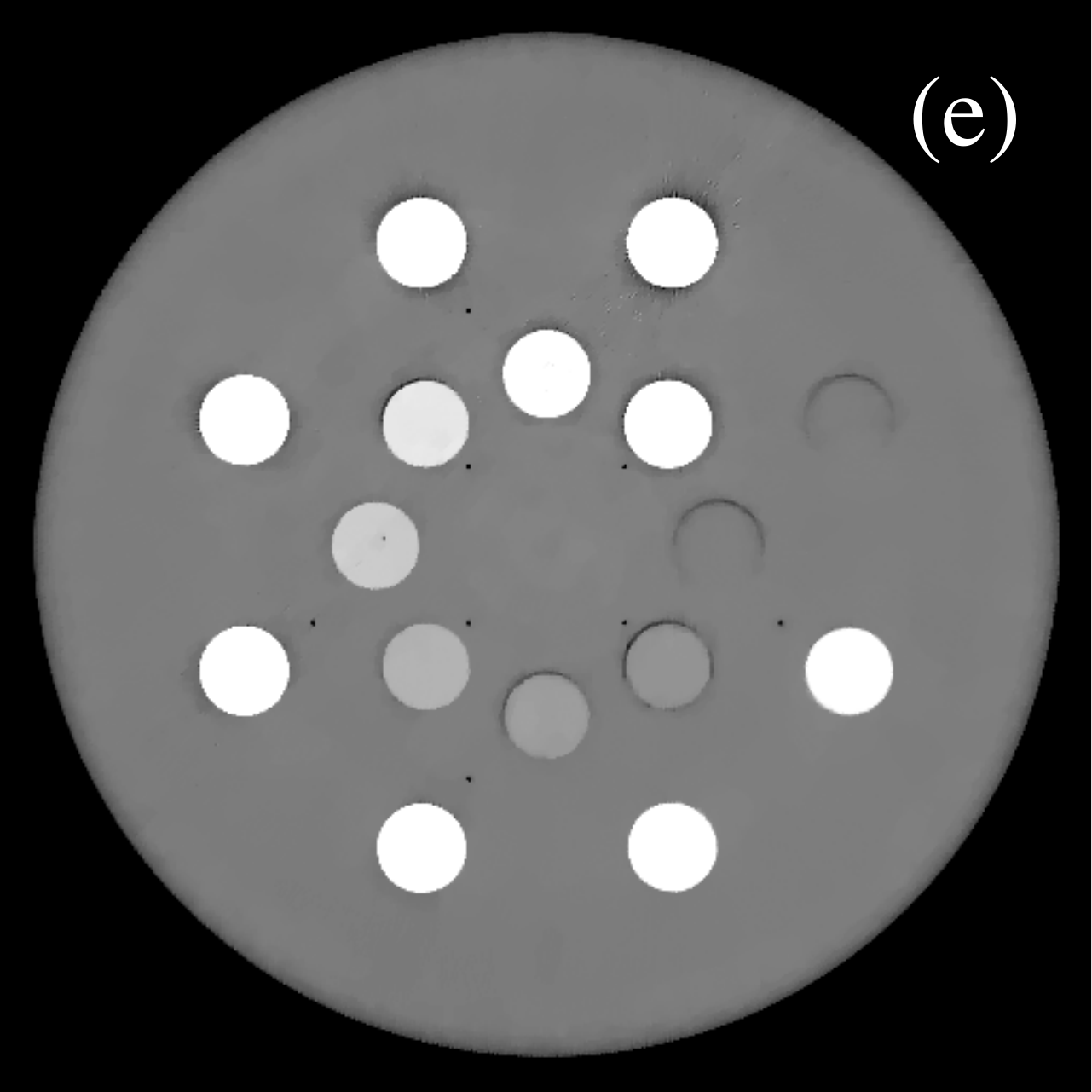}
 		\label{fig:char-realdata-gamma2-iter2000}
		}
		\subfloat[Convergent]{
 		\centering
 		\includegraphics[width=.15\textwidth]{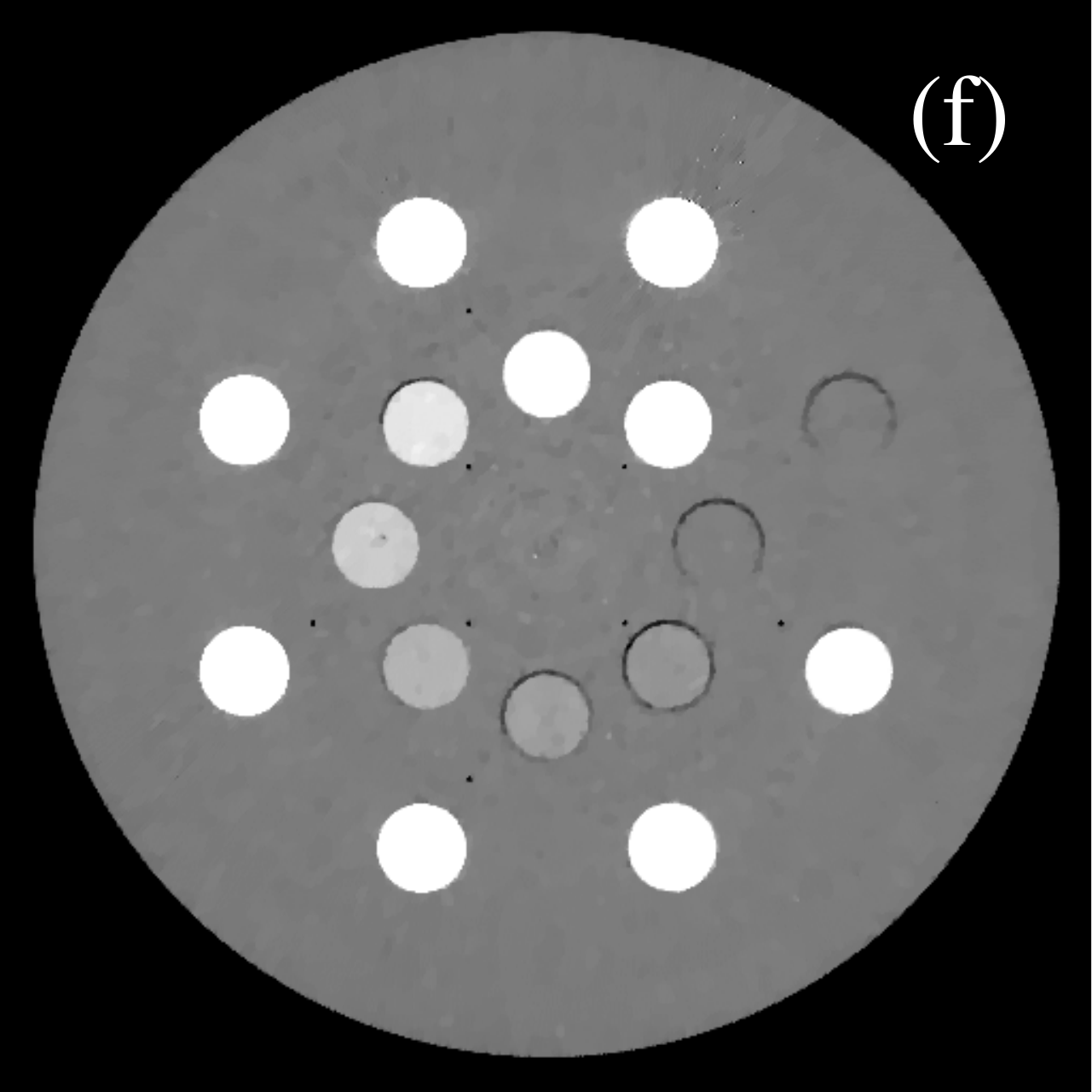}
 		\label{fig:char-realdata-gamma2-iterTemp}
		}
 	\caption{Monochromatic images at energy 100 keV obtained at iteration (a) 30, (b) 100, (c) 500, (d) 1000, and (e) 2000, respectively, and (f) the convergent reconstruction from full-scan real data by use of the NCPD algorithm. Display window: [-150, 150] HU.}
 	\label{fig:real-mono-iter}
\end{figure}

Furthermore, it may be of practical interest in inspecting how the reconstruction evolves as the iteration increases, especially in the form of monochromatic images. We show in Fig.~\ref{fig:real-mono-iter} monochromatic images at energy 100 keV reconstructed with $\gamma=436.6$ at iterations 30, 100, 500, 1000, and 2000, along with the reconstruction obtained when the convergence conditions in Eq. \eqref{eq:conv-conds-nlinear-2} are achieved. 

\section{Results: monochromatic image reconstruction from short-scan data}\label{sec:real-short-data-recon}

\begin{figure}[htb]
 	\centering
    \subfloat{
 		\centering
 		\includegraphics[width=.15\textwidth, trim={25, 25, 25, 25}, clip]{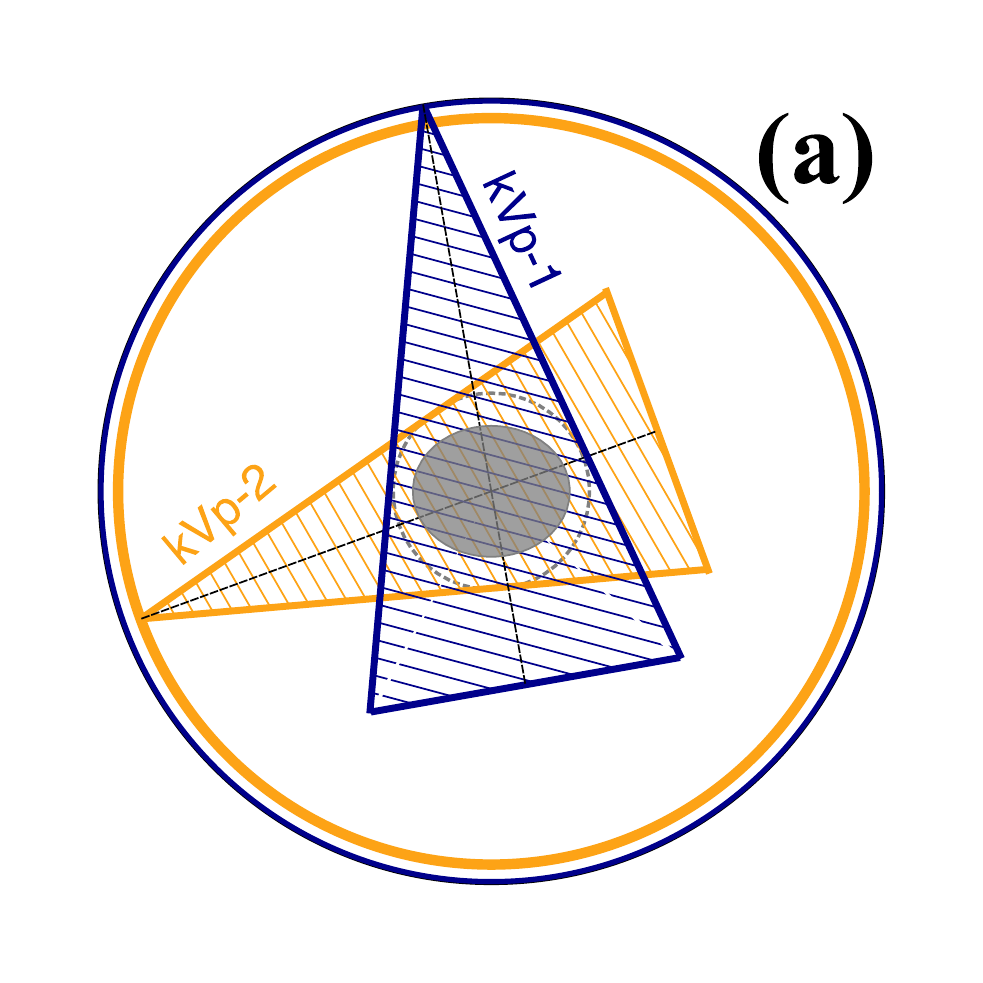}
 		\label{fig:full}
		}%
	\subfloat{
		\centering
 		\includegraphics[width=.15\textwidth, trim={25, 25, 25, 25}, clip]{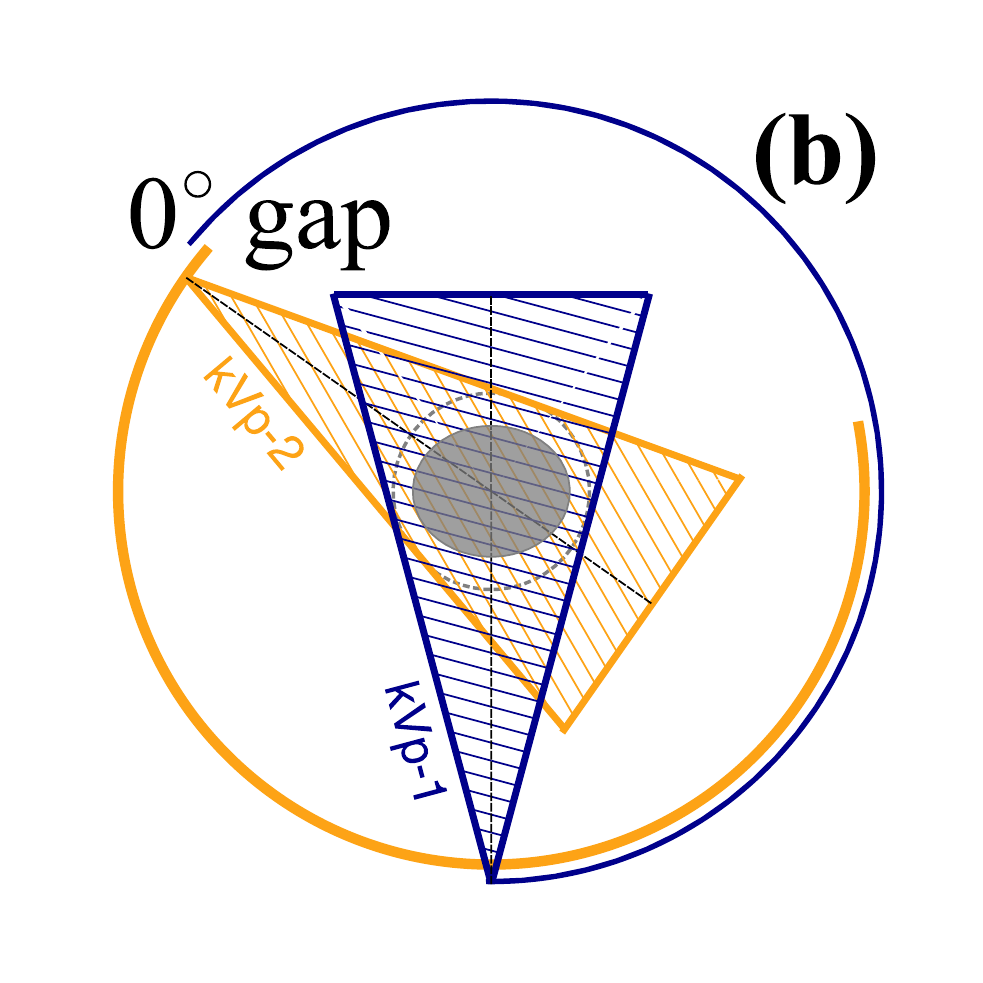}
 		\label{fig:short-0}
		}
	\subfloat{
 		\centering
 		\includegraphics[width=.15\textwidth, trim={25, 25, 25, 25}, clip]{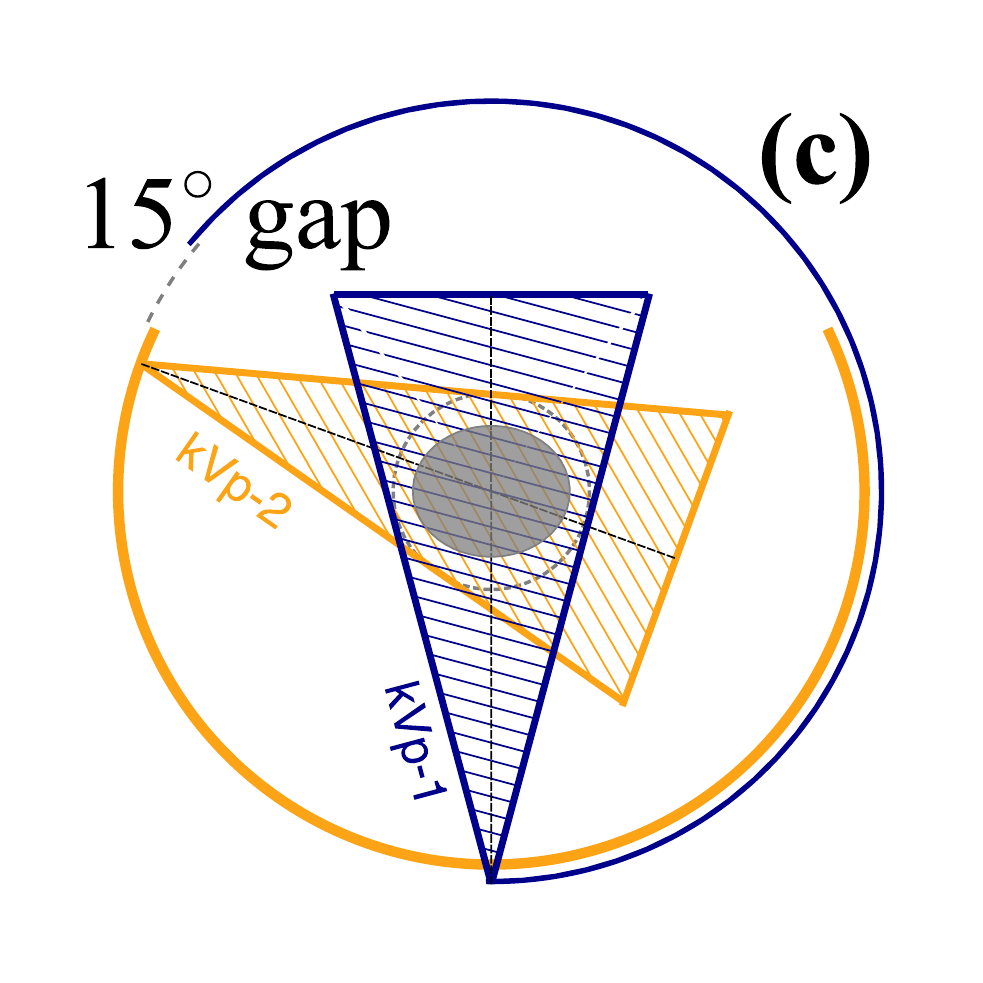}
 		\label{fig:short-15}
}
 	\caption{Illustration of (a) the full-scan configuration and short-scan configurations with (b) $0^\circ$ (b) and (c) $15^\circ$ angular gaps, with the two circular curves depicting scanning angular ranges of two distinct kVp spectra. }
 	\label{fig:short-scan-schemes}
 \end{figure}

The NCPD algorithm may be exploited for investigating accurate image reconstruction (or inversion of the non-linear data model in Eq. \eqref{eq:g-NL-vec}) from a portion of the full-scan data corresponding to data collected with a non-standard scanning configuration of practical interest. In this work, we consider a specific non-standard scanning configuration, simply referred to as the short-scan configuration in dual-energy CT. In this configuration, two regular short-scans are performed each of which is with a distinct spectrum. The two different spectra can be obtained, e.g., through a slow-kV switch technique, and the switching time defines the scanning angular gap between the end of the first regular short scan and the start of the second regular short scan, as illustrated in Fig. \ref{fig:short-scan-schemes}. 

First, from the full-scan simulated-noiseless data in Section~\ref{sec:sim-noiseless-data}, we extract two sets of short-scan simulated-noiseless data, corresponding to short-scan configurations with an angular gap of 0$^\circ$ or 15$^\circ$, mimicking an instantaneous or 15$^\circ$-delayed kV-switch operation.  
In this simulation study, truth-monochromatic image $\mathbf{f}_{m'}^{[\rm truth]}$ is known because truth image $\mathbf{b}^{[\rm truth]}$ is known, and so is its TV (i.e., truth-TV) value that is used as the value of parameter $\gamma$ in the study.

\begin{figure}[htb]
 	\centering
		\subfloat[Full]{
 		\centering
 		\includegraphics[width=.15\textwidth]{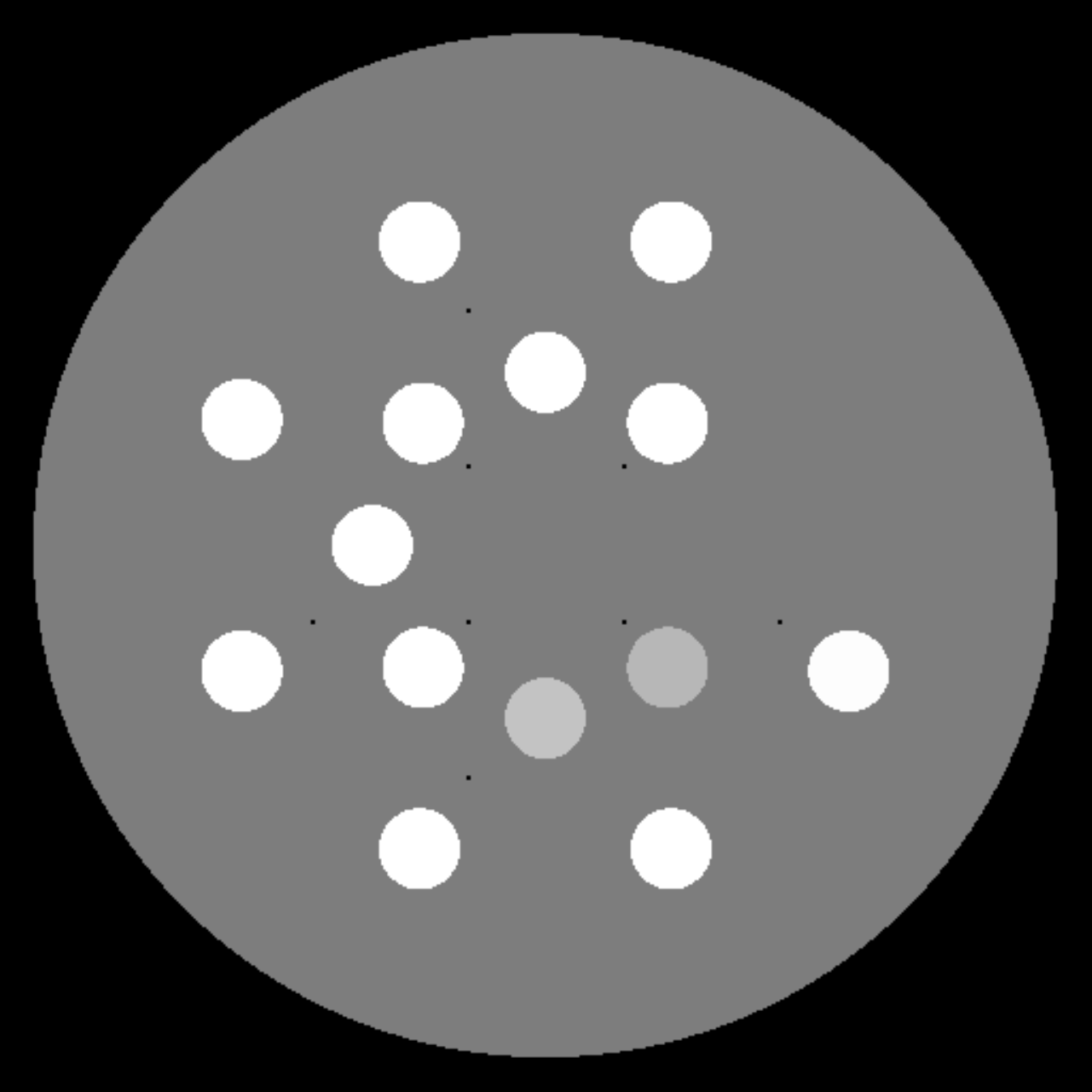}
 		\label{fig:ref}
		}
		\subfloat[Short - 0$^\circ$]{
 		\centering
 		\includegraphics[width=.15\textwidth]{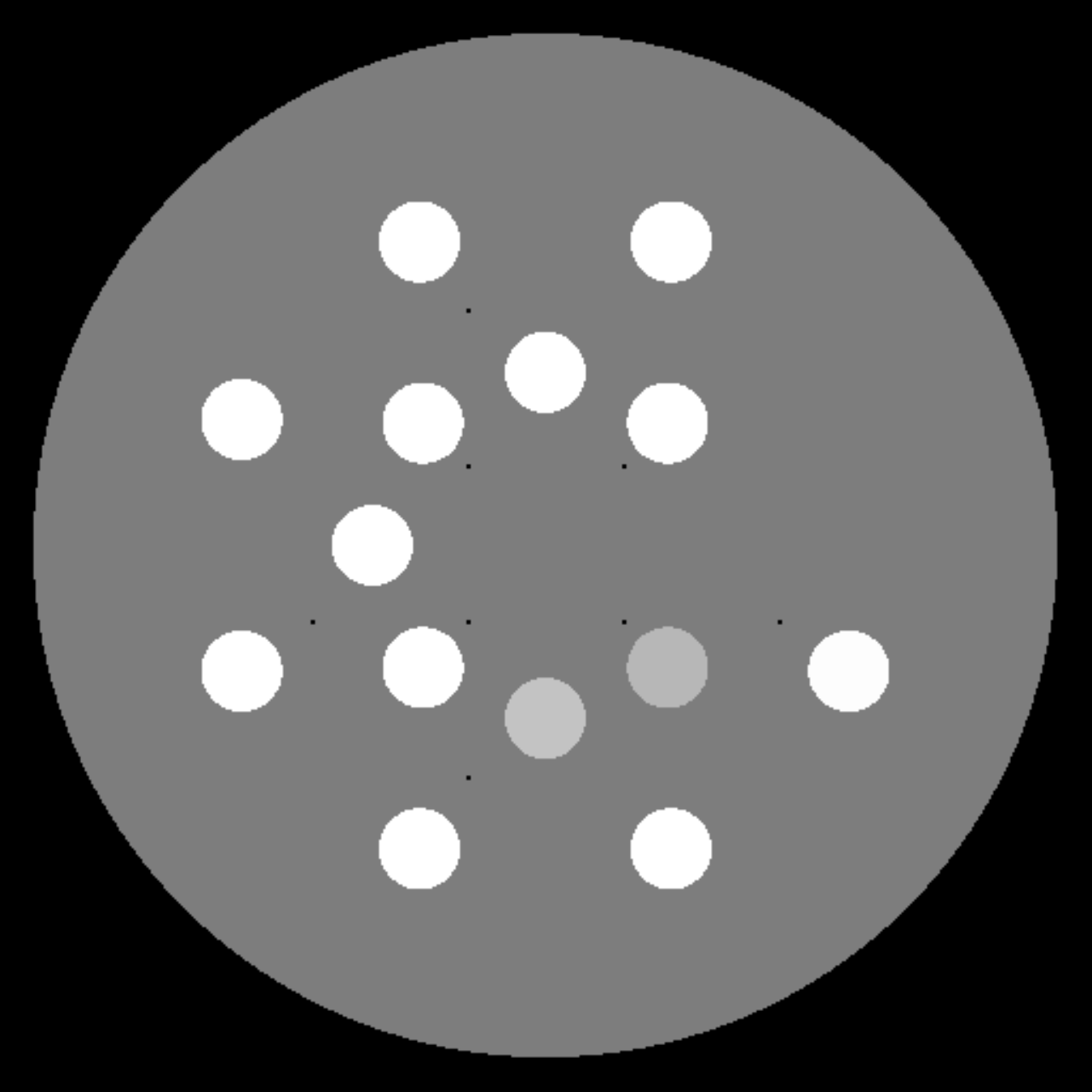}
 		\label{fig:ncpd}
		}
        \subfloat[Short - 15$^\circ$]{
 		\centering
 		\includegraphics[width=.15\textwidth]{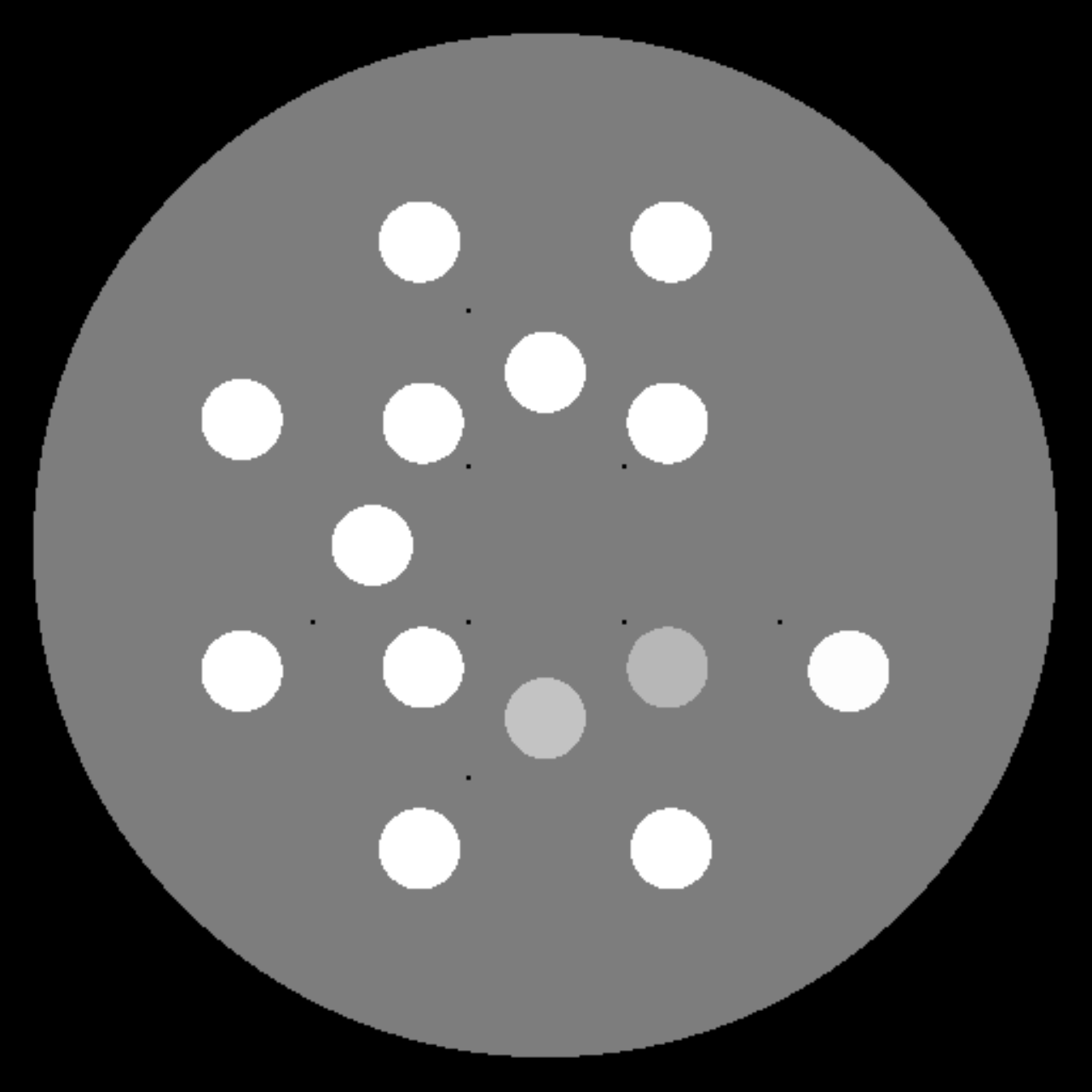}
 		\label{fig:ln}
		}
		\vspace{-5pt}
		\subfloat{
 		\centering
 		\includegraphics[width=.15\textwidth, trim={160 100 50 190}, clip]{figures/rslt_for_manuNCCP/verification_full_p6/nonlinear_single_mono100keV.pdf}
 		\label{fig:ref-roi}
		}
		\subfloat{
 		\centering
 		\includegraphics[width=.15\textwidth, trim={160 100 50 190}, clip]{figures/rslt_for_manuNCCP/verification_full_p6/gap0_mono100keV.pdf}
 		\label{fig:ncpd-roi}
		}
        \subfloat{
 		\centering
 		\includegraphics[width=.15\textwidth, trim={160 100 50 190}, clip]{figures/rslt_for_manuNCCP/verification_full_p6/gap15_mono100keV.pdf}
 		\label{fig:ln-roi}
		}
 	\caption{Monochromatic (top row) and their ROI (bottom row) images at energy 100 keV reconstructed from (a) the full-scan simulated-noiseless data and from short-scan simulated-noiseless data acquired with an angular gap of (b) 0$^\circ$ or (c) 15$^\circ$. 
 	The rectangle ROI is depicted in Fig. \ref{fig:sim-noiseless}. Display window: [-150, 150] HU.}
 	\label{fig:noiseless-disk-short-recons}
\end{figure}

For each set of short-scan simulated-noiseless data, using the NCPD algorithm, we reconstruct the basis images by using the NCPD algorithm,
and then obtain the monochromatic image at 100 keV.
As shown in columns (b) $\&$ (c)  of Fig. \ref{fig:noiseless-disk-short-recons}, the  short-scan images are visually comparable to that obtained with the full-scan data in column (a) of Fig.~\ref{fig:noiseless-disk-short-recons}, showing the potential of the NCPD algorithm for enabling the short-scan configuration. This verification study with noiseless consistent data on the new, non-standard short-scan configurations also surveys the data sampling conditions of the configurations and guides designing the configurations of promise before moving on to the characterization studies with inconsistent data. It sets ``best case scenarios'' and performance upper-bar for the upcoming study with real data.

\begin{figure}[htb]
 	\centering 
		\subfloat[Full]{
 		\centering
 		\includegraphics[width=.15\textwidth]{figures/rslt_for_manuNCCP/characterization/realdata_r/gamma45_mono100keV.pdf}
 		\label{fig:ref}
		}
		\subfloat[Short - 0$^\circ$]{
 		\centering
 		\includegraphics[width=.15\textwidth]{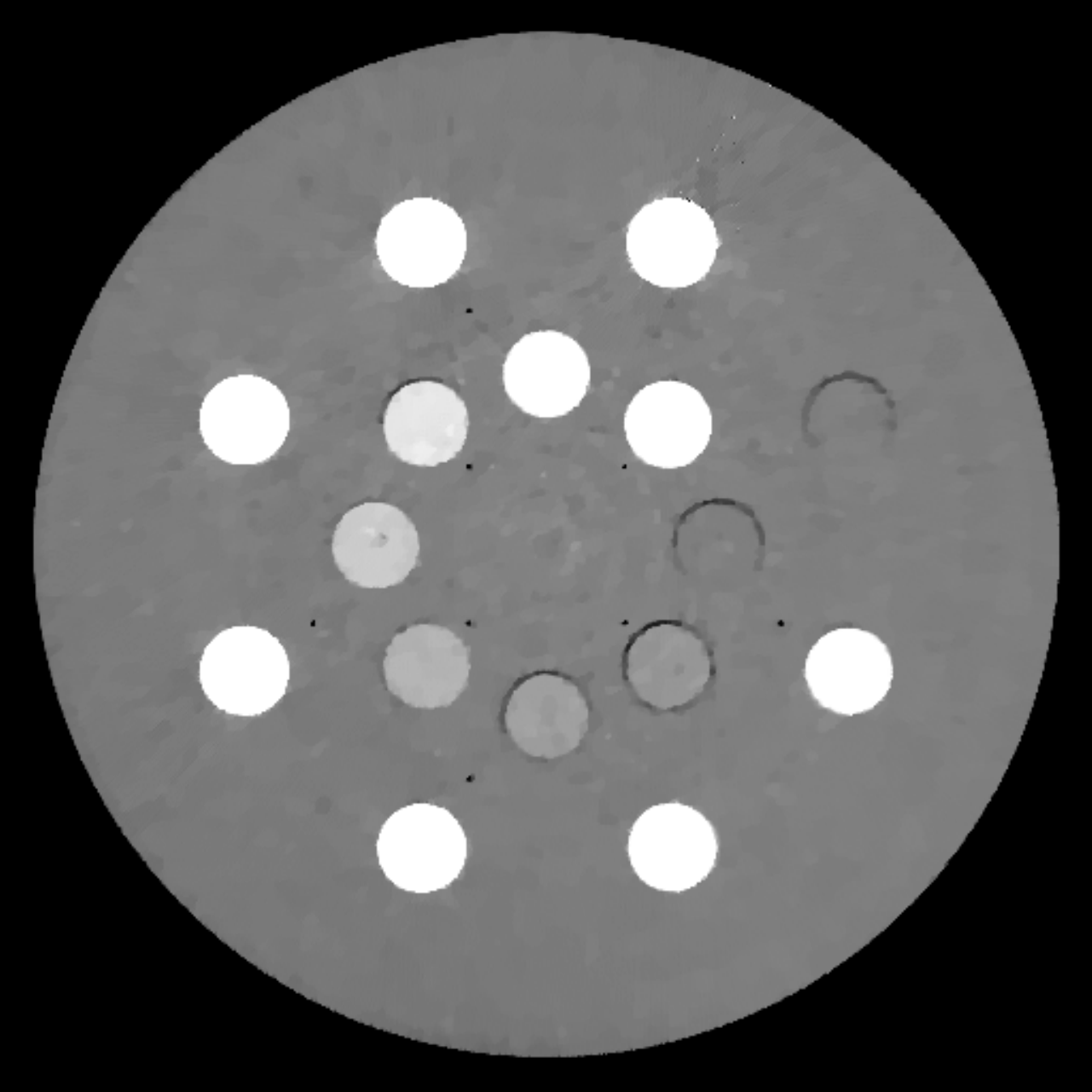}
 		\label{fig:ncpd}
		}
        \subfloat[Short - 15$^\circ$]{
 		\centering
 		\includegraphics[width=.15\textwidth]{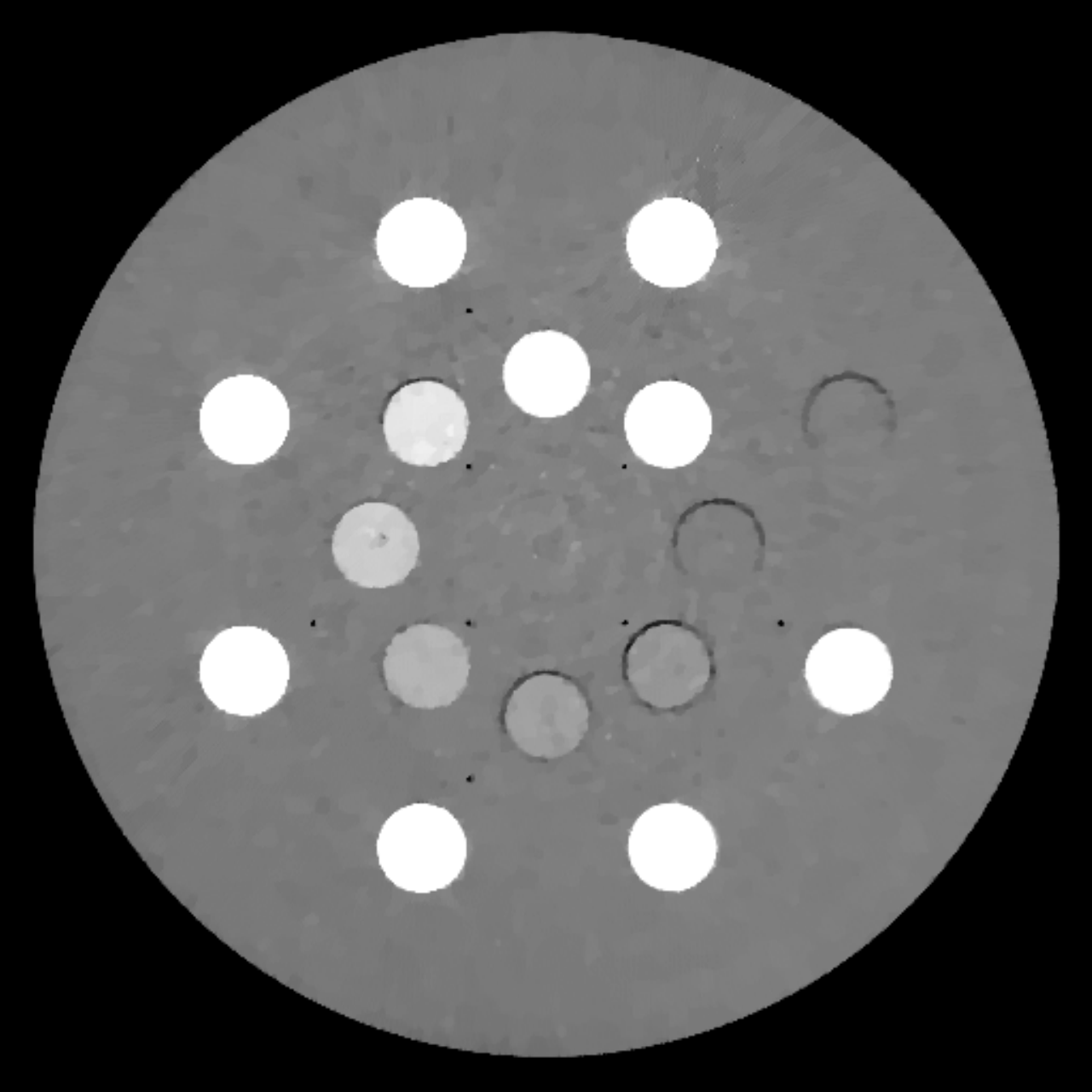}
 		\label{fig:ln}
		}
\vspace{-5pt}
		\subfloat{
 		\centering
 		\includegraphics[width=.15\textwidth, trim={160 100 50 190}, clip]{figures/rslt_for_manuNCCP/characterization/realdata_r/gamma45_mono100keV.pdf}
 		\label{fig:ref-roi}
		}
		\subfloat{
 		\centering
 		\includegraphics[width=.15\textwidth, trim={160 100 50 190}, clip]{figures/rslt_for_manuNCCP/short/short_mono100keV.pdf}
 		\label{fig:ncpd-roi}
		}
        \subfloat{
 		\centering
 		\includegraphics[width=.15\textwidth, trim={160 100 50 190}, clip]{figures/rslt_for_manuNCCP/short/short-gap15_mono100keV.pdf}
 		\label{fig:ln-roi}
		}
 	\caption{Monochromatic (top row) and their ROI (bottom row) images at energy 100 keV reconstructed from (a) the full-scan real data and from short-scan real data acquired with an angular gap of (b) 0$^\circ$ or (c) 15$^\circ$. The rectangle ROI is depicted in Fig. \ref{fig:sim-noiseless}. Display window: [-150, 150] HU.}
 	\label{fig:real-disk-short-recons}
\end{figure}

Second,  from the full-scan real data in Section~\ref{sec:real-data-recon}, we extract two sets of short-scan real data, for the two short-scan configurations in Fig. \ref{fig:short-scan-schemes}. In this real-data study, we use the approach discussed above to selecting $\gamma=436.6$. 
We display in columns (b) $\&$ (c)  of Fig. \ref{fig:real-disk-short-recons} monochromatic images at 100 keV for the two short-scan configurations. It can be observed that they are visually comparable to that obtained with the full-scan data in column (a) of Fig. \ref{fig:real-disk-short-recons}, demonstrating that the NCPD algorithm can enable the short-scan configuration in dual-energy CT. 

\section{Discussion}
We have developed the NCPD algorithm directly to invert the non-linear data model in spectral CT.  Considering the heuristic and iterative nature of the NCPD algorithm, we have designed and performed verification studies numerically to demonstrate that the NCPD algorithm can solve the non-convex optimization program and under appropriate data condition, can invert the non-linear data model in spectral CT. While the NCPD algorithm is adapted heuristically from the CPD algorithm that is developed on a mathematically exact footing, its heuristic component is limited only to the adaptation step. As such, the NCPD algorithm appears to possess desirable features:  it becomes the  CPD algorithm in the absence of the spectral effect, and thus the CPD algorithm provides a useful benchmark of the NCPD convergence property; and it can enable prototyping of different scanning configurations, thus possibly yielding insights into designs of scanning configurations of practical workflow interest in spectral CT. Therefore, the NCPD algorithm may be exploited for helping design image-reconstruction algorithms of practical value.

We have reconstructed images by using the NCPD algorithm from simulated and real data. The study results reveal that the NCPD algorithm can correct effectively for the non-linear beam-hardening artifacts observed in images reconstructed with an algorithm for a linear data model. In the real-data study, we have also demonstrated that the involved parameters such as $\gamma$ can impact image reconstruction, as expected, in the presence of data components such as noise that are inconsistent with the non-linear data model upon which the NCPD algorithm is based. We have also carried out additional studies with different energy spectra and with numerical and physical phantoms of various levels of anatomy complexity of interest. While results of these studies are not shown, observations  consistent with those presented were obtained.

The NCPD algorithm can accommodate data in which some of the rays are not  measured with the full set of multiple spectra,
and can thus allow for enabling spectral CT with non-standard scanning configurations, as illustrated in the study examples of short-scan dual-energy CT. The NCPD algorithm can also readily be applied to spectral data acquired with photon-counting or sandwiched detectors. In the case of photon-counting detector, spectrum $q_{sj m}$ in the non-linear model in Eq.~\eqref{eq:g-disc} would denote the response function within energy channel $s$ for incoming photons at energy $m$ and ray $j$. Ideally, the response function would have its support within the energy channel. Practically, due to physical factors such as charge sharing, the response function would have spill-overs to neighboring channels. As a result, direct algorithms such as the NCPD algorithm that deal directly with the non-linear data model in Eq. (2) may be sufficiently flexible to take into consideration such non-idealities.

As shown in Eq. \eqref{eq:opt-constr-NL}, the constraints on image TV and non-negativity are applied to monochromatic image $\mathbf{f}_{m'}$, instead of the basis images themselves, at an energy specified by energy bin $m'$. While being assumed to be independent of spectral energy  (see Eq.~\eqref{eq:basis-decomposition}), the basis images reconstructed through solving Eq. \eqref{eq:opt-constr-NL} are possibly to be dependent on the choice of energy level $m'$.  It remains to be investigated as to how constraints on monochromatic images at different or
combined energies in Eq.~\eqref{eq:opt-constr-NL} impact the basis images and consequently monochromatic images obtained.

We have based the NCPD algorithm upon the specific CPD algorithm. Additional algorithms may be developed for solving non-convex optimization programs by basing upon other algorithms 
for solving the convex optimization programs. Again, whether such algorithms can accurately solve the non-convex optimization programs and invert the non-linear data model considered needs to be investigated.

The NCPD algorithm can be extended potentially to address other non-convex optimization programs in spectral CT. It would be a future research topic of significance to evaluate the practical utility of the NCPD algorithm in rigorous assessment studies. The general approach taken, and the development of the CPD and NCPD algorithms, may also yield insights into solving non-convex optimization programs \citep{ochs2014ipiano, valkonen2014primal},
and inverting non-linear data models \citep{liu2019optimization, pan2015optimization}, in other imaging technologies and applications.

\section{Conclusion}
We have developed an algorithm, referred to as the NCPD algorithm, to reconstruct images through solving a non-convex optimization program for the non-linear data model in spectral CT.  Results of simulated- and real-data studies suggest that the NCPD algorithm can yield accurate reconstruction from full-scan data and that it has the potential to enable non-standard scanning configuration such as the short-scan configurations in spectral CT. The work may also yield insights into solving non-convex optimization programs and inverting non-linear data models that arise in other imaging modalities.

 \appendix

\section{Derivation of the new CPD instance of the general PD algorithm} \label{sec:app-a}
We consider a linear data model 
\begin{equation}\label{eq:linear-data-model}
\mathbf{g}^{[L]}(\mathbf{b})= \mathcal{H} \mathbf{b} + \Delta \mathbf{g}_{\rm c},
\end{equation}
 and a convex optimization program 
\begin{equation} \label{eq:opt-constr-ln}
\begin{aligned}
	{\mathbf{b}}^\star &= \argmin_\mathbf{b}
    D_{\mathbf{g}^{[L]}}(\mathbf{b}) \\
    &\mathrm{s.\,t.} \,\,
   || ( |\nabla \mathbf{f}_m'(\mathbf{b}) | )||_1
    \leq \gamma \,\,{\&}
   \,\, \mathbf{f}_m'(\mathbf{b}) \succeq \!0,
\end{aligned}
\end{equation}
where $\Delta \mathbf{g}_{\rm c}$ denotes a constant data vector independent of $\mathbf{b}$,  
\begin{equation}\label{eq:data-div-L}
\begin{aligned}
D_{\mathbf{g}^{[L]}}(\mathbf{b}) & \!=\!  || \mathbf{g}^{[\mathcal{M}]}  - \mathbf{g}^{[L]}(\mathbf{b})  ||_2^2/2 \\
&\! =\!  ||  (\mathbf{g}^{[\mathcal{M}]} - \Delta  \mathbf{g}_c) - \mathcal{H}\, \mathbf{b}  ||_2^2/2,
\end{aligned}
\end{equation}
and notations and symbols are consistent with those in Section~\ref{Sec:methods}.
It can be observed that the linear term in Eq. \eqref{eq:linear-data-model} is identical to the linear component of the non-linear model in Eq. \eqref{eq:g-NL-vec} and that the form of the convex optimization program in Eq.~\eqref{eq:opt-constr-ln} is identical to that of the non-convex optimization program in Eq.~\eqref{eq:opt-constr-NL} except for the difference between $\Delta \mathbf{g}_{\rm c}$ in the former and $\Delta \mathbf{g}^{[NL]}(\mathbf{b})$ in the latter. It is the similarity observed between the two cases that motivates the development of the NCPD algorithm by adapting the CPD algorithm that is derived below for solving the convex optimization program in Eq.~\eqref{eq:opt-constr-ln}.
 
While the general PD algorithm can mathematically exactly solve the convex optimization program in Eq.~\eqref{eq:opt-constr-ln}, a direct, brute-force implementation of some of the key operations, which are expressed only in mathematically compact forms (e.g., the proximal operations,) in the general PD algorithm can often lead to ambiguity issues concerning not only computational accuracy and efficiency but also additional parameters required, thus compromising their original theoretical rigor and applicability. Also, the general PD algorithm provides little concrete knowledge of its convergence and efficiency properties when solving a specific convex optimization program. Therefore, we develop the CPD algorithm, as a new instance of the general PD algorithm, by deriving analytic, close-form solutions to the proximal operations specific to the convex optimization program in Eq.~\eqref{eq:opt-constr-ln} so that the solution can readily be implemented accurately and efficiently.   

The convex optimization program in Eq.~\eqref{eq:opt-constr-ln} can be fitted into a primal minimization~\citep{chambolle_first-order_2010, sidky_convex_2012} as
\begin{equation}\label{eq:primal}
	\min_{\mathbf{x}} \,\, \{ F(\mathcal{K}\mathbf{x}) + G(\mathbf{x}) \},
\end{equation}
where, with ${\mathbf{g}'}={\mathbf{g}^{[\mathcal{M}]}}-{\Delta \mathbf{g}_c}$, 
\begin{equation} \label{eq:PD-assoc}
\begin{aligned}
    F(\mathcal{K}\mathbf{x}) &= F(\mathbf{u}, \mathbf{v}, \mathbf{w})
    =
    F_1(\mathbf{u}) +
    F_2(\mathbf{v}) +
    F_3(\mathbf{w}), \\
	F_1(\mathbf{u}) &= \frac{1}{2} || \mathbf{u} - \mathbf{g}'||_2^2, \,\,\,F_3(\mathbf{w}) = \delta_P ( \mathbf{w} ),\\
	F_2(\mathbf{v}) &= \delta_{\mathrm{diamond}(\alpha_{m'} \gamma)} (| \mathbf{v}
|), \,\,\, G( \mathbf{x} ) = 0,
\end{aligned}
\end{equation}
\vskip -.2cm
\begin{equation} \label{eq:PD-assoc-1}
\begin{aligned}
	\mathbf{x} &= \mathbf{b}, \quad
	\mathbf{y} = \mathcal{K}_{m'}\mathbf{b} = \begin{pmatrix}
											\mathbf{u} \\
											\mathbf{v} \\
											\mathbf{w}
	                                    \end{pmatrix}, \quad
	                                   \mathcal{K}_{m'} = \begin{pmatrix}
					\mathcal{H} \\
					\alpha_{m'} \mathcal{U}_{m'} \\
					\beta_{m'} \mathcal{V}_{m'}
	              \end{pmatrix}, \\
	\mathbf{u} &= \mathcal{H}\mathbf{b}, \quad
	\mathbf{v} = \alpha_m \mathcal{U}_{m'} \mathbf{b}, \quad
	\mathbf{w} = \beta_m \mathcal{V}_{m'} \mathbf{b},	
\end{aligned}
\end{equation}
 $\mathbf{x}$ or $\mathbf{b}$ is the primal variable; $\mathbf{y}$ is the splitting variable that consists of three independent components, $\mathbf{u}$, $\mathbf{v}$, and $\mathbf{w}$, obtained by applying  matrix $\mathcal{K}_{m'}$ to $\mathbf{b}$; matrices $\mathcal{U}_{m'}=(\mu_{m 1} \nabla, \mu_{m 2} \nabla, ..., \mu_{m K} \nabla)$ and $\mathcal{V}_{m'}=(\mu_{m 1} \mathcal{I}, \mu_{m 2} \mathcal{I}, ..., \mu_{m K} \mathcal{I})$; $\alpha_{m'} = || \mathcal{H} ||_2 / || \mathcal{U}_m ||_2, \beta_m = ||  \mathcal{H}  ||_2 / || \mathcal{V}_m ||_2$. Here, $\alpha_{m'}$ and $\beta_{m'}$ are multiplicative factors for the TV and non-negativity constraints in Eq. (18), which are computed in this work as the ratio of the matrix norms to scale and balance the magnitude of the matrices in stacked matrix $\mathcal{K}_{m'}$ \citep{sidky_convex_2012}.
Further, $\delta_\mathrm{diamond(\alpha\gamma)}(\mathbf{x})$ and $\delta_P(\mathbf{x})$ are indicator functions in the form of 
\begin{equation}
	\delta_S(\mathbf{x}) =
	\begin{cases}
		0	\quad & \mathbf{x} \in S \\
		\infty \quad & \mathbf{x} \notin S
	\end{cases},
\end{equation}
where $S$ is a convex set. Here $\mathrm{diamond}(\alpha\gamma)$ and $P$ are two sets defined as $\mathrm{diamond}(\alpha\gamma) = \{\mathbf{x} | \,\, || \mathbf{x} ||_1 \le \alpha\gamma\}$ and $P = \{ \mathbf{x} | \,\, x_i \ge 0, \forall i\}$. Similarly, $\delta_\mathbf{0}(\mathbf{x})$ is also an indicator function for a set of zero vectors, defined as $\{ \mathbf{x} | \,\, x_i = 0, \forall i\}$.

Using $\mathbf{p}$, $\mathbf{q}$, and $\mathbf{r}$ to denote the respective dual variables of $\mathbf{u}$, $\mathbf{v}$, and $\mathbf{w}$, we can obtain $F^*$ and $G^*$, the convex conjugate functions  of $F$ and $G$, as
\begin{equation} \label{eq:f-conj}
\begin{aligned}
	F^*(\mathbf{p}, \mathbf{q}, \mathbf{r}) &= F^*_1(\mathbf{p}) +
F^*_2(\mathbf{q}) + F^*_3(\mathbf{r}), \\
	G^*(\mathbf{s}) &= \delta_\mathbf{0} (\mathbf{s}),
\end{aligned}
\end{equation}
where
\begin{equation} \label{eq:fcomponent-conj}
\begin{aligned}
	F^*_1(\mathbf{p}) &= \frac{1}{2} || \mathbf{p} ||_2^2 +
{\mathbf{g}'}^\top \mathbf{p}, \\
	F^*_2(\mathbf{q}) &= \alpha_{m'}\gamma || (| \mathbf{q} |) ||_\infty, \\
	F^*_3(\mathbf{r}) &= \delta_P ( -\mathbf{r} ).
\end{aligned}
\end{equation}
The dual optimization of Eq. \eqref{eq:primal} takes the general form of
\begin{equation}
\max_{\mathbf{y}} \,\, \{ -F^*(\mathbf{y}) - G^*(-\mathcal{K}^\top \mathbf{y})
\},
\end{equation}
which can be written explicitly here as
\begin{equation} \label{eq:opt-dual}
\begin{aligned}
    \max_{\mathbf{p},\mathbf{q},\mathbf{r}}
	& - \frac{1}{2} || \mathbf{p} ||_2^2 - {\mathbf{g}'}^\top
\mathbf{p}
	- \alpha_{m'} \gamma || (| \mathbf{q} |) ||_\infty
	- \delta_P ( -\mathbf{r} ) \\
	& - \delta_\mathbf{0}
	\left( -\mathcal{H}^\top \mathbf{p}
		- \alpha_{m'} \mathcal{U}_{m'}^\top  \mathbf{q}
		- \beta_{m'} \mathcal{V}_{m'}^\top \mathbf{r}
			\right).
\end{aligned}
\end{equation}
The difference between the primal and dual objectives in Eqs.~\eqref{eq:opt-constr-ln} and~\eqref{eq:opt-dual}, without the indicator functions, is referred to as the conditional PD (cPD) gap \citep{sidky_convex_2012}, which is given by
\begin{equation} \label{eq:cpd}
\begin{aligned}
	\mathrm{cPD}(\mathbf{b}, \mathbf{p}, \mathbf{q})
	& = \frac{1}{2} || \mathbf{g}' - \mathcal{H}\mathbf{b}||_2^2
	+ \frac{1}{2} || \mathbf{p} ||_2^2 \\
	 & +  {\mathbf{g}'}^\top \mathbf{p} + \alpha_{m'} \gamma || (| \mathbf{q} |) ||_\infty.
\end{aligned}
\end{equation}

A proximal mapping of function $F^*(\mathbf{p})$ is defined as 
\begin{equation} \label{eq:prox}
	\prox_\sigma[F^*](\mathbf{p})
 = \argmin_\mathbf{p'}
	\left\{
		F^*(\mathbf{p}')
		+
		\frac{|| \mathbf{p} - \mathbf{p'} ||_2^2}{2\sigma}
	\right\}.
\end{equation}
For deriving the CPD algorithm of our interest, we first determine the proximal mappings of $F^*$ and $G$. Noticing Eq.~\eqref{eq:f-conj}, the proximal mapping of $F^*$ is given by 
\begin{equation} \label{eq:prox-F}
\begin{aligned}
	\prox_\sigma[F^*](\mathbf{p},\mathbf{q},\mathbf{r})
& =  \prox_\sigma[F_1^*](\mathbf{p}) 
+ \prox_\sigma[F_2^*](\mathbf{q}) \\
& + \prox_\sigma[F_3^*](\mathbf{r}),
\end{aligned}
\end{equation}
where
\begin{equation} \label{eq:prox-F1}
	\prox_\sigma[F_1^*](\mathbf{p})
= \frac{\mathbf{p} - \sigma \mathbf{g}' }{1+\sigma},
\end{equation}
\begin{equation} \label{eq:prox-F2}
	\prox_\sigma[F_2^*](\mathbf{q})
	=\mathbf{q} - \sigma \frac{\mathbf{q}}{ |\mathbf{q}| }
	\mathrm{POLB}_{\alpha_{m'} \gamma}(
	\frac{ |\mathbf{q} | }{\sigma} ),
\end{equation}
\begin{equation} \label{eq:prox-F3}
	\prox_\sigma[F_3^*](\mathbf{r})
= \mathrm{neg}(\mathbf{r})
	\equiv
	\begin{cases}
		r_i 	& \quad \text{if } r_i \leq 0\\
		0 		& \quad \text{if } r_i > 0 \\
    \end{cases}.
\end{equation}
The proximal mapping of $G$ is given as
\begin{equation} \label{eq:prox-G}
	\prox_\tau[G](\mathbf{x})
= \mathbf{x}.
\end{equation}
\noindent \paragraph{The CPD algorithm and its pseudo codes} 
Substituting the proximal mappings derived above into the corresponding update steps in the general PD algorithm~\citep{sidky_convex_2012}, we obtain the CPD algorithm with the pseudo-code in Algorithm~\ref{alg:algo} for solving the optimization problem in Eq.~\eqref{eq:opt-constr-ln}, where $\mathrm{POLB}_{\gamma_0}(\mathbf{x})$ 
indicates a $\ell_2$-projection onto the $\ell_1$-ball of size $\gamma_0$~\citep{duchi2008efficient}; neg($\mathbf{x}$) denotes a non-positivity thresholding on each component of $\mathbf{x}$. 

\vskip -.1cm
\begin{algorithm}
\caption{Pseudo-code of the CPD algorithm}
\label{alg:algo}
\begin{algorithmic}[1]
\State $\alpha_{m'} \leftarrow ||\mathcal{H}||_2/||\mathcal{U}_{m'}||_2$;
$\beta_{m'} \leftarrow ||\mathcal{H}||_2/||\mathcal{V}_{m'}||_2$;
\State $L_{m'} \leftarrow ||\mathcal{K}_{m'}||_2$; $\sigma_{m'} \leftarrow
1/L_{m'}$; $\tau_{m'}
\leftarrow 1/L_{m'}$; $\theta \leftarrow 1$
\State Initialize $\mathbf{b}^{(0)}, \mathbf{p}^{(0)}, \mathbf{q}^{(0)}$ and
$\mathbf{r}^{(0)}$ to zero
\State $\bar{\mathbf{b}}^{(0)} \leftarrow \mathbf{b}^{(0)}$
\Repeat
	\State $\displaystyle \mathbf{p}^{(n+1)} = \frac{\mathbf{p}^{(n)} -
\sigma_{m'} ( \mathbf{g}^{[\mathcal{M}]}- \Delta\mathbf{g}_c -\mathcal{H} \bar{\mathbf{b}}^{(n)})} {1+\sigma_{m'}} $
	\State $\mathbf{q}'^{(n)} = \mathbf{q}^{(n)} + \sigma_{m'} \alpha_{m'}
\sum_k\mu_{m k}\nabla \bar{\mathbf{b}}_{k}^{(n)}$
	\State $\displaystyle \mathbf{q}^{(n+1)}
		\!\!=\! \mathbf{q}'^{(n)}
		\!\!-\! \sigma_{m'} \frac{\mathbf{q}'^{(n)}}{|\mathbf{q}'^{(n)}|}
		\mathrm{POLB}_{\alpha_{m'}\gamma}(
\frac{|\mathbf{q}'^{(n)}|} {\sigma_{m'}} ) $
	\State $\mathbf{r}^{(n+1)} = \mathrm{neg}(\mathbf{r}^{(n)} +
\sigma_{m'}{\beta_{m'}}(\sum_k\mu_{m k}\bar{\mathbf{b}}_{k}^{(n)}
))$
	\State $\mathbf{b}^{(n+1)}
	= \mathbf{b}^{(n)}
	- \tau_{m'}
	\mathcal{K}_{m'}^\top \begin{pmatrix} \mathbf{p}^{(n+1)}\\ \mathbf{q}^{(n+1)} \\ \mathbf{r}^{(n+1)}\end{pmatrix}
	$
	\State $\bar{\mathbf{b}}^{(n+1)} = \mathbf{b}^{(n+1)} + \theta
(\mathbf{b}^{(n+1)} - \mathbf{b}^{(n)})$
\Until{convergence conditions are satisfied}
\end{algorithmic}
\end{algorithm}

\noindent \paragraph{General convergence conditions} We devise general convergence conditions of the CPD algorithm as
\begin{equation}\label{eq:conv-conds-linear-2}
\begin{aligned}
	 d\widetilde{D}_{\mathbf{g}}^{(n)} \rightarrow 0
     \quad
	\widetilde{D}_\text{TV}^{(n)}\rightarrow 0
	 \quad
	 d\widetilde{D}_\mathbf{b}^{(n)} \rightarrow 0,
\end{aligned}
\end{equation}
\vskip -.6cm
\begin{equation}\label{eq:conv-conds-linear-1}
\begin{aligned}
   \widetilde{\text{cPD}}^{(n)} \rightarrow 0
	 \quad  
	\widetilde{\text{T}}^{(n)}\rightarrow 0
	 \quad
	\widetilde{\text{S}}^{(n)}\rightarrow 0,
	\end{aligned}
\end{equation}
as $n\rightarrow \infty$, where the dimensionless metrics are defined as
\begin{equation}\label{eq:Dg-Dtv-Df-metrics}
\begin{aligned}
  	d\widetilde{D}_{\mathbf{g}}^{(n)}
    &  ={\vert D_{\mathbf{g}^{[L]}}(\mathbf{b}^{(n)})-D_{\mathbf{g}^{[L]}}(\mathbf{b}^{(n-1)})\vert}/{||\mathbf{g}^{[\mathcal{M}]}||_2}\\
    \widetilde{D}_\text{TV}^{(n)}
	&= {|\,|| ( |\nabla \mathbf{f}_{m'}'(\mathbf{b}^{(n)}) | )||_1-\gamma|}/{\gamma}\\
	d\widetilde{D}_\mathbf{b}^{(n)}
&= {||\mathbf{b}^{(n)}-\mathbf{b}^{(n-1)}||_2}/{||\mathbf{b}^{(n-1)}||_2},
\end{aligned}
\end{equation}
\begin{equation}\label{eq:normal-PD-T-S-metrics}
\begin{aligned}
 	\hskip -.3cm
 	\widetilde{\text{cPD}}^{(n)}=\frac{{\text{cPD}}^{(n)}}{{\text{cPD}}^{(1)}},
\,\,\,
	\widetilde{\text{T}}^{(n)}=\frac{{\text{T}}^{(n)}}{{\text{T}}^{(1)}},
	\,\,\,
	{\rm and}
\,\,\,
	\widetilde{\text{S}}^{(n)}=\frac{{\text{S}}^{(n)}}{{\text{S}}^{(1)}},
\end{aligned}
\end{equation}
conditional primal-dual (cPD) gap ${\text{cPD}}^{(n)}$~\citep{xia2016optimization},  transversality ${\text{T}}^{(n)}$~\citep{hiriart2013convex}, and dual residual  ${\text{S}}^{(n)}$~\citep{goldstein2013adaptive} are given by
\begin{equation}\label{eq:PD-T-S-metrics}
\begin{aligned}
 	{\text{cPD}}^{(n)}	&\!= {\text{cPD}}(\mathbf{b}^{(n)}, \mathbf{p}^{(n)}, \mathbf{q}^{(n)}),\\
	{\text{T}}^{(n)}
	&\!=\!  ||\mathcal{H}^\top \mathbf{p}^{(n)}
		+ \alpha_{m'} \mathcal{U}_{m'}^\top  \mathbf{q}^{(n)}
		+ \beta_{m'} \mathcal{V}_{m'}^\top \mathbf{r}^{(n)}||_2 ,\\
	{\text{S}}^{(n)}
	&\!=\! ||\frac{1}{\sigma}\!\!\begin{pmatrix}
                            \!\mathbf{p}^{(n)}\! -\! \mathbf{p}^{(n-1)}\! \\
							\!\mathbf{q}^{(n)}\! - \!\mathbf{q}^{(n-1)}\!\\
							\!\mathbf{r}^{(n)}\! -\! \mathbf{r}^{(n-1)}\!
                        \end{pmatrix}
      \!-\!\mathcal{K}_{m'} (\!\mathbf{b}^{(n)}\! -\! \mathbf{b}^{(n-1)}\!) ||_2.
\end{aligned}
\end{equation}
The conditions on $d\widetilde{D}_{\mathbf{g}}^{(n)}$ and $d\widetilde{D}_\mathbf{b}^{(n)}$ check the change rates of the data-divergence and reconstructed image, whereas conditions on $\widetilde{\text{cPD}}^{(n)}$, $\widetilde{\text{T}}^{(n)}$, $\widetilde{\text{S}}^{(n)}$, and $\widetilde{D}_\text{TV}^{(n)}$ form a set of first-order optimality conditions~\citep{sidky_convex_2012,hiriart2013convex} for solving Eq.~\eqref{eq:opt-constr-ln}.

\noindent \paragraph{Data-convergence condition}
For measured data consistent with the linear data model, i.e., $\mathbf{g}^{[\mathcal{M}]}=\mathbf{g}^{[L]}(\mathbf{b}^{[\rm truth]})$, we can devise an additional convergence condition as 
\begin{equation}\label{eq:conv-conds-linear-3}
\begin{aligned}
	\widetilde{D}_{\mathbf{g}}^{(n)}
		& ={D_{\mathbf{g}^{[L]}}(\mathbf{b}^{(n)})}/{||\mathbf{g}^{[\mathcal{M}]}||_2}
		\rightarrow 0,
\end{aligned}
\end{equation}
as $n\rightarrow \infty$.
Clearly, for a case in which measured data are inconsistent with the data model, i.e., $\mathbf{g}^{[\mathcal{M}]}\neq\mathbf{g}^{[L]}(\mathbf{b}^{[\rm truth]})$, this condition cannot be used because data term $D_{\mathbf{g}^{[L]}}(\mathbf{b}^{(n)})$ is generally non-zero.
It remains unknown, however, as to whether it can accurately invert the linear data model in Eq. \eqref{eq:linear-data-model}. 

\noindent \paragraph{Image-convergence condition}
The convergence conditions discussed above are used for demonstrating that the CPD algorithm can solve the convex optimization program in  Eq. \eqref{eq:opt-constr-ln}. However, evidence remains to be established that the CPD algorithm can invert the linear data model in Eq. \eqref{eq:linear-data-model}. We seek to establish the evidence under the conditions that $\mathbf{g}^{[\mathcal{M}]}= \mathbf{g}^{[L]}(\mathbf{b})$ and that $\mathbf{b}^\mathrm{[truth]}$ is known, as the CPD algorithm is developed based upon the linear data model. Therefore, we devise an image-convergence condition as 
\begin{equation}\label{eq:conv-conds-linear-image}
   	\widetilde{D}_\mathbf{b}^{(n)}
		= {||\mathbf{b}^{(n)}-\mathbf{b}^{\text{[truth]}}||_2}/{||\mathbf{b}^{[\text{truth}]}||_2}
		\rightarrow 0,
\end{equation}
as $n\rightarrow \infty$.

\section{Numerical demonstration of the CPD algorithm} \label{sec:app-b}

While the CPD algorithm is known mathematically exactly to solve the convex optimization program in Eq.~\eqref{eq:opt-constr-ln}, it remains to be demonstrated to solve the program by achieving the general convergence conditions in Eqs. \eqref{eq:conv-conds-linear-2} and \eqref{eq:conv-conds-linear-1}, and additional convergence conditions in Eqs.~\eqref{eq:conv-conds-linear-3}  and \eqref{eq:conv-conds-linear-image} if measured data are consistent with the data model. This would (1) ensure that the CPD algorithm (containing the proximal operation in a closed-form with a finite number of algebraic calculation derived) is implemented correctly, (2) yield concrete knowledge about the convergence property of the CPD algorithm, and (3) provide an important benchmark of the convergence property for the heuristic NCPD algorithm adapted from the CPD algorithm.

\subsection{The CPD algorithm solving the convex optimization program} 

Using the scanning configuration described in Section~\ref{sec:simulation-configuration} above and the water and bone basis images of a numerical disk phantom of size 128$\times$128 shown in columns (a) and (c) in Fig. \ref{fig:veri-truth-image}, we generate data $\mathbf{g}^{[\mathcal{M}]}=\mathbf{g}^{[L]}(\mathbf{b}^{[\rm truth]})$ with the linear data model in Eq. \eqref{eq:linear-data-model} at 160 views uniformly distributed over 2$\pi$. We have assumed $\Delta\mathbf{g}_c=\mathbf{0}$. 
Also, truth-monochromatic image $\mathbf{f}_{m'}^{[\rm truth]}$ is known because truth image $\mathbf{b}^{[\rm truth]}$ is known, and so is its  TV (i.e., truth-TV) value.

\begin{figure}[htb]
 	\centering
 	\subfloat{
 		\centering
 		\includegraphics[width=.15\textwidth, trim={15 15 12 0}, clip]{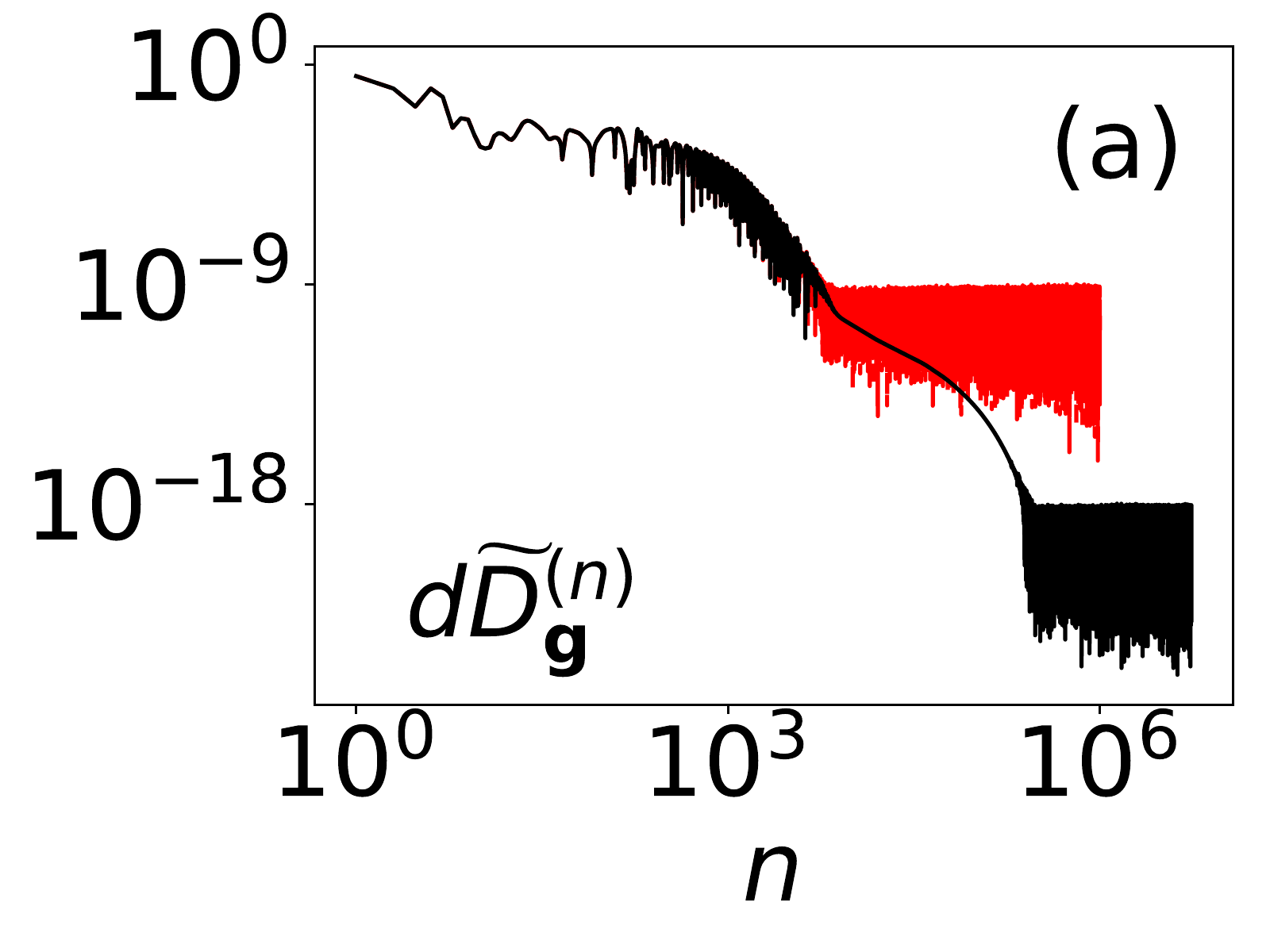}
 		\label{fig:veri-PD-data-der}
		}%
	\subfloat{
		\centering
 		\includegraphics[width=.15\textwidth, trim={15 15 12 0}, clip]{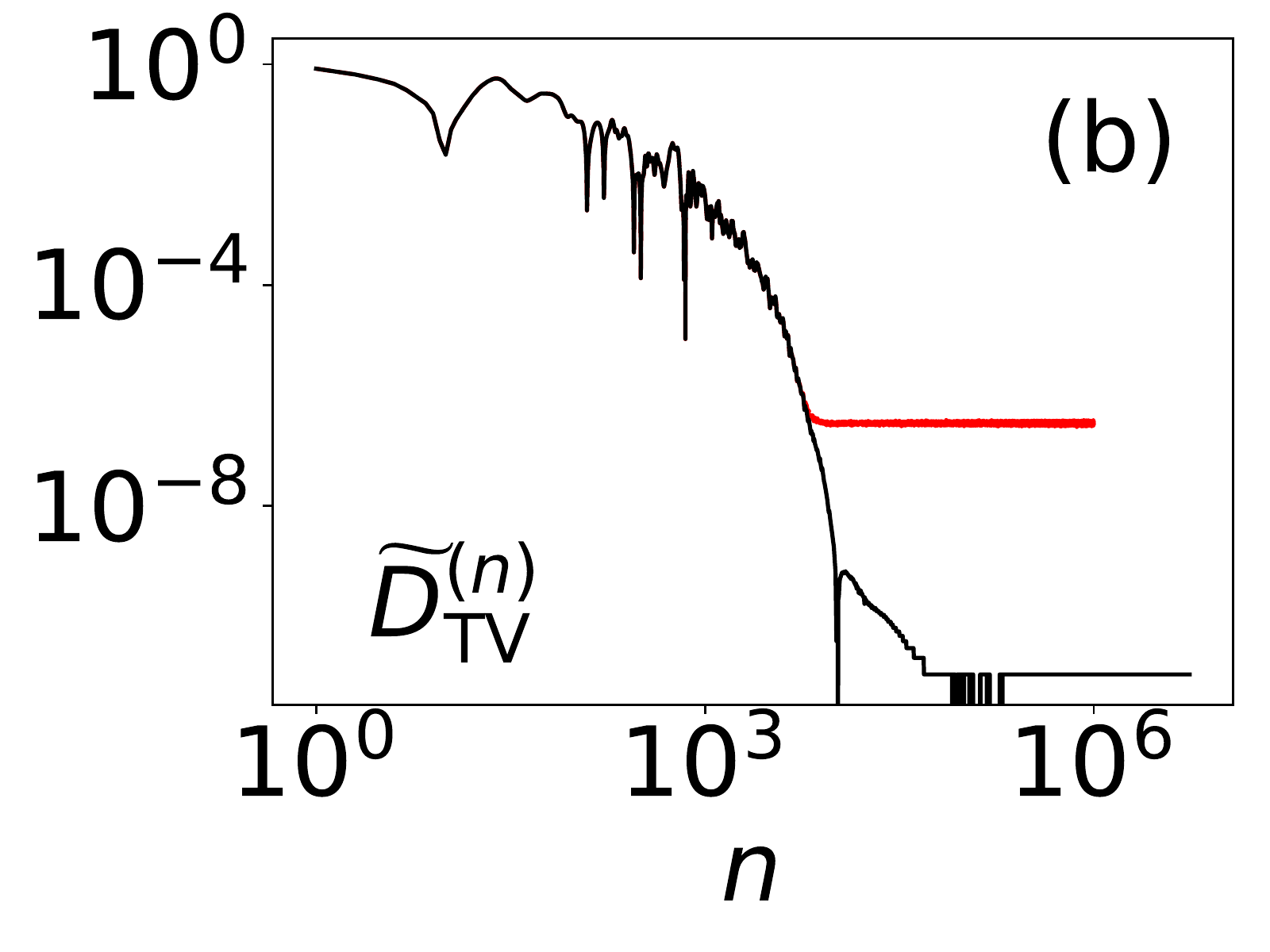}
 		\label{fig:veri-PD-tv}
		}
	\subfloat{
 		\centering
 		\includegraphics[width=.15\textwidth, trim={15 15 12 0}, clip]{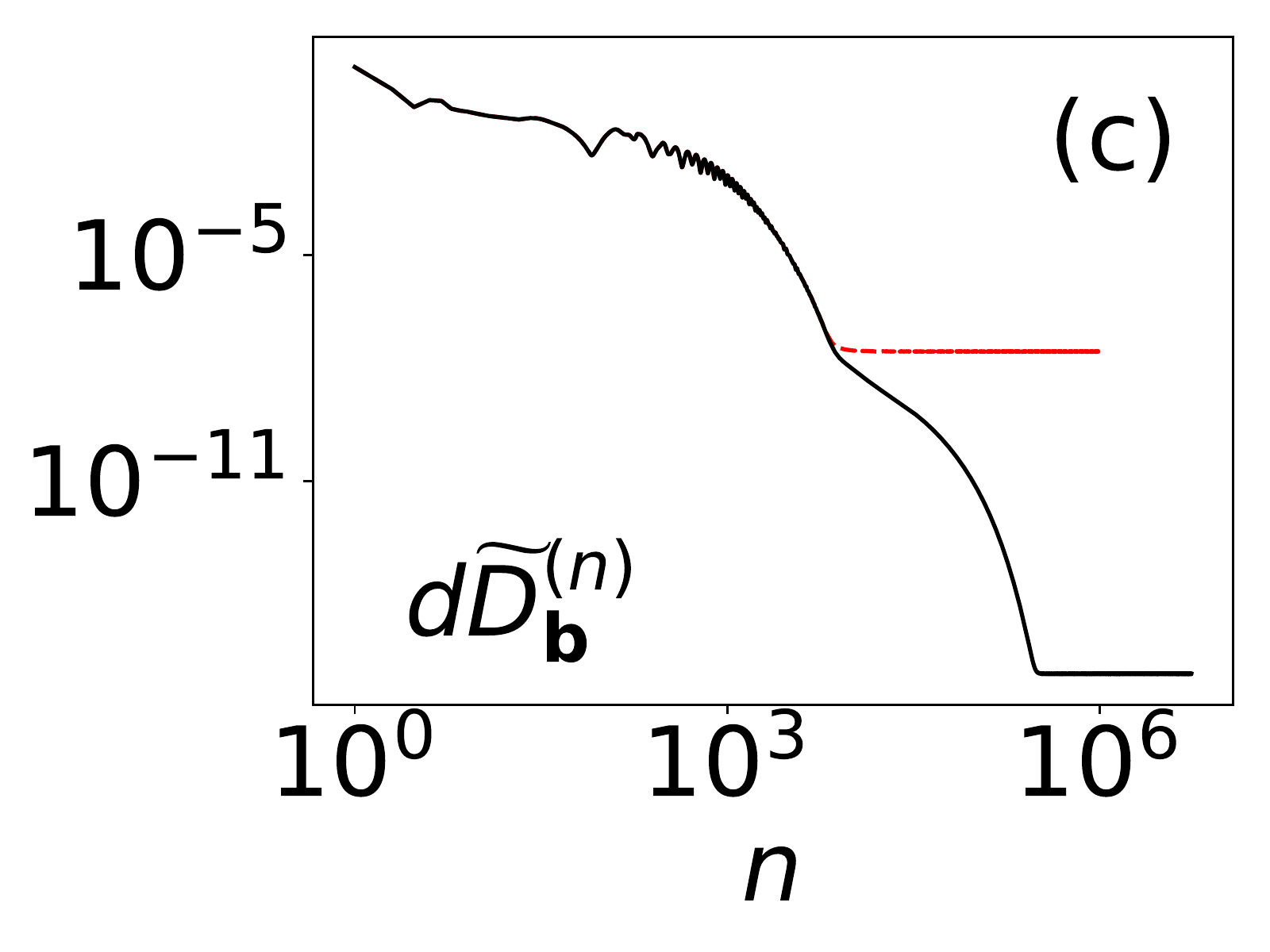}
 		\label{fig:veri-PD-img-der}
		}
	\\\vspace{-5pt}
	\subfloat{
 		\centering
 		\includegraphics[width=.15\textwidth, trim={15 15 12 12}, clip]{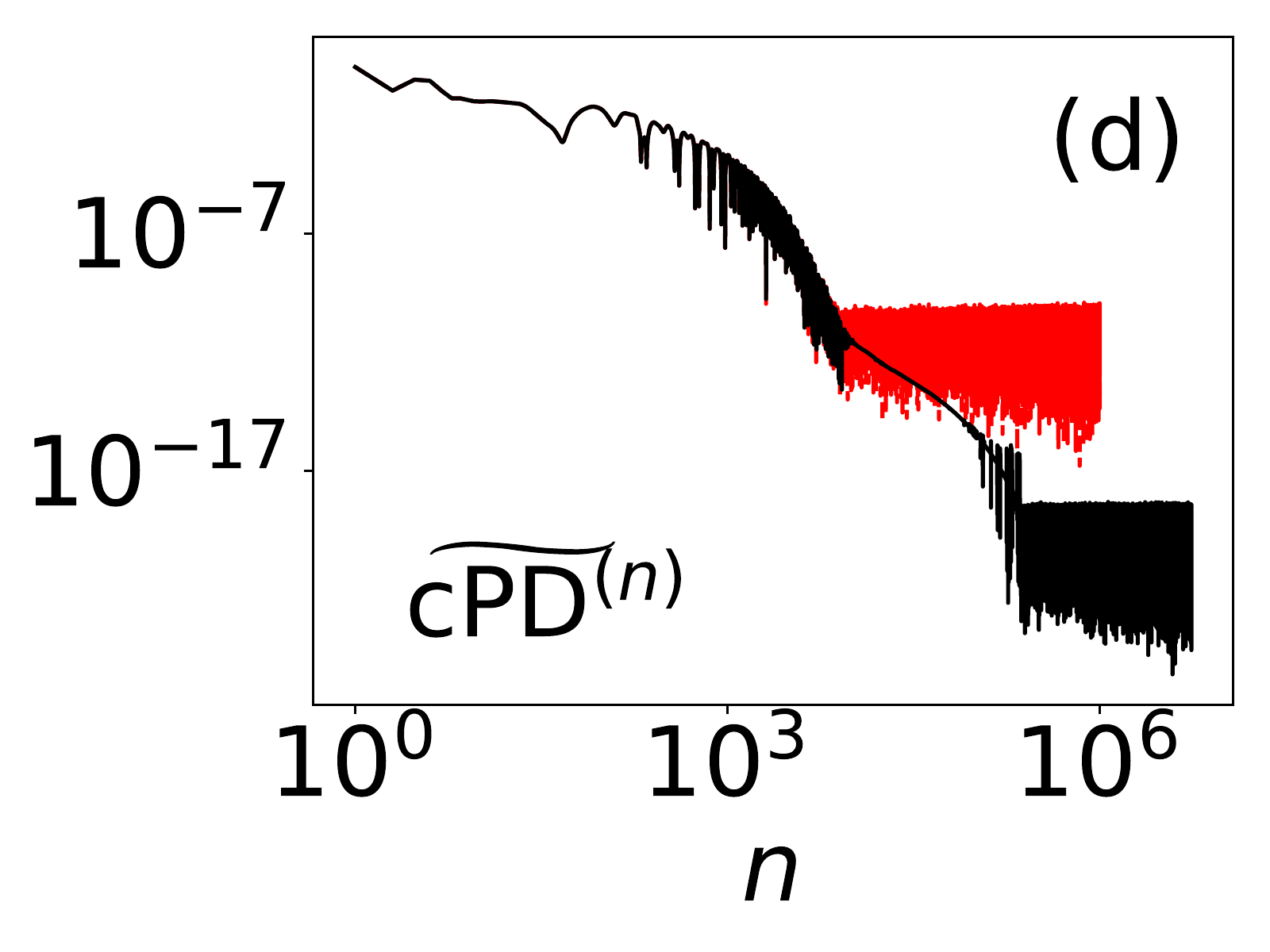}
 		\label{fig:veri-PD-cpd}
		}
	\subfloat{
 		\centering
 		\includegraphics[width=.15\textwidth, trim={15 15 12 12}, clip]{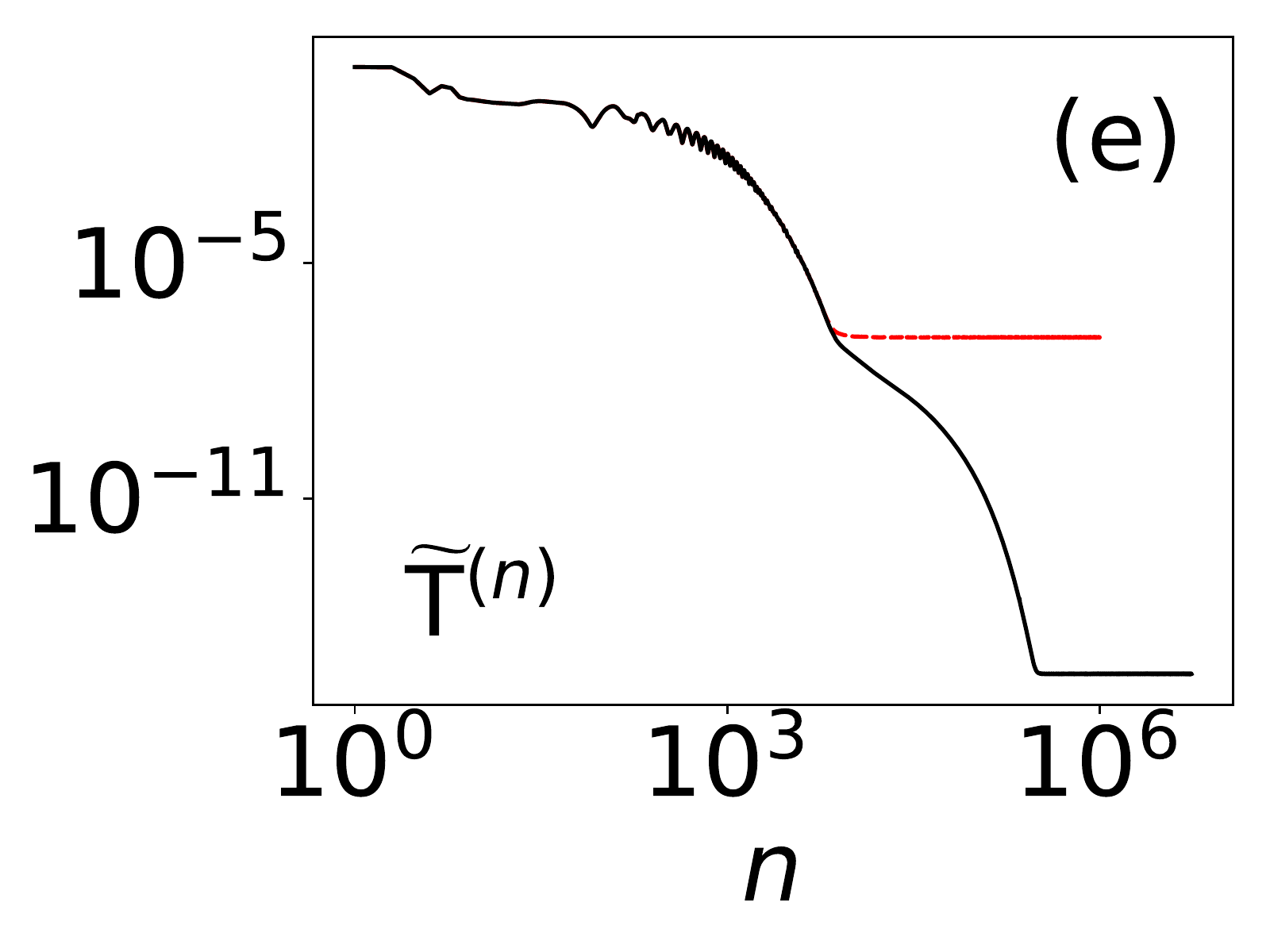}
 		\label{fig:veri-PD-transversality}
		}%
    \subfloat{
		\centering
 		\includegraphics[width=.15\textwidth, trim={15 15 12 12}, clip]{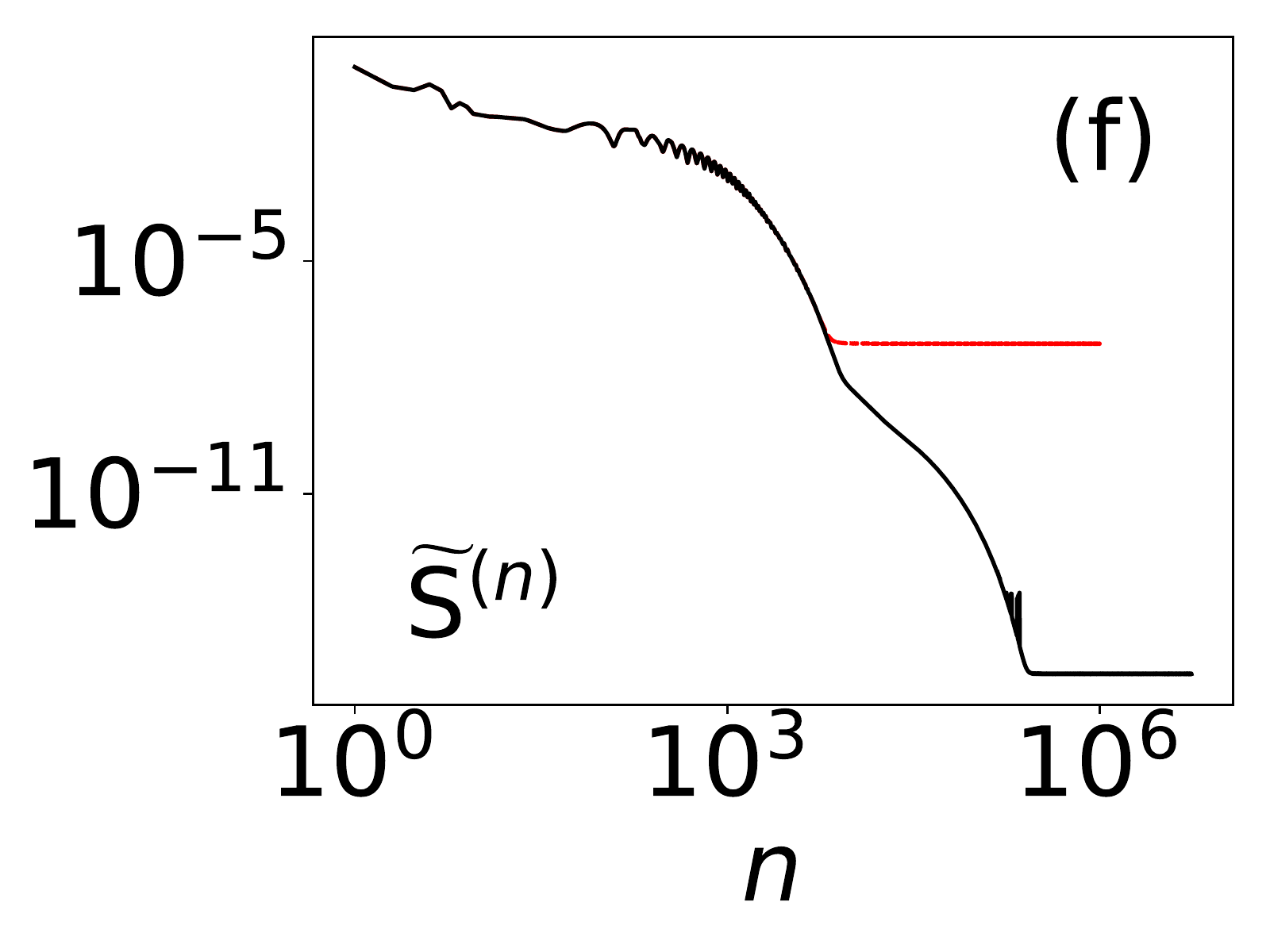}
 		\label{fig:veri-PD-splitting}
		}
\\\vspace{-5pt}
		\subfloat{
 		\centering
 		\includegraphics[width=.15\textwidth, trim={15 15 12 0}, clip]{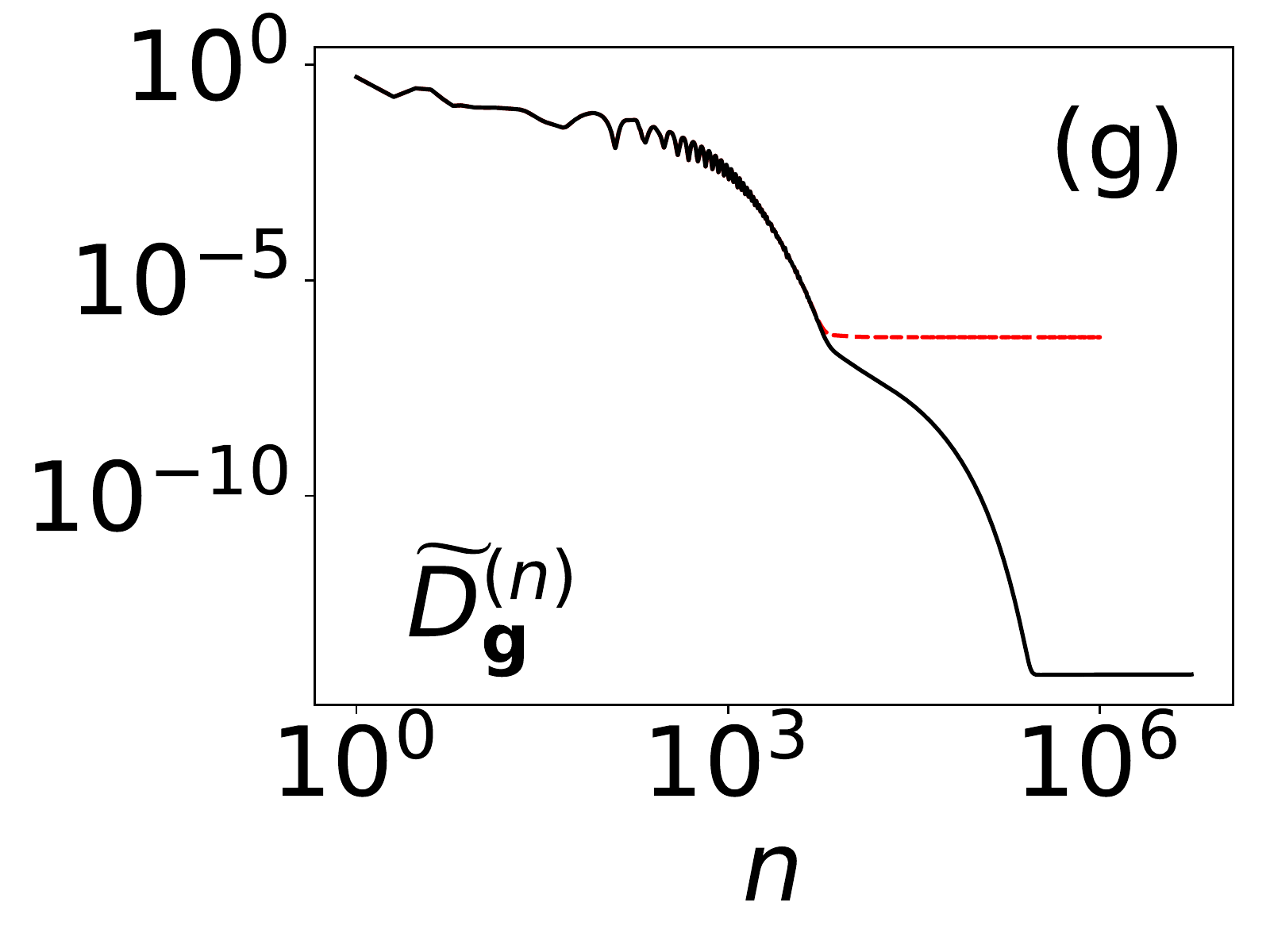}
 		\label{fig:veri-PD-data}
		}
	\subfloat{
 		\centering
 		\includegraphics[width=.15\textwidth, trim={15 15 12 12}, clip]{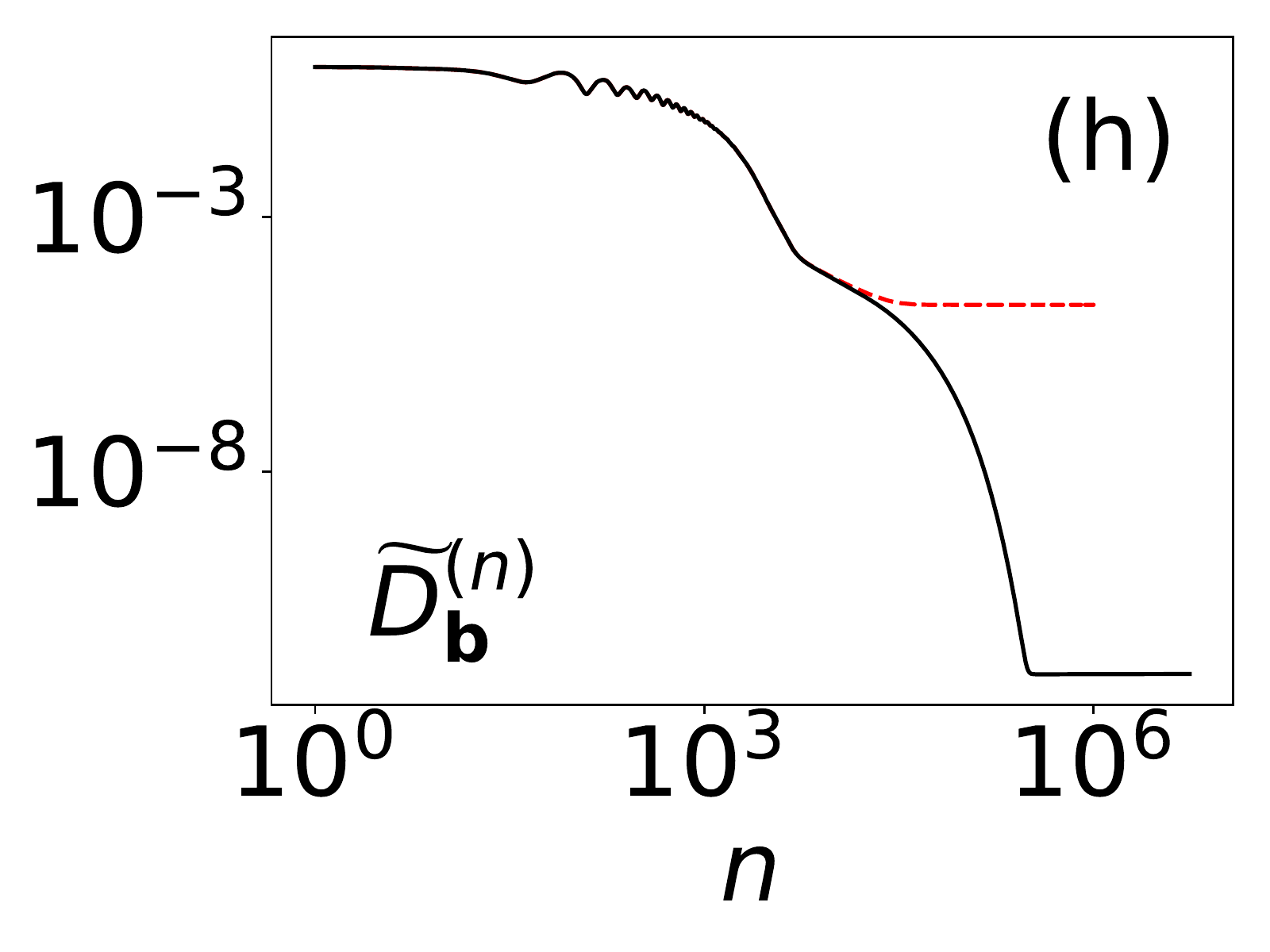}
 		\label{fig:veri-PD-img}
		}
 	\caption{Convergence metrics  
	(a) $d\widetilde{D}_{\mathbf{g}}^{(n)}$,
	(b) $\widetilde{D}_\mathrm{TV}^{(n)}$,
	(c) $d\widetilde{D}_\mathbf{b}^{(n)}$
	(d) $\widetilde{\text{cPD}}^{(n)}$,
	(e) $\widetilde{\text{T}}^{(n)}$,
	(f) $\widetilde{\text{S}}^{(n)}$,
	(g) $\widetilde{D}_{\mathbf{g}}^{(n)}$,
	and (h) $\widetilde{D}_\mathbf{b}^{(n)}$
	of the CPD algorithm, as functions of iteration $n$, from consistent data, obtained with single (dashed, red) and double (solid, dark) computer precision.
}
 	\label{fig:veri-PD}
\end{figure}

Applying the CPD algorithm, along with $\gamma$ that is the truth-TV value, to consistent data $\mathbf{g}^{[\mathcal{M}]}$, we compute the convergence conditions in Eqs. \eqref{eq:conv-conds-linear-2}, \eqref{eq:conv-conds-linear-1}, and \eqref{eq:conv-conds-linear-3} and plot them as functions of iteration $n$ in panels (a)-(c), (d)-(f), $\& $ (g) of Fig.~\ref{fig:veri-PD}, respectively. It can be observed that the consistently decaying convergence curves (dashed, red) obtained demonstrate that the CPD algorithm can solve the non-convex optimization program in Eq.~\eqref{eq:opt-constr-ln} in terms of achieving the convergence conditions. 

We have also performed a study with inconsistent data $\mathbf{g}^{[\mathcal{M}]}$ that are obtained by addition of Poisson noise to consistent data $\mathbf{g}^{[L]}(\mathbf{b}^{[\rm truth]})$ used in the verification study above. As a result, Eq. \eqref{eq:conv-conds-linear-3} can no-longer be used as a convergence condition in the study in which measured data (i.e., noisy data here) are inconsistent with the linear data model, i.e., $\mathbf{g}^{[\mathcal{M}]}\neq\mathbf{g}^{[L]}(\mathbf{b}^{[\rm truth]})$. The study results, which are not shown here, again demonstrate that the CPD algorithm can solve the non-convex optimization program in Eq. \eqref{eq:opt-constr-ln} in terms of achieving the general convergence conditions in Eqs. \eqref{eq:conv-conds-linear-2} and \eqref{eq:conv-conds-linear-1}.

\subsection{The CPD algorithm inverting the linear data model} 
While the CPD algorithm is demonstrated above to solve the convex optimization program  in Eq. \eqref{eq:opt-constr-ln} above, it remains to be shown, however, as to whether it can accurately invert the linear data model in Eq. \eqref{eq:linear-data-model}, and as to what data conditions under which it can. The verification study below is thus to demonstrate that under appropriate data condition, the CPD algorithm can invert the linear data model in Eq. \eqref{eq:linear-data-model} in terms of achieving Eq. \eqref{eq:conv-conds-linear-image}.

In panel (h) of Fig. \ref{fig:veri-PD}, we display $\widetilde{D}_\mathbf{b}^{(n)}$ (dash, red) computed from reconstructed images $\mathbf{b}^{(n)}$ as a function of iteration $n$. It can be observed that metric $\widetilde{D}_\mathbf{b}^{(n)}$  decreases consistently until the four-byte floating-point precision is reached, thus demonstrating numerically that under the full-scan data condition, the CPD algorithm can invert the linear data model in Eq. \eqref{eq:linear-data-model}.

\subsection{Numerical demonstration of achieving convergence conditions} 
It should be noted that in Fig.~\ref{fig:veri-PD}, the convergence curves (dashed, red) decay until around $\sim 10^4$ iterations where we believe that four-byte floating-point precision (i.e., single floating-point precision) is reached. To confirm this observation, we then perform the same study, but with eight-byte floating-point precision (i.e., double floating-point precision.) It can be observed that the convergence curves (solid, dark) continue their decaying trends passing $\sim 10^4$ iterations until reaching the double floating-point precision around $\sim 10^6$ iterations. The convergence curves are plotted on log-log scales for revealing unambiguously not only their values, but also equally importantly their decaying trends. The result verifies that the convergence conditions are achieved up to the single floating-point precision around $\sim 10^4$ iterations. We have also conducted the studies in Section~\ref{sec:convergence-verification-study} for the NCPD algorithm with the double floating-point precision, and results similar to those observed for the CPD algorithm have been obtained.

\section*{Acknowledgements}
This work is supported in part by NIH R01 Grant Nos. EB026282 and EB023968, and the Grayson-Jockey Club Research.

\bibliographystyle{elsarticle-harv} 
\bibliography{journals-abrv,IEEEabrv,spectral_algo}





\end{document}